\documentclass[11pt]{article}
\usepackage{amsfonts,amsmath}
\usepackage{amssymb}
\usepackage{hyperref}
\usepackage{xcolor}
\usepackage{bm}

\usepackage[titletoc,title]{appendix}

\numberwithin{equation}{section}

\usepackage{amsthm,mathtools}
\usepackage{icomma}

\renewcommand{\a}{\alpha}
\newcommand{\w}{\omega}
\newcommand{\D}{\mathbb{D}}
\newcommand{\dd}{\mathrm{d}}
\newcommand{\E}{\mathcal{E}}
\newcommand{\B}{\mathcal{B}}
\newcommand{\V}{\mathcal{V}}

\renewcommand{\b}{\beta}
\renewcommand{\r}{\rho}
\renewcommand{\l}{\lambda}
\newcommand{\s}{\sigma}

\newcommand{\z}{\zeta}

\newcommand{\tr}{\vartriangleright}
\newcommand{\ie}{{\it i.e.} }

\def\mathclap#1{\text{\hbox to 0pt{\hss$\mathsurround=0pt#1$\hss}}}
\newcommand{\fint}[1]{\int\limits_\mathclap{{\hphantom{#1} #1}}}

\newcommand{\tO}{\tilde{\Omega}}

\begin{document}

\begin{flushright}
\vspace{1mm}
 FIAN/TD/10--25\\
\vspace{-1mm}
\end{flushright}\vspace{1cm}

\begin{center}
{\large\bf Quadratic Corrections to the Higher-Spin Equations by the\\ Differential Homotopy Approach}

\vspace{1 cm}

{\bf P.T.~Kirakosiants\textsuperscript{1}, D.A.~Valerev\textsuperscript{1} and M.A.~Vasiliev\textsuperscript{1,2}}\\

\vspace{0.5 cm}
 \textsuperscript{1}I.E. Tamm Department of Theoretical Physics, Lebedev Physical Institute, \\ Leninsky prospect 53, 119991, Moscow, Russia

\vspace{1 cm}
 \textsuperscript{2}Moscow Institute of Physics and Technology, \\ Institutsky lane 9, 141700, Dolgoprudny, Moscow region, Russia\\

\end{center}

\begin{abstract}

The recently proposed differential homotopy approach to the analysis of nonlinear 
higher-spin theory is developed. The Ansatz is extended to the form applicable in the second order of the perturbation theory and general star-multiplication formulae are derived. The relation of the shifted homotopy and differential homotopy formalisms is worked out.
Projectively-compact spin-local quadratic (anti)holomorphic vertices in the one-form  sector of higher-spin equations are obtained within the differential homotopy formalism.

\end{abstract}

\newpage
\tableofcontents
\newpage

\section{Introduction}

\subsection{General background}
The theory of higher-spin (HS) fields is one of fundamental areas of
 theoretical physics for many years. In 1936, Dirac in the seminal work \cite{Dirac} proposed equations for free HS fields later investigated by various authors including Sakata and Yukawa \cite{Sakata_Yukawa}, Fierz and Pauli \cite{Fierz, Pauli_Fierz}, Hagen and Singh \cite{Hagen_Singh_I, Hagen_Singh_II}, and others. For instance, Fierz \cite{Fierz} used Dirac's equations to determine the quantum statistics of particles of different spins. In the joint paper \cite{Pauli_Fierz} Fierz and Pauli  formulated equations for HS particles in an external electromagnetic field. They also examined the massless limits of these equations, showing that for spin 1 they reduce to Maxwell equations while for spin 2 they match the linearized Einstein equations.

Rarita and Schwinger described spin 3/2 fields in \cite{Rarita_Schwinger},
Fradkin formulated a theory of spin 5/2 and HS fermions  \cite{Fradkin},  while  Schwinger considered spin 3 \cite{Schwinger}. Massive fields of all spins
in four-dimensional space-time were described by Singh and Hagen
\cite{Hagen_Singh_I, Hagen_Singh_II}.
In 1978, Fronsdal \cite{Fronsdal}
found gauge symmetry of free  massless HS  equations, derived the action generating these equations of all spins, and proposed a generalized Gupta program: the search for nonlinear HS theory that would reduce to the known free (linear) cases in the weak-field limit. For the massless fields the gauge symmetry serves as the main guiding principle.

In the sixties and seventies of the XX century  no-go theorems had already cast doubt on the consistency of interacting HS gauge theory. Among these was Weinberg's low-energy theorem \cite{Weinberg}, which imposes stringent constraints on the coupling constants of soft particles. For spins greater than two, these constraints become so restrictive that the couplings must vanish, leading Weinberg to conclude that  HS particles could exist but not interact at low energies. Analogously,
the Coleman-Mandula theorem \cite{Coleman_Mandula} and its supersymmetric
Haag-Lopuszanski-Sohnius  extension \cite{Haag} restricted the symmetries of the $S$-matrix in the $4d$ Minkowski space, ruling out Fronsdal's gauge symmetries. Additionally, the Aragone-Deser argument \cite{Aragon} prohibits minimal gravitational coupling for HS fields. A detailed review of these results and their implications for HS gauge theory can be found in \cite{BBS}.

Fortunately, the assumptions underlying these no-go theorems can be circumvented. Nontrivial interactions of massless HS particles with gravity and themselves
on an anti-de Sitter (AdS) background were first constructed in  \cite{Fradkin_1, Fradkin_2}. Crucially, available no-go arguments relied on the implicit assumption that a HS theory can be formulated in the flat Minkowski space.
AdS space, with its non-zero cosmological constant, is fundamentally curved.
 A key observation was that the HS vertex in presence of gravity was non-minimal (with the number of field derivatives growing with spin), thus evading the Aragone-Deser obstruction.

In 1992, full nonlinear HS equations fulfilling Fronsdal's program were formulated in \cite{Vasiliev:1992av}. These equations allow perturbative derivation of interaction vertices. The question of vertex locality is one of the central issues in understanding this theory. The commutation relations of the spin-$s$ HS algebra generators $T_{s_i}$ have the structure \cite{Fradkin:1986ka,Fradkin:1987ah},
\begin{equation}
    \left[T_{s_1}, T_{s_2}\right] \sim T_{s_1 + s_2 - 2} + T_{s_1 + s_2 - 4} + \dots + T_{|s_1-s_2| + 2}\,.
    \label{comm_eqs}
\end{equation}
As a result, introducing a spin $s > 2$ field  necessitates an infinite tower of HS fields as was originally pointed out in \cite{Berends:1984wp}, \cite{Berends:1984rq}. This complicates the definition of locality: while local theories with finitely many fields require vertices with a finite number of derivatives, theories involving infinitely many fields demand a more refined approach. For fields $\left\{ \phi_{s_i} \right\}$ (where $s_i$ denotes spin), governed by equations with coefficients $a^{n_1, \dots, n_k}$,
\begin{equation}
    \sum_{k=0, l=1}^{\infty} a^{n_1, \dots, n_k} (s_0, s_1, \dots, s_l)\partial_{n_1} \dots \partial_{n_k} \phi_{s_1} \dots \phi_{s_l} = 0,
\end{equation}
an interaction is called local if there exists a $K$ such that $\forall k > K$, the coefficients $a^{n_1, \dots, n_k}$ vanish for  any spins $s_i$.
The space-time spin-locality further requires that for each set of spins, there exists a $K(s_0, \dots, s_l)$ such that $\forall k > K(s_0, \dots, s_l)$, the coefficients $a^{n_1, \dots, n_k}(s_0, \dots, s_l)$ vanish. The definitions of spinor spin-local \cite{GelfondVasiliev_1910} and projectively compact \cite{Vasiliev:2022med} vertices are currently among the most useful locality concepts in HS theory.

The situation is further complicated by field redefinitions that can render seemingly non-local vertices local. Development of shifted homotopy approach \cite{Gelfond:2018vmi}-\cite{Gelfond:2021two} made it possible to achieve significant progress in the construction of lower order interacting vertices. In particular, locality required a specific shift-parameter $\beta \rightarrow -\infty$ limit. Based on this limit, a new {generating} system of nonlinear HS holomorphic equations that leads to local vertices in all orders has been proposed by Didenko \cite{Didenko} {(see also \cite{Korybut:2025vdn}; all vertices at the unconstrained level for the bosonic theory in any dimension were derived in \cite{Didenko:2024zpd})}. Being distinct from the original system \cite{Vasiliev:1992av}, this one is anticipated to emerge as its $\beta \rightarrow -\infty$ limit, although the precise connection is yet unknown. Independently, the way of constructing local vertices in the self-dual (also referred to as chiral) theory has been proposed in \cite{Sharapov:2022wpz, Sharapov:2022nps}.

However, a question of locality of the full nonlinear system  with both holomorphic and antiholomorphic sectors is still open. Extension of the shifted homotopy approach to higher orders is both not general enough and technically complicated. Its appropriate generalization is provided by the differential homotopy approach \cite{Vasiliev:2023yzx}, that postpones  the integration over homotopy parameters to the very last step , \ie when calculating the interaction vertex. In this approach the shift and homotopy parameters  are treated as coordinates of the {\it space of homotopy parameters} integration over which is performed in the end. Such a generalization extends the space of solutions of the perturbative equations hence providing alternative ways for exploring  locality of the HS theory. In \cite{Vasiliev:2023yzx}  efficiency of this approach was
demonstrated in the analysis of the first-order corrections to the equations on  one-forms  and  second-order corrections to the equations on zero-forms.

The aim of this paper is to extend this approach to the next order  to derive quadratic corrections to the equations on the one-forms in the holomorphic sector of the theory. This will demand an appropriate generalization of the
differential homotopy approach to higher orders. In particular, it will be shown that the shifted homotopy approach is a particular case of the differential homotopy one. Also, we propose a modification of the Ansatz
of \cite{Vasiliev:2023yzx} appropriate for the second order computations and
derive new star-multiplication formulae applicable at any order.
One of the important advantages of the differential homotopy approach is that it allows one to obtain the final
results with arbitrary homotopy parameters including $\beta$ introduced in \cite{Didenko:2019xzz}
 directly without using auxiliary statements like $Z$-dominance Lemma of \cite{Gelfond:2018vmi} {(see also \cite{Didenko:2022eso})} that was heavily used in \cite{Didenko:2020bxd, Gelfond:2021two}.

The interaction vertices evaluated in this paper were previously derived using shifted homotopy approach in \cite{Didenko:2019xzz} where locality was achieved  in the $\beta \rightarrow -\infty$ limit. In our computations the locality
is also achieved  in the appropriate $\beta \rightarrow -\infty$ limit, where it coincides with the result of \cite{Didenko:2019xzz}.  The differential homotopy approach is much more general that the shifted homotopy one and is anticipated to be most appropriate for the analysis of higher orders both in the holomorphic (self-dual) sector of the HS theory and beyond. It has a number of advantages including automatic accounting of Schouten identities and the ability to control desired properties, such as locality of vertices.
The aim of this paper is to illustrate how the differential contracting homotopy approach works in the previously studied cases to elaborate further its relevant properties extending those presented in \cite{Vasiliev:2023yzx} for the lower-order corrections. At higher orders the fundamental Ansatz needs further modifications elaborated in  Section \ref{2}, which is crucial for the analysis of quadratic corrections in the sector of field equations on the one-form HS gauge potentials.   In addition,  we find a number of new formulae like the star multiplication formulae elaborated in Section \ref{Any} for any number of product factors of the form of fundamental Ansatz. These formulae are anticipated to be important for the analysis of higher-order corrections.

The rest of the paper is organized as follows.

After a brief introduction to HS theory and the definition of the system of equations to be solved, Section 2 describes the Ansatz used to seek second-order perturbative solutions and derives its key properties. Specifically, it provides computationally efficient formulae for the star products.

Section 3 presents all calculations leading to quadratic interaction vertices in the holomorphic sector of one-forms. A concluding discussion of the results in Section 4 is followed by technical appendices:
\begin{itemize}
    \item Appendix A details star-product formulae for arbitrary canonical functions.
    \item Appendix B proves that any shift homotopy can be reformulated in terms of differential ones.
    \item Appendix C compares $B_2$ fields yielding projectively compact quadratic vertices in holomorphic zero-form sector via two approaches: shifted vs. differential homotopies.
    \item Appendix D derives the expression for the auxiliary field $W_2^{(2)}$.
    \item Appendix E demonstrates the $\beta \to -\infty$ limiting procedure for the interaction vertex.
\end{itemize}

\subsection{Higher-spin equations}

A nonlinear system of equations of motion for massless fields of all spins in $AdS_4$
studied in this paper is formulated as follows
\cite{Vasiliev:1992av}. \par
The fields propagate in a four-dimensional space-time with local coordinates $x$.
In addition, the space of four-component spinors $Y_A$ =$(y_{\alpha}, \bar{y}_{\dot{\alpha}})$ and  $Z_A$ =$(z_{\alpha}, \bar{z}_{\dot{\alpha}})$, where $\alpha, \dot{\alpha}  \in  \{ 1, 2\}$, is introduced.   De Rham differential is introduced both on the $x$ space,  $\dd_x =\dd x^n\frac{\partial}{\partial x^n}$, and on the $Z$ space,  $\dd_Z =\theta^A\frac{\partial}{\partial Z^A}$, where $\dd x^n$ and $\theta^A = \left(\theta^\a, \bar{\theta}^{\dot{\a}} \right)$ are anticommuting differentials,
\begin{equation}
    \{\dd x^m, \dd x^n\}= 0, \qquad \{\theta_A, \theta_B\} = 0, \qquad \{\theta_A, \dd x^n\} = 0.
\end{equation}
The Klein operators $K$ = $k,\bar{k}$ are defined to act as follows:
\begin{equation}
    \{k, y_{\alpha}\}= \{k, z_{\alpha}\}= 0, \quad [k, \bar{y}_{\dot{\alpha}}] = [k, \bar{z}_{\dot{\alpha}}] = 0, \quad k^2 = 1, \quad [k, \bar{k}]=0.
\end{equation}
(Analogously for $\bar{k}$.) A symplectic form is defined on the spinor spaces
\begin{equation}
    \epsilon_{AB} = (\epsilon_{\alpha \beta}, \epsilon_{\dot{\alpha} \dot{\beta}}), \quad  \epsilon_{\alpha \beta}= - \epsilon_{\beta \alpha}, \quad \epsilon _{\dot{\alpha} \dot{\beta}} = -\epsilon_{\dot{\beta} \dot{\alpha}}, \quad \epsilon_{1 2} = 1.
\end{equation}
The spinor indices are lowered and raised according to the following rule
\begin{equation}
    A_{\alpha} = A^{\beta}\epsilon_{\beta \alpha}, \quad A^{\alpha} = \epsilon^{\alpha \beta}A_{\beta}, \quad A_{\dot{\alpha}} = A^{{\dot\beta}}\epsilon_{{\dot\beta} {\dot\alpha}}, \quad A^{\dot{\alpha}} = \epsilon^{\dot{\alpha} {\dot\beta}}A_{{\dot\beta}}.
\end{equation}
The fields are described in terms of one-forms $W(Z, Y, K|x) = W_n(Z, Y, K|x) \dd x^n$, $S(Z, Y, K|x) = S_A(Z, Y, K|x)\theta^A$ and zero-forms $B(Z, Y, K|x)$. Note that the variables $Z$ were introduced to build interaction and the physical fields are $Z-$independent. So, physical fields are described by one-forms $\omega(Y, K|x) = W(0, Y, K|x)$ such that
\begin{equation}
    \omega(Y,K|x) = \sum_{j=0}^{1} \sum_{n=0,m=0}^{\infty} \frac{1}{2n!m!} \omega_{\alpha_1, \dots, \alpha_n, \dot{\alpha}_1, \dots, \dot{\alpha}_m}^{(j)}(x) y^{\alpha_1} \dots y^{\alpha_n} \bar{y}^{\dot{\alpha}_1} \dots  \bar{y}^{\dot{\alpha}_m} (k \bar{k})^j ,
    \label{omega_expan}
\end{equation}
where the components with $n+m = 2(s-1)$ correspond to the spin $s$ field (the components with $n = m$ describe the integer spin frame-like field, while those with $n = m \pm 1$ describe the half-integer spin frame-like field; the other components describe their derivatives); and zero-forms $C(Y, K|x) = B(0, Y, K|x)$,
\begin{equation}
    C(Y,K|x) = \sum_{j=0}^{1} \sum_{n=0,m=0}^{\infty} \frac{1}{2n!m!} C_{\alpha_1, \dots, \alpha_n, \dot{\alpha}_1, \dots, \dot{\alpha}_m}^{(j,1-j)}(x) y^{\alpha_1} \dots y^{\alpha_n} \bar{y}^{\dot{\alpha}_1} \dots  \bar{y}^{\dot{\alpha}_m} k^j \bar{k}^{1-j} \,.
    \label{C_expan}
\end{equation}
The components with $n-m = \pm 2s$ correspond to the spin $s$ field (components with $n = 0$ or $m = 0$ in the case of $s = 0$ and $s=\frac{1}{2}$ describe the
dynamical matter fields  while other components are expressed in terms of derivatives of the dynamical fields). Note that this dictionary is valid when, at the linear approximation, field equations correspond to the so-called Central On-Shell Theorem, which encodes the equations for free HS fields. The condition for the validity of this theorem at the first order is a necessary one for the solution of the nonlinear system \eqref{base_eq1}-\eqref{base_eq5}, that restricts homotopy procedure at this order. The full system of nonlinear HS equations is
 \cite{Vasiliev:1992av}
\begin{subequations}
\label{base_all}
\begin{align}
    &\dd_x W + W \star W = 0, \label{base_eq1} \\
    &\dd_x B + W\star B - B \star W = 0, \label{base_eq2} \\
    &\dd_x S + S \star W + W \star S= 0, \label{base_eq3} \\
    &S \star B - B \star S = 0, \label{base_eq4} \\
    &S \star S = i(\theta^{\alpha} \theta_{\alpha} + \bar{\theta}^{\dot{\alpha}} \bar{\theta}_{\dot{\alpha}} + B \star (\eta \gamma + \bar{\eta} \bar{\gamma}))\,,\label{base_eq5}
\end{align}
\end{subequations}
where $\eta$ is a complex parameter, the star-product operation $\star$ is introduced as follows:
\begin{equation}
    f(Z,Y) \star g(Z, Y) = \int f(Z+S, Y+S)g(Z-T, Y+T) e^{i S_A T^A} \; \dd^4 S \; \dd^4 T
    \label{star}
\end{equation}
and
\begin{equation}
    \begin{gathered}
        \gamma = e^{iz_{\alpha}y^{\alpha}} \theta^{\alpha} \theta_{\alpha} k, \quad f(Z, Y| \theta) \star \gamma = \gamma \star f(Z, Y| \theta), \\
        \bar{\gamma} = e^{i\bar{z}_{\dot{\alpha}}\bar{y}^{\dot{\alpha}}}\bar{\theta}^{\dot{\alpha}} \bar{\theta}_{\dot{\alpha}} \bar{k}, \quad f(Z, Y| \theta)  \star  \bar{\gamma} = \bar{\gamma}  \star  f(Z, Y| \theta).
    \end{gathered}
\end{equation}
Throughout this paper, we adopt the following normalization convention for the spinor variable integration measure:
\begin{equation}
    \dd^2 q := \frac{\dd q^1 \; \dd q^2}{2\pi},
\end{equation}
which leads, for example, to
\begin{equation}
    \int \dd^2s \; \dd^2t\; e^{is_\a t^\a} = 1.
\end{equation}
\par
The analysis of the system \eqref{base_eq1}-\eqref{base_eq5} is performed perturbatively. We will denote the order of perturbation by the lower subscript of the field (e.g., $S_1, W_2$, {\it etc.}). The vacuum solution is
\begin{equation}
    B_0 = 0, \quad S_0 = \theta^{\alpha} z_{\alpha} + \bar{\theta}^{\dot{\alpha}}\bar{z}_{\dot{\alpha}}.
    \label{vac}
\end{equation}
A remarkable feature of $S_0$ is that
\begin{equation}
    [S_0, f(Z)]_ \star  = -2i \dd_Z f(Z).
    \label{d_z-prop}
\end{equation} \par
Substitution of  \eqref{vac} into equations gives $W_0 = \omega(Y,K|x), B_1= C(Y,K|x)$. Further analysis reproduces the complete system of nonlinear equations for the physical HS fields:
\begin{equation}
    \dd_x \omega + \omega \star \omega = \Upsilon(\omega \omega C) +\Upsilon(\omega \omega C C) + \dots
    \label{eq_omega}
\end{equation}
\begin{equation}
    \dd_x C + \omega \star C - C \star \omega = \Upsilon(\omega C C) +\Upsilon(\omega C C C) + \dots
    \label{eq_C}
\end{equation}
within the expansion in powers of $C$. \par
The consistency of the full system along with the property \eqref{d_z-prop} imply that the equations, that contain $S$, have the form
\begin{equation}
    \dd_Z f(Z, Y, \theta) = g(Z, Y, \theta)
    \label{eq_dd}
\end{equation}
with some $\dd_Z$ closed $g$,
\begin{equation}
    \dd_Z g(Z, Y, \theta) = 0\,.
\end{equation}
The general solution to such equation has the form
\begin{equation}
    f(Z, Y, \theta) = \triangle_0 g(Z, Y, \theta) + \dd_Z \epsilon + h,
\end{equation}
where $h$ belongs to $\dd_Z$-cohomology and
\begin{equation}
    \triangle_0 g(Z, Y, \theta) = Z^A \frac{\partial}{\partial \theta^A} \int_0^1 \frac{\dd t}{t} g(tZ, Y, t\theta).
    \label{conv_homotopy}
\end{equation}
The choice of different $\dd_Z$-cohomology representatives $h$ leads to different interaction vertices in \eqref{eq_omega} and \eqref{eq_C}. The zero-dimensional cohomology consists of functions independent of $Z$ and thus depending only on $Y$, while the first cohomology is trivial. That is why differences in the choice of $h$ correspond to  field redefinitions. It turns out that some solutions lead to vertices with an infinite number of derivatives over spinor variables $Y$, which eventually leads to an infinite number of derivatives over the space-time variables.
(See, {\it e.g.}, \cite{Vasiliev:2023yzx}.)
\par
This raised a problem of  developing an approach that allows one to find solutions to equations of the form \eqref{eq_dd}, leading to spin-local vertices. To this end the so-called shifted homotopy method was developed in \cite{Gelfond:2018vmi, Didenko:2018fgx}, in which the solution
to the equation \eqref{eq_dd} was presented in the form
\begin{equation}
    \triangle_a g(Z, Y,  \theta) = \int_0^1 (Z+a)^A \frac{\partial}{\partial \theta^A} \frac{\dd t}{t} g(tZ-(1-t)a, Y, t\theta),
    \label{shifted}
\end{equation}
where $a$ is any $Z-$independent shift. The approach was further developed with the introduction of the $\beta$-parameter \cite{Didenko:2019xzz,Didenko:2020bxd}, leading to the construction of spin-local vertices in lower orders of the perturbation theory
\cite{Didenko:2018fgx,Didenko:2019xzz,Gelfond:2021two}.
However, a generalization
of these results to higher orders was not straightforward. In particular, it was speculated that the analysis would simplify the selection of shift parameters at a certain step if they would  depend on the  parameters such as the integration variable $t$ of the previous steps, which is not feasible in the shifted homotopy approach since the integration has already been performed.\par
\section{Differential homotopy}
\subsection{The space of homotopy parameters}
\label{sec:Chapter1} \index{Chapter1}
In the differential homotopy  approach of \cite{Vasiliev:2023yzx}, the shift parameters $a$ and integration variables like $t$ in \eqref{shifted} are treated as coordinates of some space $M=\{h^a\}$. This space comprises two parts:
\begin{enumerate}
    \item A spinorial part, whose coordinates are spinors.
    \item A homotopy part, whose coordinates are scalar homotopy parameters.
\end{enumerate}
Crucially, the homotopy part must be compact to ensure the convergence of relevant expressions. Moreover, in practice, this homotopy part turns out to be contractible. Note, that integration variables $s_\a$ and $t_\a$ in \eqref{star} now belong to $M$, which means integrals over them are preserved until the final stage of computations. For this reason, we will employ the integral kernel of the previously introduced star product \eqref{star}
\begin{equation}
    f(Z,Y) * g(Z, Y) =  f(Z+S, Y+S)g(Z-T, Y+T) e^{i S_A T^A} \; \dd^4 S \; \dd^4 T
    \label{star_kernel}
\end{equation}
and use \eqref{star_kernel} instead of \eqref{star} in our computations.

The differential $\dd_h =\dd h^a\frac{\partial}{\partial h^a}$ in $M=\{h^a\}$ anticommutes with $\theta^A$. The total differential is
\begin{equation}
    \dd = \dd_Z + \dd_h.
\end{equation}\par
Since $h^a$ belong to a contractible space, all closed $h-$differential forms of positive degrees whose integral vanishes are exact. Closed zero-forms are
$h^a$ independent functions.\par
Equations \eqref{eq_dd} are generalized  to
\begin{equation}
    \dd f(Z, Y, h, \theta, \dd h) = g(Z, Y, h, \theta, \dd h),
    \label{eq_dd_ext}
\end{equation}
with the consistency condition
\begin{equation}
    \dd g(Z, Y, h, \theta, \dd h) = 0.
\end{equation}\par
{Following \cite{Vasiliev:2023yzx}, we use the integral symbol solely to control the order of integration variables, where a change in order (equivalent to choosing the orientation of $M$) determines the sign of an expression. Integration itself is performed only at the last step when the interaction vertex is calculated. In this convention we have the following property}
\begin{equation}
    \int_M \equiv \int_{h^1} \dots \int_{h^k} \equiv \int_{h^1 \dots h^k} = \mathrm{sign}(\sigma) \int_{\sigma(h^1) \dots \sigma(h^k)},
\end{equation}
where $\sigma$ is a permutation of $k$ symbols. Additionally, the order of integral signs and differentials can be arranged arbitrarily with the commutation rules
\begin{equation}
    \dd \int_{h^1} = - \int_{h^1} \dd \quad \Rightarrow \quad \dd \int_{h^1 \dots h^k} = (-1)^k  \int_{h^1 \dots h^k} \dd\,.
\end{equation}
Analogously, for any $p$-form $\lambda$
\begin{equation}
    \lambda \int_{h^1 \dots h^k} = (-1)^{kp}  \int_{h^1 \dots h^k} \lambda.
    \label{commutation_with_integr}
\end{equation}

Also, it should be stressed that the integral of a differential form can be nonzero only if its degree coincides with the dimension of the integration cycle. This will play an important role in the sequel.

\subsection{Weak terms}\label{subsec:weak_terms}

A differential form $F(Z,Y,h)$, understood to have implicitly assumed dependence on $\theta$ and $\dd h$, is called {\it weak} iff
\begin{equation}
\label{weak}
    \int_{\{h\}} F = 0,
\end{equation}
where \{h\} represents the space of homotopy parameters and additional spinor variables.
We write $F \cong G$ when $F$ and $G$ differ by a weak term, meaning their integrals over ${h}$ coincide.
It is easy to see that weak terms are closed under total differential and star products. Indeed, if a form $F$ is weak ({\it i.e.}, satisfies \eqref{weak}), then its derivative $\dd F$ is also weak by the Stokes' theorem,
\begin{equation}
    \int_{\{h\}} \dd_h F = 0\,,
    \label{eq:d_tF=0}
\end{equation}
since
\begin{equation}
    \int_{\{h\}} \dd F = \int_{\{h\}} (\dd_Z F + \dd_h F) = \int_{\{h\}} \dd_Z F = (-1)^{|\{h\}|} \dd_Z \int_{\{h\}} F = 0\,,
    \label{eq:d_weak_term}
\end{equation}
where $|\{h\}|$ counts the number of integration variables $h^a$.

Analogously, given a weak form $F(Z,Y,h_1)$ and an arbitrary form $G(Z, Y, h_2)$ such that its homotopy parameters $\{h_2^a\}_a$ do not contain any of $h_1^a$, their star product $F*G$ is weak. This follows from
\begin{multline}
    \int_{\{h_1\}} \int_{\{h_2\}} \int_{S T} (F * G) (Z, Y, h_1, h_2) = \\
    = \int_{\{h_1\}} \int_{\{h_2\}} \int_{S T} \dd^4 S \; \dd^4 T \; F(Z+S, Y+S, h_1) G(Z-T, Y+T, h_2) e^{iS_A T^A} = \\
    = (-1)^{|\{h_1\}| \cdot |\{h_2\}|}\int_{\{h_2\}} \int_{S T} \dd^4 S \; \dd^4 T \; \left(\int_{\{h_1\}} F(Z+S, Y+S, h_1) \right) G(Z-T, Y+T, h_2) e^{iS_A T^A} = 0,
\end{multline}
where $|\{h_i\}|$ is the number of parameters in the $i$-th group.

Also, let us notice that there are some terms that can be easily identified as weak. At least they are the ones
which represent the properties of a standard star product. For example, a star commutator of a form $F$ with central element $\gamma$ is weak, just like the associator of any three forms
\begin{equation}
    [F, \gamma]_* =
    \left( F*\gamma - \gamma*F\right) \cong 0;
    \label{eq:gamma_com}
\end{equation}
\begin{equation}
    (F * G) * H - F * (G * H) \cong 0.
    \label{eq:assoc}
\end{equation}
Previously, the {\it l.h.s.} of \eqref{eq:gamma_com} and \eqref{eq:assoc} were identically zero, but now they vanish only in the weak sense,
as the equality requires integration over auxiliary spinor variables $S, T$ to be satisfied.

\subsection{Fundamental Ansatz}
As shown in \cite{Vasiliev:2023yzx}  (see also text around \eqref{S1}),
before solving equation for $S_2$, it is convenient to represent all encountered functions, starting from $S_1$, in the form
\begin{equation}
    f \equiv f(g_1,\dots,g_n) = \fint{p r u v  \tau \rho \beta \sigma^l}\mu_0 \mu(\tau, \rho, \beta, \sigma^l)\dd \Omega^2  \E(\Omega) g_1(r_1)\bar{*}\dots \bar{*}g_n(r_n)k,
    \label{anz}
\end{equation}
where $l \in \overline{1, n}$, $g_l$ are functions of spinor variables $r^{\alpha}_l$  (e.g., $C(r)$ or $\w(r)$) with implicit dependence on $\bar{y}, K,$ and space-time coordinates $x$;
$\tau, \rho, \beta$ and $\sigma^l$ are integration parameters restricted to a compact domain by the measure factor of $\mu(\tau, \rho, \beta, \sigma^l)$,
\begin{equation}
    \mu_0 = \prod_{i=1}^{n}(\dd^2p_i \dd^2r_i) \dd^2 u \dd^2 v\,,
    \label{mu0_anz}
\end{equation}
\begin{equation}
    \begin{gathered}
        \Omega^{\alpha} = \tau z^{\alpha}-(1-\tau)(p^{\alpha}(\sigma)- \beta v^{\alpha} + \rho(y^{\alpha}+p_+^{\alpha}+u^{\alpha})),
    \end{gathered}
\end{equation}
\begin{equation}
    \dd \Omega^2 = \dd \Omega^{\alpha} \dd \Omega_{\alpha},
\end{equation}
\begin{equation}
    \begin{gathered}
        {\E}({\Omega})  = \exp i \left({\Omega}_{\alpha}(y+p_++u)^{\alpha}+u_{\alpha} v^{\alpha} - \sum_{i<j} p_{i\alpha} p_j^{\alpha} - \sum_{i=1}^n p_{i\alpha} r_i^{\alpha}  \right),
    \end{gathered}
\end{equation}
\begin{equation}
    p_+^{\alpha} = \sum_{i=1}^n p_i^{\alpha}, \quad p^{\alpha}(\sigma) = \sum_{i=1}^n p_i^{\alpha} \sigma^i,
\end{equation}
$\bar{*}$ denotes the $*$-product with respect to antiholomorphic spinor variables $\bar{y}_{\dot{\alpha}}$ on which $g_l$ depend.

Functions of the form (\ref{anz}) will be referred to as having canonical form or simply {\it canonical}.
\eqref{anz} shares the following remarkable property. Since the index $\alpha$ takes two values and $\dd\Omega_\a$ is a one-form, $\dd\Omega_\a  \dd \Omega_{\b} \dd \Omega_\gamma = 0$. Therefore, the action of $\dd$ on $\E$ preceded by $\dd\Omega^2$ does not contribute, \ie $\dd$ only acts on the measure factors like $\mu$:
\begin{equation}
    \dd f_{\mu} = (-1)^Nf_{\dd \mu},
    \label{d_measure}
\end{equation}
where index $\mu$ denotes the measure in \eqref{fundament} and $N$ is the number of integration parameters.
\par

\subsubsection{Star product of two canonical functions}
When solving the equation for $S_2$, one encounters a star product of the expressions of the form \eqref{anz}. Hence it is useful to have a formula for such a product. It has the form
\begin{multline}
    f_1(g_1,\dots,g_n)*f_2(g_{n+1},\dots,g_{n+m}) = (-1)^{m+(1+m)deg(\mu_1)+(1+m+deg(\mu_2))\sum_{i=1}^n deg(g_i)}\times\\\times \fint{p r u v s t \tau_1 \rho_1 \beta_1 \sigma_1^j \tau_2 \rho_2 \beta_2 \sigma_2^j} \mu_0^{(12)} \D(\sigma_1^{n+m}-(1-\beta_1))\dots \D(\sigma_1^{n+1}-(1-\beta_1))\D(\sigma_2^n-(1-\beta_2))\dots  \D(\sigma_2^1-(1-\beta_2)) \times \\ \times \mu_1(\tau_1, \rho_1, \beta_1, \sigma_1^l) \mu_2(\tau_2, \rho_2, \beta_2, -\sigma_2^l) \dd \Omega_1^2 \dd \Omega_2^2  \E(\Omega_1, \Omega_2) g_1(r_1)\bar{*}\dots \bar{*}g_{n+m}(r_{n+m}),
    \label{anz_mult}
\end{multline}
where
\begin{equation}
    \begin{gathered}
        \Omega_1^{\alpha} =\tau_1 z^{\alpha} - (1-\tau_1)(p^{\alpha} (\sigma_1) + (1-\beta_1) s^{\alpha} - \beta_1 v_1^{\alpha} + \rho_1(y^{\alpha}+p_+^{\alpha}+s^{\alpha}+u_1^{\alpha})), \\
        \Omega_2^{\alpha} =\tau_2 z^{\alpha} - (1-\tau_2)(p^{\alpha} (\sigma_2) - (1-\beta_2) t^{\alpha} - \beta_2 v_2^{\alpha} + \rho_2(y^{\alpha}-p_+^{\alpha}+t^{\alpha}+u_2^{\alpha})),
    \end{gathered}
\end{equation}
\begin{multline}
    {\E}({\Omega}_1, {\Omega}_2)  = \exp i \left({\Omega}_{1\alpha}(y+s+p_++u_1)^{\alpha} + {\Omega}_{2\alpha}(y+t-p_++u_2)^{\alpha} + s_{\alpha} t^{\alpha} +\right. \\ \left. + (s-t + p_+)_{\alpha}y^{\alpha}+u_{1\alpha}v_1^{\alpha}+u_{2\alpha}v_2^{\alpha}  - \sum_{i<j} p_{i\alpha} p_j^{\alpha} - \sum_{i=1}^{n+m} p_{i\alpha} r_i^{\alpha} \right),
\end{multline}
\begin{equation}
    \quad p_+^{\alpha} = \sum_{i=1}^{n+m} p_i^{\alpha}, \quad \mu_0^{(12)} = \prod_{i=1}^{n+m}(\dd^2p_i \dd^2r_i) \dd^2 u_1 \dd^2 v_1 \dd^2 u_2 \dd^2 v_2 \dd^2 s \dd^2 t,
\end{equation}
\begin{equation}
    \text{and} \quad \D(x) = \delta(x) \dd x.
\end{equation}
We point out that while $\Omega_1$ in $f_1$ initially depended only on $\{\sigma_1^i\}_{i=1}^n$,
the remaining variables $\{\sigma_1^i\}_{i>n}$ were absent in $f_1$. However, these
variables appear in $\Omega_1$ when considering the star product $f_1 * f_2$ in
\eqref{anz_mult}, where they all become expressed in terms of $1-\beta_1$. An analogous situation occurs for $\{\sigma_2^i\}_{i\leq n}$
in $\Omega_2$. It is also worth noting that the variables $s_\a$ and $t_\a$ of the star product are included in the set of coordinates $M$, so integration over them is postponed until the final step of the vertex calculations.
\par
Equality  \eqref{anz_mult} can be obtained as follows. Consider $f_1(g_1,\dots,g_n)*f_2(g_{n+1},\dots,g_{n+m})$. The labels 1 and 2 refer to $\Omega$, $\s$, $\b$, $\r$, $\tau$ and $p$, $u$, $v$ from $f_1$ and $f_2$, respectively.
Straightforward application of the star product \eqref{star_kernel} yields
\begin{equation}
    \begin{gathered}
        \Omega_{1\alpha} \rightarrow \tau_1 (z_{\alpha}+s_{\alpha}) - (1-\tau_1)(p_{\alpha} (\sigma_1) - \beta_1 v_{1\alpha} + \rho_1(y_{\alpha}+p_{(1)+\alpha}+s_{\alpha}+u_{1\alpha})) =  \Omega_{1\alpha}' + s_{\alpha}, \\
        \Omega_{2\alpha} \rightarrow \tau_2 (z_{\alpha}-t_{\alpha}) - (1-\tau_2)(p_{\alpha} (-\sigma_2) -  \beta_2 v_{2\alpha} + \rho_2(y_{\alpha}-p_{(2)+\alpha}+t_{\alpha}+u_{2\alpha})) = \Omega_{2\alpha}' - t_{\alpha},
    \end{gathered}
\end{equation}
where
\begin{equation}
    \begin{gathered}
         \Omega_{1\alpha}'= \tau_1 z_{\alpha} - (1-\tau_1)(p_{\alpha} (\sigma_1) + s_{\alpha} - \beta_1 v_{1\alpha} + \rho_1(y_{\alpha}+p_{(1)+\alpha}+s_{\alpha}+u_{1\alpha})), \\
        \Omega_{2\alpha}'  = \tau_2 z_{\alpha} - (1-\tau_2)(p_{\alpha} (-\sigma_2) - t_{\alpha}-  \beta_2 v_{2\alpha} + \rho_2(y_{\alpha}-p_{(2)+\alpha}+t_{\alpha}+u_{2\alpha}));
    \end{gathered}
\end{equation}
\begin{multline}
        {\E}({\Omega})  \rightarrow \exp i \big[(\Omega_{1\alpha}' + s_{\alpha})(y+s+p_{(1)+}+u_1)^{\alpha} + (\Omega_{2\alpha}' - t_{\alpha})(y+t-p_{(2)+}+u_2)^{\alpha} + s_{\alpha} t^{\alpha} +
        \\
        +u_{1 {\alpha}}v_1^{\alpha}+u_{2 {\alpha}}v_2^{\alpha} -\sum_{1\leq i < j \leq n} p_{i\a}p_j^\a - \sum_{n < i < j \leq n+m} p_{i\a}p_j^\a - \sum_{i=1}^{n+m} p_{i{\alpha}} r_i^{\alpha}  \big] =
        \\
        = \exp i \big[\Omega_{1\alpha}'(y+s+p_{(1)+}+u_1)^{\alpha} + \Omega_{2\alpha}'(y+t-p_{(2)+}+u_2)^{\alpha}+  s_{\alpha} p_{(1)+}^{\alpha}+t_{\alpha} p_{(2)+}^{\alpha}+
        \\
        + (s-t)_{\alpha} y^{\alpha}+s_{\alpha} t^{\alpha}+u_{1 {\alpha}}(v_1-s)^{\alpha}+u_{2 {\alpha}}(v_2+t)^{\alpha} -\sum_{1\leq i < j \leq n} p_{i\a}p_j^\a - \sum_{n < i < j \leq n+m} p_{i\a}p_j^\a - \sum_{i=1}^{n+m} p_{i{\alpha}} r_i^{\alpha}  \big],
\end{multline}
where $p_{(1)+\alpha} = \sum_{i=1}^{n} p_{i{\alpha}}, \quad p_{(2)+\alpha} = \sum_{i=n+1}^{n+m} p_{i{\alpha}}$.
The shifts $v_{1\alpha} \rightarrow v_{1\alpha} + s_{\alpha}, v_{2\alpha} \rightarrow v_{2\alpha} - t_{\alpha}$ and then $s_{\alpha} \rightarrow s_{\alpha} + p_{(2)+\alpha}$ and $ t_{\alpha} \rightarrow t_{\alpha} - p_{(1)+\alpha}$ along with the change $\sigma_2^l \rightarrow -\sigma_2^l$ yield \eqref{anz_mult}. The sign factor arises as a consequence of the last change of variables and the interchange of measures, the sign of the integral, and the factors $g_i$.

\subsubsection{Star product of any number of canonical functions}
\label{Any}
Remarkably, \eqref{anz_mult} can be generalized to the star product of any number of canonical product factors
\begin{equation}
    \label{FJ}
    F_J := f_1(g_{1,1}, \dots, g_{1, n_1})*\ldots *f_J(g_{J, 1}, \dots, g_{J, n_J})
\end{equation}
with
\begin{equation}
    \label{phii}
    f_j(g_{j,1}, \dots, g_{j, n_j}) =  \fint{p_j r_j u_j v_j \tau_j \rho_j \b_j \s_j} \dd^2 u_j \dd^2 v_j \dd^2 p_j \dd^2 r_j \mu_j(\tau_j, \rho_j, \beta_j, \s_j) \dd \Omega_j^2
    {{\E}(\Omega_j)}   g_{j,1}(r_{j,1})\overline{*}\dots\overline{*}g_{j,n_j}(r_{j,n_j}) k\,,
\end{equation}
\begin{equation}
    \label{Expi}
    {{\E}(\Omega_j)} = \exp i \left({\Omega}_{j\alpha}(\tilde{q}_j+u_j)^{\alpha}+u_{j\alpha} v_j^{\alpha}  - \sum_{i=1}^{n_j}p_{j, i\alpha} r_{j, i}^{\alpha} -\sum_{1\leq i<i'\leq n_j}p_{j,i\a}p_{j,i'}^\a   \right),
\end{equation}
\begin{equation}
    \tilde{q}_j^{\alpha} := y^{\alpha} + p_{(j)+}^\a, \qquad p_{(j)+}^\a = \sum_{i=1}^{n_j}p_{j,i}^\a,
\end{equation}
\begin{equation}
    {\Omega}_{j\alpha} = \tau_j z_{\alpha}-(1-\tau_j)(p_{j\alpha}(\sigma_j)- \beta_j v_{j\alpha} + \rho_j(\tilde{q}_{j\alpha}+u_{j\alpha}))
\end{equation}
(no summation over $j$).
\par
Introducing the integration variables $s_i^{\alpha}, t_i^{\alpha}$ associated with
the star products one obtains (for the proof see \ref{App_A:sec_1})
\begin{equation}
    F_J = \prod_{j=1}^{J} \Bigg( \fint{p_j r_j u_j v_j s_j t_j \tau_j \rho_j \b_j \s_j} \dd^2 u_j \dd^2 v_j \dd^2 p_j \dd^2 r_j \dd^2 s_j \dd^2 t_j \mu_j(\tau_j, \rho_j, \beta_j, \s_j) \dd \tilde{\Omega}_j^2
    g_{j,1}(r_{j,1})\overline{*}\dots\overline{*}g_{j,n_j}(r_{j,n_j}) \Bigg) \delta^2(t_1) \delta^2(s_J)  \E_J k^J,
    \label{star_mult_any}
\end{equation}
where the integral sign is divided into separate parts which can be assembled on the left side, while simultaneously considering the appropriate sign factors in accordance with \eqref{commutation_with_integr}, 
\begin{multline}
    \E_J = \exp i \sum_{j=1}^J \left[ (s_{j\alpha
    } + s_{j-1 \alpha}) t_j^{\alpha} + (s_{j\alpha
    }-t_{j\alpha
    })(q_j^{\alpha}+u_j^{\alpha}) + u_{j\alpha
    }v_j^{\alpha
    } + \vphantom{\sum_i}\right.
    \\ + \tilde{\Omega}_{j\alpha}(q_j^{\alpha}+u_j^{\alpha} + s_j^{\alpha
    }+t_j^{\alpha
    })
    -\left. \sum_i p_{j,i\alpha} r_{j,i}^{\alpha} -\sum_{i<i'}p_{j,i\a}p_{j,i'}^\a   \right],
    \label{Ej}
\end{multline}
\begin{equation}
    \tilde{\Omega}_{j\alpha} = \tau_j z_{\alpha}-(1-\tau_j)((-1)^{j+1}p_{j\alpha}(\sigma_j)- \beta_j v_{j\alpha} + s_{j\alpha} - t_{j\alpha} + \rho_j(q_{j\alpha}+u_{j\alpha}+s_{j\alpha}+t_{j\alpha})),
    \label{omega_j}
\end{equation}
\begin{equation}
    q_j^{\alpha} := y^{\alpha} + (-1)^{j+1}p_{(j)+}^\a,
    \label{qj}
\end{equation}and $\prod$ also implies $\bar{*}$-product between groups $g_{j,1}(r_{j,1})\overline{*}\dots\overline{*}g_{j,n_j}(r_{j,n_j})$ with consecutive $j$'s.
(Here and in the sequel, objects with indices beyond the prescribed range are assumed to be zero.)

\par It is important that the homotopy parameters $\tau_j, \sigma_j, \rho_j$ and $\beta_j$ enter \eqref{star_mult_any} only via $\Omega_j$ thus
preserving the fundamental property that $\dd$ acts only on the measure for each product factor.
\par
In the $J=2$ case  one can shift $v_1^{\alpha} \rightarrow v_1^{\alpha} + s_1^{\alpha}, v_2^{\alpha} \rightarrow v_2^{\alpha} - t_2^{\alpha}$, $s_1^{\alpha} \rightarrow s_1^{\alpha}+p_{(2)+}^{\alpha}$, $t_2^{\alpha} \rightarrow t_2^{\alpha}-p_{(1)+}^{\alpha}$ bringing \eqref{star_mult_any} to \eqref{anz_mult}.

As shown in \ref{App_A:sec_2}, \eqref{Ej}, \eqref{omega_j} can be reduced to the following form
\begin{multline}
    \E_J  = \exp i \sum_{j=1}^J \left[(s_{j\a} + s_{j-1 \a})t_j^\a + \tO_{j\a}(s_j^\a + t_j^\a) - |J+1|_2\sum_{j=1}^J y_\a p_{(j)+}^\a -   \sum_i p_{j,i\alpha} r_{j,i}^{\alpha} \right] - \\
    -\sum_{j<j'}p_{(j)+\a}p_{(j')+}^\a-\sum_j \sum_{i<i'}p_{j,i\a}p_{j,i'}^\a -v_{k\a} v_{k+1}^\a,
    \label{Ej17}
\end{multline}
\begin{multline}
    \tO_i^\a = \tau_i z^\a - (1-\tau_i)\Big[(-1)^{i+1}p_i^\a(\s_i) - \b_i\big( (\delta_{i,k} + \delta_{i,k+1})v_{i}^\a + (-1)^i (\tilde{n}_i^\a - n_i^\a)
    + c_i(s_i^\a + t_{i+1}^\a) - b_i(t_i^\a + s_{i-1}^\a)\big) +\\+  s_i^\a - t_i^\a +(-1)^i (\tilde{n}_i^\a - n_i^\a) + (c_i - b_i)q_i^\a + \r_i(s_i^\a + t_i^\a + (-1)^i n_J^\a)\Big],
    \label{oi5}
\end{multline}
where
\begin{equation}
    \delta_{i, j} = \begin{cases}
        1, &i=j,\\
        0, &i\neq j,
    \end{cases}
    \qquad |J|_2 = \begin{cases}
        0, & \text{ when $J$ is even}\\
        1, & \text{ when $J$ is odd}
    \end{cases}
    \label{modulo}
\end{equation}
and
\begin{equation}
    c_i = \begin{cases}
        1, &i \leq k, \\
        0, &i > k,
    \end{cases}
    \qquad
    b_i = \begin{cases}
        0, &i \leq k, \\
        1, &i > k
    \end{cases}
    \label{ab-def}
\end{equation}
with some $1\leq k < J$,
\begin{equation}
    n_i^\a := \sum_{l=1}^i(-1)^l q_l^\a, \quad \tilde{n}_j^\a = n_J^\a - n_{j-1}^\a\,.
    \label{n_i}
\end{equation}

Let us stress that, since the dependence on $k$ resulted from
a particular change of variables  \eqref{v}, \eqref{ab-def}, it cannot affect the final result.
This implies that the dependence on $k$ in  $\tilde \Omega_i$ \eqref{oi5} is a kind of
artificial and can be chosen in any convenient way.

\subsubsection{Second-order Ansatz}
\label{2}
To solve equations in the second order starting from $S_2$ we introduce a generalized second-order Ansatz, that results from the multiplication \eqref{anz_mult} of two canonical functions \eqref{anz}, followed by the replacement of $z^{\alpha}$ by $\Omega_{12}^{\alpha}$ along with introducing $u_{12}^{\alpha}, v_{12}^{\alpha}$:
\begin{equation}
    g \equiv g(g_1,\dots,g_n) = \fint{p_l r_l u_j v_j  \tau_j \rho_i \beta_j \sigma_j^l a_i}\mu_0 \mu(\tau_j, \rho_i, \beta_j, \sigma_j^l, a_i)\dd \Omega_1^2 \dd \Omega_2^2 \E(\Omega_1, \Omega_2) g_1(r_1)\bar{*}\dots \bar{*}g_n(r_n)\,,
    \label{fundament}
\end{equation}
with $l \in \overline{1, n}, j \in \{ 1, 2, 12 \}, i \in \{ 1, 2 \}$ and
\begin{equation}
    \begin{gathered}
        \Omega_{1}^\a =\tau_1 \Omega_{12}^\a - (1-\tau_1)(p^{\alpha} (\sigma_1) + a_1 s^{\alpha} - \beta_1 v_{1}^\a+ \rho_1(y^{\alpha}+p_+^\a+s^{\alpha}+u_{1}^\a+u_{12}^\a)), \\
        \Omega_{2}^\a =\tau_2 \Omega_{12}^\a - (1-\tau_2)(p^{\alpha} (\sigma_2) - a_2 t^{\alpha} - \beta_2 v_{2}^\a + \rho_2(y^{\alpha}-p_+^\a+t^{\alpha}+u_{2}^\a+u_{12}^\a)),
    \end{gathered}
    \label{omegas}
\end{equation}
\begin{equation}
        \Omega_{12}^\a = \tau_{12} z^{\alpha}-(1-\tau_{12})(p^{\alpha}(\sigma_{12}) -\beta_{12}v_{12}^\a ),
        \label{Omega12}
\end{equation}
\begin{equation}
    \mu_0= \prod_{i=1}^{n}(\dd^2p_i \dd^2r_i) \dd^2 u_1 \dd^2 v_1 \dd^2 u_2 \dd^2 v_2 \dd^2 s \dd^2 t \dd^2 u_{12} \dd^2 v_{12},
    \label{mu0}
\end{equation}
\begin{multline}
    {\E}({\Omega}_1, {\Omega}_2)  = \exp i \Big({\Omega}_{1\alpha}(y+s+p_++u_1+u_{12})^{\alpha} + {\Omega}_{2\alpha}(y+t-p_++u_2+u_{12})^{\alpha} + s_{\alpha} t^{\alpha} + \\  + (s-t + p_+)_{\alpha}(y+u_{12})^{\alpha}+u_{1\alpha}v_1^{\alpha}+u_{2\alpha}v_2^{\alpha} +u_{12\alpha}v_{12}^{\alpha} - \sum_{i<j} p_{i\alpha} p_j^{\alpha} - \sum_{i=1}^{n} p_{i\alpha} r_i^{\alpha} \Big).
    \label{exponenta}
\end{multline}\par
In the cases where the integration over $a_j$ is not specified we will set $a_i = 1-\beta_i$. Recall that the measure $\mu$ in \eqref{fundament} is compact. Measure $\mu_0$ \eqref{mu0} contains the differentials of the spinor coordinates as well as the measure \eqref{mu0_anz}. Hereafter, we will refer to $\mu_0$ as a measure along all spinor coordinates of the space $M$.\par
By virtue of Taylor expansion
$g_k(x)=e^{x^{\alpha}\frac{\partial}{\partial r^{\alpha}}}g_k(r)|_{r=0}$,
the $p_k$ and $r_k$ integrations can be replaced by a differential realization
\begin{equation}
    p_{k\alpha} = -i \frac{\partial}{\partial r_k^{\alpha}}
    \label{differ}
\end{equation}
followed by the substitution $r_{k}^{\alpha} = 0$. Such differential representation of $p_{k\a}$ changes the measure \eqref{mu0} to
\begin{equation}
    \mu_0 = \dd^2 u_1 \dd^2 v_1 \dd^2 u_2 \dd^2 v_2 \dd^2 s \dd^2 t \dd^2 u_{12} \dd^2 v_{12}.
    \label{tmu0}
\end{equation}

Note that the Ansatz \eqref{anz} of \cite{Vasiliev:2023yzx} was developed in the process of solving equations on $S_1, W_1, B_2$. Its structure facilitates calculations in the lower orders, and in particular, property \eqref{d_measure}, which is derived from the Schouten identity, plays a significant role in this process. The relevance of the new Ansatz \eqref{fundament}-\eqref{exponenta} arises from the solution of equations on $S_2, W_2$, that contain the products of two canonical functions. It also exhibits several remarkable properties, including \eqref{d_measure}. Furthermore, the introduction of an
auxiliary variable $\Omega_{12}$ facilitates the tracking of zero terms in the calculations.
The proposed second-order Ansatz  \eqref{fundament}-\eqref{exponenta} allows us to obtain all standard results. Specifically, the parameter $\beta_{12}$ plays a key role in the construction of local interacting vertices and recovers the limiting contracting homotopy \cite{Didenko:2019xzz} from which these vertices originate. Furthermore, the freedom it provides suggests a higher-order extension beyond the previously considered cases.
\par

\subsection{Operations}
In this section, we consider the operations with \eqref{fundament}, which will be used below to solve the system of equations for HS fields. Formulae obtained below generalize those of \cite{Vasiliev:2023yzx} but they do not boil down to direct application of \cite{Vasiliev:2023yzx} along with multiplication \eqref{anz_mult} due to the presence of $\Omega_{12}^{\alpha}$ \eqref{Omega12}.

\par
Firstly consider the replacement of the factor of $g_k$ in \eqref{fundament}
by $g_k*f$, where $f$ is a $z_{\alpha}$-independent function of $y_{\alpha}$,
\begin{equation}
    \int_{p_k r_k}e^{-ip_{k\alpha}r_k^{\alpha}}(g_k*f)(r_k) = \int_{pruv} g(r_k+u)\bar{*}f(r_k+v) e^{i(u_{\alpha}v^{\alpha}-p_{k\alpha}r_k^{\alpha})}\,.
\end{equation}
In the integral we change  $v_{\alpha}$ to $v_{\alpha}-r_{k\alpha}$, then $r_{k\alpha}$ to $r_{k\alpha} - u_{\alpha}$. After that $v$ should be renamed as $r_{f}$ and $u$ --- as $-p_{f}$. This yields
\begin{equation}
    \int_{p_k r_k}e^{-ip_{k\alpha}r_k^{\alpha}}(g_k*f)(r_k) = \int_{p_k r_k p_{f} r_{f}} g(r_k)\bar{*}f(r_{f}) e^{i(-p_{f\alpha}r_{f}^{\alpha}+p_{f\alpha}r_k^{\alpha}-p_{k\alpha}r_k^{\alpha}-p_{k\alpha}p_{f}^{\alpha})}.
\end{equation}
Now change $p_{k\alpha}$ to $p_{f\alpha}+p_{k\alpha}$ to obtain
\begin{equation}
    \int_{p_k r_k}e^{-ip_{k\alpha}r_k^{\alpha}}(g_k*f)(r_k) = \int_{p_k r_k p_{f} r_{f}} g(r_k)\bar{*}f(r_{f}) e^{i(-p_{f\alpha}r_{f}^{\alpha}-p_{k\alpha}r_k^{\alpha}-p_{k\alpha}p_{f}^{\alpha})}
    \label{d_x}
\end{equation}
in accordance with the Ansatz \eqref{fundament}. Hence, the star-multiplication  of one of the functions $g_k$ by a function $f(y)$ boils down to adding integration over a new parameters $\sigma_{j}^{f}$ that enters  the integral through $\Omega_j$ and with $\delta(\sigma_{j}^{f} - \sigma_{j}^{k})$ ($j \in \{1, 2, 12 \}$) since  $p_{k\alpha}$ was replaced by $p_{f\alpha}+p_{k\alpha}$,
\begin{multline}
    g(g_1,\dots, g_k*f,\dots,g_n) = \fint{p_l r_l u_j v_j  \tau_j \rho_i \beta_j \sigma_j^l a_i \sigma_1^{f} \sigma_2^{f} \sigma_{12}^{f}} \D(\sigma_{12}^{f} - \sigma_{12}^{k}) \D(\sigma_{2}^{f} - \sigma_{2}^{k})\D(\sigma_{1}^{f} - \sigma_{1}^{k})\mu_0 \mu(\tau_j, \rho_i, \beta_j, \sigma_j^l, a_i) \dd \Omega_1^2 \dd \Omega_2^2 \times \\ \times \E(\Omega_1, \Omega_2) g_1(r_1)\bar{*}\dots \bar{*} g(r_k)\bar{*}f(r_{f})\bar{*}\dots \bar{*} g_n(r_n).
    \label{star-property_in_right}
\end{multline}\par
Analogously, one can obtain
\begin{multline}
    g(g_1,\dots, f*g_k,\dots,g_n) = \fint{p_l r_l u_j v_j  \tau_j \rho_i \beta_j \sigma_j^l a_i \sigma_1^{f} \sigma_2^{f} \sigma_{12}^{f}}\D(\sigma_{12}^{f} - \sigma_{12}^{k}) \D(\sigma_{2}^{f} - \sigma_{2}^{k})\D(\sigma_{1}^{f} - \sigma_{1}^{k}) \mu_0 \mu(\tau_j, \rho_i, \beta_j, \sigma_j^l, a_i) \dd \Omega_1^2 \dd \Omega_2^2 \times \\ \times \E(\Omega_1, \Omega_2) g_1(r_1)\bar{*}\dots \bar{*} f(r_{f}) \bar{*} g(r_k)\bar{*}\dots \bar{*} g_n(r_n).
    \label{star-property_in_left}
\end{multline}\par
The next operation is the star-multiplication of an expression of the form \eqref{fundament} by some $z_{\alpha}$-independent function $\phi(y)$.
It is not hard to see that
\begin{multline}
    g*\phi(y) = \fint{p_l r_l u_j v_j  \tau_j \rho_i \beta_j \sigma_j^l a_i \sigma_1^{\phi} \sigma_2^{\phi} \sigma_{12}^{\phi}} \D(\sigma_{12}^{\phi}+1-\beta_{12})\D(\sigma_2^{\phi}+(1-\beta_2))\D(\sigma_1^{\phi}-2 a_1 + (1-\beta_1)) \mu_0 \mu(\tau_j, \rho_i, \beta_j, \sigma_j^l, a_i)\times \\ \times \dd \Omega_1^2 \dd \Omega_2^2 \E(\Omega_1, \Omega_2) g_1(r_1)\bar{*}\dots \bar{*}g_n(r_n)\bar{*} \phi(r_{\phi}),
    \label{product_r}
\end{multline}
\begin{multline}
    \phi(y)*g = (-1)^{(n+deg(\mu))deg(\phi)}\fint{p_l r_l u_j v_j  \tau_j \rho_i \beta_j \sigma_j^l a_i \sigma_1^{\phi} \sigma_2^{\phi} \sigma_{12}^{\phi}} \D(\sigma_{12}^{\phi}+1-\beta_{12})\D(\sigma_2^{\phi}-2a_2+(1-\beta_2))\D(\sigma_{1}^{\phi}+(1-\beta_{1})) \mu_0 \mu(\tau_j, \rho_i, \beta_j, \sigma_j^l, a_i)\times \\ \times \dd \Omega_1^2 \dd \Omega_2^2 \E(\Omega_1, \Omega_2)  \phi(r_{\phi})\bar{*} g_1(r_1)\bar{*}\dots \bar{*}g_n(r_n),
    \label{product_l}
\end{multline}
where $p_{\phi}$ and $r_{\phi}$ appear in  \eqref{omegas}-\eqref{exponenta} as $p_i$ with $i = \phi$.
\par

\subsection{Symmetry property}
There is a symmetry between the indices 1 and 2 in  presence of $\D(1-\tau_1)$ or $\D(1-\tau_2)$ in \eqref{fundament}. Namely, the following equality holds true (recall that objects with label $"12"$ belong to $\Omega^{\alpha}_{12}$ \eqref{Omega12}):
\begin{multline}
    \fint{\tau_1 \tau_2 \rho_2 \beta_2 \sigma_2^k \tau_{12} a_2 \sigma_{12}^i \beta_{12}} \D(a_2-(1-\beta_2)) \D(1-\tau_1) \mu(\sigma_{12}^i, \beta_{12}, \tau_{12}, \tau_2, \rho_2, \beta_2, \sigma_2^k)\dd \Omega_1^2 \dd \Omega_2^2 \E(\Omega_1, \Omega_2) g_1(r_1)\bar{*}\dots \bar{*}g_n(r_n) =
    \\
    = -\fint{\tau_2 \tau_1 \rho_1 \beta_1 \sigma_1^k \tau_{12} a_1 \sigma_{12}^i \beta_{12}} \D(a_1-(1-\beta_1)) \D(1-\tau_2) \mu(\sigma_{12}^i, \beta_{12},\tau_{12}, \tau_1, \rho_1, \beta_1, -\sigma_1^k)\dd \Omega_1^2 \dd \Omega_2^2 \E(\Omega_1, \Omega_2) g_1(r_1)\bar{*}\dots \bar{*}g_n(r_n),
    \label{symmetry}
\end{multline}
where $\mu$ contains at least $n-1$ differentials of $\sigma_{1(2)}$, one differential of $\rho_{1(2)}$ and one differential of $\beta_{1(2)}$. There is no restriction on the degree of this form with respect to $\dd \tau_i$. In the proof, it will be shown how objects with index 1 and 2 can be swapped.

To prove \eqref{symmetry}, we first consider the simple case where $\mu$ contains the differential $\mathrm{d}\tau_{1(2)}$. Taking into account all other differentials in $\mu$, we observe that on the \textit{r.h.s.} $
\mathrm{d}\Omega_1^2 \cong \mathrm{d}\Omega_{12}^\alpha (\dots)_\alpha$,
where the weak equality $\cong$ implies equality up to terms, that do not contribute to the integral.
Thus, in the presence of $\D(1-\tau_2)$, we have
\begin{equation}
\mathrm{d}\Omega_1^2 \mathrm{d}\Omega_2^2 \cong \mathrm{d}\Omega_{12}^\alpha (\dots)_\alpha \mathrm{d}\Omega_{12}^\beta \mathrm{d}\Omega_{12 \beta},
\end{equation}
which vanishes due to the Schouten identity $\mathrm{d}\Omega_{12}^\alpha \mathrm{d}\Omega_{12}^\beta \mathrm{d}\Omega_{12}^\gamma = 0$. The \textit{l.h.s.} weakly vanishes for the same reason, reducing \eqref{symmetry} to the identity $0=0$.

In the absence of $\mathrm{d}\tau_{1(2)}$ in $\mu$, we decompose $\mathrm{d}\Omega_1^2$ on the \textit{r.h.s.} as follows:
\begin{equation}
    \dd \Omega_1^2 \cong 2 \dd \tau_1 (\Omega_{12}^{\alpha} + p^{\alpha} (\sigma_1) + s^{\alpha}(1-\beta_1) - \beta_1 v_{1}^{\alpha} + \rho_1(y^{\alpha}+p_+^{\alpha}+s^{\alpha}+u_{1}^{\alpha}+u_{12}^{\alpha})) \dd p_{\alpha} (\sigma_1).
\end{equation}

Since $t_{\alpha}$ does not contribute to the pre-exponential factor integration over this variable gives the delta function $\delta^2(s+\Omega_{12}+y+u_{12})$. After integration over $s_{\alpha}$ one gets
\begin{equation}
    \dd \Omega_1^2  \cong 2 \dd \tau_1 (\beta_1 \Omega_{12}^{\alpha} - (1-\beta_1)(y^{\alpha}+u_{12}^{\alpha}) + p^{\alpha} (\sigma_1) - \beta_1 v_{1}^{\alpha} + \rho_1(p_+^{\alpha}-\Omega_{12}^{\alpha}+u_{1}^{\alpha})) \dd p_{\alpha} (\sigma_1).
\end{equation}
The exponent acquires the form
\begin{multline}
    \E  = \exp i \big((\Omega_{12}-(1-\tau_1)(\beta_1 \Omega_{12}- (1-\beta_1)(y+u_{12}) + p (\sigma_1) - \beta_1 v_1))_{\alpha}(-\Omega_{12}+p_++u_1)^{\alpha}+ \\  + {\Omega}_{12\alpha}(y-p_++u_2+u_{12})^{\alpha} -\Omega_{12\alpha}(y+u_{12})^{\alpha} + p_{+{\alpha}} (y+u_{12})^{\alpha} +u_{1\alpha}v_1^{\alpha}+u_{2\alpha}v_2^{\alpha}+u_{12\alpha}v_{12}^{\alpha} +\\  - \sum_{i<j} p_{i\alpha} p_j^{\alpha} - \sum_{i=1}^n p_{i\alpha} r_i^{\alpha} \big) = \exp i \big((-(1-\tau_1)(\beta_1 \Omega_{12}- (1-\beta_1)(y+u_{12}) + p (\sigma_1) - \beta_1 v_1))_{\alpha}(-\Omega_{12}+\\+p_++u_1)^{\alpha}  + {\Omega}_{12\alpha}(u_1+u_2+u_{12})^{\alpha} + p_{+{\alpha}} (y+u_{12})^{\alpha} +u_{1\alpha}v_1^{\alpha}+u_{2\alpha}v_2^{\alpha}+u_{12\alpha}v_{12}^{\alpha}  - \sum_{i<j} p_{i\alpha} p_j^{\alpha} - \sum_{i=1}^n p_{i\alpha} r_i^{\alpha} \big).
\end{multline}
 \par
Analogously, on the {\it l.h.s.} of \eqref{symmetry} one integrates over $s_{\alpha}$ to get $\delta^2 (t-\Omega_{12}+y+u_{12})$. Then integration over $t_{\alpha}$ yields
\begin{multline}
    \dd \Omega_2^2  \cong 2 \dd \tau_2 (\beta_2 \Omega_{12 }^{\alpha} + (1-\beta_2)(y+u_{12})^{\alpha} + p^{\alpha} (\sigma_2) - \beta_2 v_{2}^{\alpha} + \rho_2(-p_{+}^{\alpha}+\Omega_{12}^{\alpha}+u_2^{\alpha})) \dd p_{\alpha} (\sigma_2) =\\ =-2 \dd \tau_2 (-\beta_2 \Omega_{12}^{\alpha} -(1-\beta_2)(y+u_{12})^{\alpha} - p^{\alpha} (\sigma_2) + \beta_2 v_2^{\alpha} + \rho_2(p_+^{\alpha}-\Omega_{12}^{\alpha}-u_2^{\alpha})) \dd p_{\alpha} (\sigma_2)
    \label{sym_right}
\end{multline}
and
\begin{multline}
    \E  = \exp i \big((\Omega_{12}-(1-\tau_2)(\beta_2 \Omega_{12} + (1-\beta_2)(y+u_{12}) + p (\sigma_2) - \beta_2 v_2))_{\alpha}(\Omega_{12}-p_++u_2)^{\alpha}  + {\Omega}_{12\alpha}(y+\\+p_++u_1+u_{12})^{\alpha} -\Omega_{12\alpha}(y+u_{12})^{\alpha} + p_{+{\alpha}} (y+u_{12})^{\alpha} +u_{1\alpha}v_1^{\alpha}+u_{2\alpha}v_2^{\alpha}+u_{12\alpha}v_{12}^{\alpha}  - \sum_{i<j} p_{i\alpha} p_j^{\alpha} - \sum_{i=1}^n p_{i\alpha} r_i^{\alpha} \big) =\\= \exp i \big((-(1-\tau_2)(-\beta_2 \Omega_{12}- (1-\beta_2)(y+u_{12}) - p (\sigma_2) + \beta_1 v_2))_{\alpha}(-\Omega_{12}+p_+-u_2)^{\alpha}+\\  + {\Omega}_{12\alpha}(u_1+u_2+u_{12})^{\alpha} + p_{+{\alpha}} (y+u_{12})^{\alpha} +u_{1\alpha}v_1^{\alpha}+u_{2\alpha}v_2^{\alpha}+u_{12\alpha}v_{12}^{\alpha}  - \sum_{i<j} p_{i\alpha} p_j^{\alpha} - \sum_{i=1}^n p_{i\alpha} r_i^{\alpha} \big).
\end{multline}
Changing a sign of $u_{2\alpha}$ and $v_{2\alpha}$ to $-v_{2\alpha}+2\Omega_{12\alpha}$, one obtains
\begin{equation}
    \dd \Omega_2^2 \cong -2 \dd \tau_2 (\beta_2 \Omega_{12}^{\alpha} - (1-\beta_2)(y+u_{12})^{\alpha} - p^{\alpha} (\sigma_2) - \beta_2 v_{2}^{\alpha} + \rho_2(p_+^{\alpha}-\Omega_{12}^{\alpha}+u_2^{\alpha})) \dd p_{\alpha} (\sigma_2),
\end{equation}
\begin{multline}
    \E  = \exp i \big[(-(1-\tau_2)(\beta_2 \Omega_{12}- (1-\beta_2)(y+u_{12}) - p (\sigma_2) - \beta_1 v_2))_{\alpha}(-\Omega_{12}+p_++u_2)^{\alpha} + {\Omega}_{12\alpha}(u_1+u_2+u_{12})^{\alpha}+\\   + p_{+{\alpha}} (y+u_{12})^{\alpha} +u_{1\alpha}v_1^{\alpha}+u_{2\alpha}v_2^{\alpha}++u_{12\alpha}v_{12}^{\alpha}  - \sum_{i<j} p_{i\alpha} p_j^{\alpha} - \sum_{i=1}^n p_{i\alpha} r_i^{\alpha} \big].
\end{multline}
Interchange of the indices 1 and 2 completes the proof (exchange of $\tau_1$ and $\tau_2$,  change of a sign of $\sigma_2$, together with the minus from \eqref{sym_right}  gives a minus on the {\it r.h.s.} of \eqref{symmetry}).

Analogous property holds true in the presence of a factor of $\D(\tau_i)\D(a_i)$
\begin{multline}
    \fint{\tau_j \rho_i \beta_i \sigma_j^k a_i} \D(a_2)\D(a_1-(1-\beta_1)) \D(\tau_1) \mu(\tau_{12}, \sigma_{12}^k, \tau_2, \rho_2, \beta_2, \sigma_2^k, \rho_1, \beta_1, \sigma_1^k)\dd \Omega_1^2 \dd \Omega_2^2 \E g_1(r_1)\bar{*}\dots \bar{*}g_n(r_n) =\\= \fint{\tau_j \rho_i \beta_i \sigma_j^k a_i}\D(a_1) \D(a_2-(1-\beta_2)) \D(\tau_2) \mu(\tau_{12}, \sigma_{12}^k, \tau_1, \rho_1, \beta_1, \sigma_1^k, \rho_2, \beta_2, -\sigma_2^k)\dd \Omega_1^2 \dd \Omega_2^2 \E g_1(r_1)\bar{*}\dots \bar{*}g_n(r_n),
    \label{symmetry-2}
\end{multline}
where $\mu$ contains $n-2$ differentials of $\sigma_1$ on the {\it l.h.s.} of the equality. \par
The proof of \eqref{symmetry-2} is analogous to that of \eqref{symmetry}. Namely, on the {\it l.h.s.} one should integrate over $t_{\alpha}$, that yields $\delta^2(s+\Omega_2+y+u_{12})$ integrated over $s_{\alpha}$. On the {\it r.h.s.} the integral over $s_{\alpha}$ gives $\delta^2(t-\Omega_1+y+u_{12})$ to be integrated over $t_{\alpha}$. As a result, swapping the indices 1 and 2, one arrives at  \eqref{symmetry-2}.

\subsection{Realization of the shifted homotopy operator}
In \cite{Didenko:2019xzz}, the $\b$-shifted contracting homotopy operator
\begin{equation}
    \Delta_{p(\s_0), \b_0} F = \fint{u^2 v^2} d^2 u d^2 v \fint{\tau} \dfrac{d\tau}{\tau} l(\tau) (z + p(\s_0) -\b_0 v)^\a \dfrac{\partial}{\partial \theta^\a} F(\Omega_0, y + u| \tau \theta) \exp i u_\gamma v^\gamma,
    \label{eq:b-shifted_hom_op}
\end{equation}
with
\begin{equation}
    l(\tau) = \theta(1-\tau)\theta(\tau),
    \label{l(tau)}
\end{equation}
\begin{equation}
\Omega_0 = \tau z - (1-\tau)\left[ p(\s_0) - \b_0 v\right]
\end{equation}
was introduced to obtain a quadratic ultralocal vertex in the one-form holomorphic sector.

 As we show now, the action of this operator can be realized in terms of the differential homotopy approach, demonstrating that the latter  is more general.
  That the two methods are not fully equivalent follows from the construction of a differential homotopy for solving the $B_2$-field equation in \cite{Vasiliev:2023yzx} which particular solution admits no shifted homotopy realization.

We define an operator $\tr_\l$ acting on differential forms in $z_\alpha$ and auxiliary homotopy variables to generate
a differential form in an extended space with additional variables $\tau, \beta, \sigma^i$
and spinors $u_\alpha, v_\alpha$:
\begin{equation}
     \left( \vphantom{\big|}\tr_{\l}F(z, y| \theta) \right) [\tau, \s^i, \b, u ,v] = d^2 u \; d^2 v \; \lambda(\s^i, \beta) l(\tau) F(\Omega, y + u | \dd \Omega ) \exp i u_\a v^\a,
     \label{eq:shift_diff_op}
\end{equation}
where $\Omega_\a = \tau z_\a - (1-\tau) \left[p_\a(\s) - \beta v_\a\right]$ while  $\l(\s^i, \b)$ is some closed differential form. To reproduce the standard results obtained by the $\beta$-shifted contracting homotopy technique one can choose the measure $\lambda$ in the form
\begin{equation}
\l(\s^i, \beta) = \l_0 (\s^i, \b) = \prod_{i} \D(\s^i - \s_0^i) \cdot \D(\beta - \beta_0),
\label{eq:hom_op_measure}
\end{equation}
where $\s_0^i, \b_0$ are constants.
It can be shown that the action of this operator,
with the  measure form \eqref{eq:hom_op_measure},
coincides with that of the $\beta$-shifted homotopy operator in a weak sense
\begin{equation}
    \fint{u^2 v^2 \tau \s^i \b } \quad \tr_{\l_0} F \cong \Delta_{p(\s_0), \b_0} F, \qquad \text{with } \l_0(\s^i, \b) = \prod_{i} \D(\s^i - \s_0^i) \cdot \D(\beta - \beta_0).
    \label{eq:hom_ops_conection_in text}
\end{equation}

Details of the proof of this statement, along with some its useful consequences
are presented in \ref{subsec:homotopy_op}.

\section{Application to  higher-spin equations}
\label{sec:Chapter2} \index{Chapter2}
\subsection{Generating system}
The  differential homotopy approach is applied  to the system \eqref{base_all} as follows \cite{Vasiliev:2023yzx}. We make the fields depend on the homotopy parameters $h^a$ and the variables $p_{\alpha}, r_{\alpha}, u_{\alpha}, v_{\alpha}$ along with $Z, Y, x$. The original fields are restored via integration over all additional variables.\par
It is useful to expand the fields ${B}$ and ${W}$ in degrees of
differential forms as follows:
\begin{equation}
    \mathcal{B} = B + \sum_{n \geq 1} {B}^{(n)}, \quad {\mathcal{W}}' = W + \sum_{n \geq 2} W^{(n)},
    \label{modific}
\end{equation}
where ${B}^{(n)}$ and $W^{(n)}$ are space-time $n$-forms.
Thus, $\mathcal{B}$ contains higher forms of  positive degrees. \par
We also introduce notation
\begin{equation}
    \mathcal{W} = {\mathcal{W}}' + S - \theta^A Z_A,
\end{equation}
where the vacuum solution \eqref{vac} is explicitly taken into account. Then the system \eqref{base_all} is replaced by
\begin{subequations}
    \label{base_eq_diff}
    \begin{align}
        (\dd_x - 2i\dd)\mathcal{B}+\mathcal{W}*\mathcal{B} - \mathcal{B}*\mathcal{W} = \V_1 \cong 0, \label{base_eq_diff1} \\
        (\dd_x - 2i\dd)\mathcal{W}+\mathcal{W}*\mathcal{W} = \V_2 \cong i \mathcal{B}*(\eta \gamma + \bar{\eta} \bar{\gamma}),\label{base_eq_diff2}
    \end{align}
\end{subequations}
where the weak equality $\cong$ implies equality up to terms, that do not contribute to the integral over additional variables. The property \eqref{d_z-prop} is also taken into account. \par
Since equations \eqref{base_eq_diff} contain space-time differential forms of different degrees, they  split into independent equations for each degree.\par
Equations \eqref{base_eq_diff} are invariant under the following gauge transformations
\begin{equation}
    \begin{gathered}
        \delta \mathcal{W} = (\dd_x - 2i\dd) \epsilon_{\mathcal{W}}+[\mathcal{W}, \epsilon_{\mathcal{W}}]_*, \quad \delta \mathcal{B} = [\mathcal{B}, \epsilon_{\mathcal{W}}]_*, \\
        \delta \mathcal{V}_2 = [\mathcal{V}_2, \epsilon_{\mathcal{W}}]_*, \qquad \delta \mathcal{V}_1 = [\mathcal{V}_1, \epsilon_{\mathcal{W}}]_*,\\
        \delta \mathcal{B} = (\dd_x - 2i\dd) \epsilon_{\mathcal{B}}+\left\{\mathcal{W} , \epsilon_{\mathcal{B}}\right\}_*, \qquad \delta \mathcal{V}_1 = [\mathcal{V}_2, \epsilon_{\mathcal{B}}]_*,
        \label{gauge}
    \end{gathered}
\end{equation}
where the gauge parameters $\epsilon_{\mathcal{W}}$ and $\epsilon_{\mathcal{B}}$ ar
 zero- and  one-forms  in $\theta$, respectively. (We use notations $[a, b]_*= a*b-b*a, \; \{a, b\}_* = a*b + b*a$.)

Note that all higher degree forms except for the first ones in the expansions \eqref{modific} do not affect the physical equations. For instance, let $B^{(1)}$ enter the field $C$ equation as follows:
\begin{equation}
    \dd_x C + \omega * C - C * \omega - 2i \dd B^{(1)} + \dots = 0\,.
\end{equation}
then the part with $B^{(1)}$ under the differential $\dd_t$ does not contribute upon integration over $M$ while that  under  $\dd_Z$ does not affect the $Z,\,\theta$-independent sector of the dynamical equations. Analogously, the equation for the one-form field $\omega$ takes the form
\begin{equation}
    \dd_x \omega + \omega * \omega- 2i \dd W^{(2)}  + \dots = 0\,.
\end{equation}

\subsection{Vacuum solution}
The vacuum solution is chosen as follows
\begin{equation}
    \mathcal{B} = 0, \quad \mathcal{W} = \omega(Y, K | x),
\end{equation}
where $\omega(Y,K|x)$ is a polynomial in $Y_A$ that verifies the flatness equation
\begin{equation}
    \dd_x \omega(Y, K | x) + \omega(Y, K | x)*\omega(Y, K | x) = 0.
\end{equation} \par
\subsection{First-order calculations}
Substitution of  the vacuum solution into \eqref{base_eq_diff} yields in the first order
\begin{equation}
    \dd_x \B_1 + \omega*\B_1 - \B_1*\omega -2i\dd \B_1 \cong 0\,.
    \label{B-first_order}
\end{equation}
The $\dd x$--independent part of  this equation implies that the zero-form part of $\B_1$ (\ie $B_1$) is $\dd$-closed. The gauge transformation in the last line of \eqref{gauge} allows us to eliminate the exact part setting $\B_1=C(Y,K|x)$. The latter is subject to a condition resulting from the degree one sector in $\dd x^n$ of \eqref{B-first_order},
\begin{equation}
    \dd_x C + \omega*C - C*\omega = 0\,.
\end{equation}\par
Equation \eqref{base_eq_diff2} allows us to find $S_1$. In the $\bar{\eta}$-independent holomorphic sector, this equation takes the form
\begin{equation}
    -2i\dd S_1 \cong i\eta C*\gamma\,.
    \label{eq:S_1}
\end{equation}
Its solution is found in the form \cite{Vasiliev:2023yzx}
\begin{equation}
    S_1 = -\dfrac{\eta}{2}\fint{\tau \rho \beta \sigma} l(\tau) \D(\sigma) \mu(\rho, \beta) \dd \Omega^2 \E(\Omega) C(r, \bar{y}, K)\big|_{r = 0} k
    \label{S1}
\end{equation}
(the $u_{i}^{\alpha}$, $v_{i}^{\alpha}$ integrations and measure $\mu_0$ are implicit), where
\begin{equation}
    \mu(\rho, \beta) = \mu(\beta)\D(\rho)\,,\qquad \int_{\beta} \mu(\beta) = 1\,.
    \label{mu}
\end{equation}
Here and below we use the representation \eqref{differ}.\par
Acting by $\dd$ on the {\it r.h.s.\,} of \eqref{S1} we get two terms. The one with $\D(\tau)$ is weak since there are no differentials left for $\dd\Omega_{\alpha}$.  Another one with $\D(1-\tau)$ reproduces the {\it r.h.s.\,} of \eqref{eq:S_1}, which is verified by direct integration.
\par

\par
$W_1$ is determined from the equation
\begin{equation}
\label{dw1}
    -2i\dd W_1 + \dd_x S_1 + [\omega, S_1]_* \cong 0\,.
\end{equation}
The result splits into a sum of two terms corresponding to the two possible orderings of the product factors --- $C\bar{*}\w$ and $\w \bar{*}C$ --- under the integral in \eqref{anz}. As shown in \cite{Vasiliev:2023yzx}, equation (\ref{dw1})  admits the following solution:
\begin{equation}
    W_1|_{\omega C} = \dfrac{i\eta}{4}\fint{ \tau \rho \beta \sigma \sigma^{\omega} } l(\tau) \D(\sigma) P(\beta-1, \sigma^{\omega}, \sigma) \mu(\rho, \beta) \dd \Omega^2 \E(\Omega) \omega(r_{\omega}, \bar{y}, K) \bar{*} C(r, \bar{y}, K)\big|_{r = 0} k\,,
    \label{W_1_wC}
\end{equation}
\begin{equation}
    W_1|_{C \omega} = \dfrac{i\eta}{4}\fint{ \tau \rho \beta \sigma \sigma^{\omega} } l(\tau) \D(\sigma) P(\sigma, \sigma^{\omega}, 1-\beta) \mu(\rho, \beta) \dd \Omega^2 \E(\Omega) C(r, \bar{y}, K) \bar{*} \omega(r_{\omega}, \bar{y}, K)\big|_{r = 0} k\,,
    \label{W_1_Cw}
\end{equation}
where
\begin{equation}
    P(x_1, \dots, x_n) := \theta(x_n-x_{n-1})\theta(x_{n-1}-x_{n-2})\dots\theta(x_2-x_1).
\end{equation}\par
The space-time two-form sector of equation \eqref{base_eq_diff2} has the form
\begin{equation}
    \dd_x \omega +\omega * \omega  + \dd_x W_1 + \omega * W_1 + W_1 * \omega  - 2i \dd W_1^{(2)} \cong 0
\end{equation}
with
\begin{multline}
    W_1^{(2)} = \dfrac{i\eta}{4}\fint{\tau \rho \beta \sigma \sigma^{\omega1} \sigma^{\omega2}} l(\tau) \D(\sigma) \mu(\rho, \beta) \dd \Omega^2 \E(\Omega)\big [ \\ P(\beta-1, \sigma^{\omega1}, \sigma^{\omega2}, \sigma, 1-\beta) \omega(r_{\omega 1}, \bar{y}, K) \bar{*} \omega(r_{\omega 2}, \bar{y}, K) \bar{*} C(r, \bar{y}, K) + \\+ P(\beta-1, \sigma^{\omega1}, \sigma, \sigma^{\omega2}, 1-\beta) \omega(r_{\omega 1}, \bar{y}, K) \bar{*} C(r, \bar{y}, K) \bar{*}  \omega(r_{\omega 2}, \bar{y}, K)+ \\+ P(\beta-1, \sigma, \sigma^{\omega1}, \sigma^{\omega2}, 1-\beta) C(r, \bar{y}, K) \bar{*} \omega(r_{\omega 1}, \bar{y}, K) \bar{*}  \omega(r_{\omega 2}, \bar{y}, K)  \big] \big|_{r = 0} k
    \label{W_1,2}
\end{multline}
and $Z$-independent contribution to the physical equations
\begin{multline}
    \dd_x \omega +\omega * \omega = \dfrac{\eta}{4i}\fint{\tau \rho \beta \sigma \sigma^{\omega1} \sigma^{\omega2}} \D(\tau) \D(\sigma) \mu(\rho, \beta) \dd \Omega^2 \E(\Omega)\big [ \\ P(\beta-1, \sigma^{\omega1}, \sigma^{\omega2}, \sigma, 1-\beta) \omega(r_{\omega 1}, \bar{y}, K) \bar{*} \omega(r_{\omega 2}, \bar{y}, K) \bar{*} C(r, \bar{y}, K) + \\+ P(\beta-1, \sigma^{\omega1}, \sigma, \sigma^{\omega2}, 1-\beta) \omega(r_{\omega 1}, \bar{y}, K) \bar{*} C(r, \bar{y}, K) \bar{*} \omega(r_{\omega 2}, \bar{y}, K)+ \\+ P(\beta-1, \sigma, \sigma^{\omega1}, \sigma^{\omega2}, 1-\beta) C(r, \bar{y}, K) \bar{*} \omega(r_{\omega 1}, \bar{y}, K) \bar{*} \omega(r_{\omega 2}, \bar{y}, K)  \big] \big|_{r = 0} k\,.
    \label{D_omega}
\end{multline}\par
Equations \eqref{B-first_order} and \eqref{D_omega} reproduce the linearized equations for the HS fields \cite{Vasiliev:1988sa} or, equivalently, Central On-Shell Theorem. Note that the choice of the solution \eqref{S1} with $\D(\sigma)$ is dictated by the condition that the linearized equations \cite{Vasiliev:1988sa} contain $C(0,\bar{y}, K|x)$ on the {\it r.h.s.\,}, that is why there should be no dependence on $y_{\alpha}$, which is guaranteed by the absence of $p_{\alpha} y^{\alpha}$ in the exponent in \eqref{D_omega}.

\subsection{Second-order calculations}
In the second order, calculations start with the equation
\begin{equation}
    2i\dd B_2 \cong [S_1, C]_*,
\end{equation}
that can be solved as follows \cite{Vasiliev:2023yzx}:
\begin{equation}
    B_{2sh} = \dfrac{\eta}{4i}\fint{\tau \rho \beta \sigma^1 \sigma^2} l(\tau) P(-\overrightarrow{\sigma}, \sigma^1, \sigma^2, \overrightarrow{\sigma}) \D (\sigma^2-\sigma^1-\overrightarrow{\sigma}) \mu(\rho, \beta) \dd \Omega^2 \E(\Omega) C(r_1, \bar{y}, K) \bar{*}   C(r_2, \bar{y}, K) \big|_{r_i = 0}k
    \label{B2sh}
\end{equation}
or
\begin{multline}
    B_{2pc} = \dfrac{\eta}{4i}\fint{\tau \rho \beta \sigma^1 \sigma^2} l(\tau) P(-\overrightarrow{\sigma}, \sigma^1, \sigma^2, \overrightarrow{\sigma})\D(\rho+\sigma^1) \D (\sigma^2-\sigma^1-\overrightarrow{\sigma}) (\dd \Omega^2 + 2i(1-\beta)\D(\sigma^2)\dd \tau) \times \\ \times \mu(\beta)  \E(\Omega) C(r_1, \bar{y}, K) \bar{*}   C(r_2, \bar{y}, K) \big|_{r_i = 0}k\,,
    \label{B2}
\end{multline}
where
\begin{equation}
    \overrightarrow{\sigma} := 1-\beta\,.
\end{equation}\par
In \ref{App_C}, we show that $B_{2 pc} - B_{2 sh}$ reproduces the shift $\delta B_2$ found in \cite{Didenko:2019xzz} to reach the vertex with the minimal number of derivatives. This is the new result of the paper. In the sequel we will use  $B_2 = B_{2 pc}$.
\par
Note also that Ansatz \eqref{B2} is appropriate for the following reasons. The factor of $\D(\sigma^2-\sigma^1-\overrightarrow{\sigma})$ ensures that the coefficient
in front of  $p_{1\alpha} p_{2}^{\alpha}$ in the exponent vanishes at $\tau = 0$, which
implies by the $Z$-dominance lemma \cite{Gelfond:2018vmi} spin-locality of the corrections to physical vertex. Indeed, the part of the exponent affecting this coefficient at $\tau = 0$ takes the form
\begin{equation}
    \exp i\big[ -(p(\sigma)-\beta v)_{\alpha}(p_++u)^{\alpha} + u_{\alpha}v^{\alpha} \big].
\end{equation}
Upon integration over $u_{\alpha}$ and $v_{\alpha}$ we obtain that the coefficient in front of $p_{1\alpha} p_{2}^{\alpha}$ is
\begin{equation}
    \frac{\sigma^2-\sigma^1}{1-\beta}-1,
\end{equation}
that is zero for the measure factor in \eqref{B2}. This property of spin-locality is valid for both  \eqref{B2sh} and \eqref{B2}. The shift of $\rho$ leading to \eqref{B2} is motivated by the requirement of projective compactness. In fact, due to the factors of $\D(\sigma^2-\sigma^1-\overrightarrow{\sigma})$ and $\D(\rho+\sigma^1)$ there is $p^\a_1+p^\a_2 - (y^\a+p^\a_1+p^\a_2) = -y^\a$ in front of $\sigma^1$ in $\Omega$. That is why $\dd \sigma^1$ appears in $\dd \Omega^2$ with a factor $y^\a$. The same is true for the contribution of $\dd \sigma^2$ to $\dd \Omega^2$. As was shown in \cite{Vasiliev:2022med}, such a factor leads to the projectively-compact vertex in the equation on zero-forms, characterised by the two related properties: it contains the minimal possible number of derivatives because the replacement of $y^\a$ by $p_i^a$ brings one more space-time derivative into the vertex and that the concepts of spin-locality in the spinor variables and space-time coordinates are equivalent (projective compactness).
So, the $y^\a$ dependence of the vertex ensures both of these properties providing a general principle.

The shift of $\rho$, and, as a result, the appearance of an additional term without $\dd\Omega^2$ allows one to achieve projectively-compact quadratic vertex in \eqref{eq_C} \cite{Vasiliev:2023yzx}.\par
Now consider the equation for $\dd_x C$:
\begin{equation}
    \dd_x C + [\omega, C]_* + [W_1, C]_* + \dd_x B_2 + [\omega, B_2]_* - 2i \dd B_2^{(1)} \cong 0,
    \label{d_x_C}
\end{equation}
where \cite{Vasiliev:2023yzx}
\begin{equation}
    B_2^{(1)} = \frac{1}{2i} \int_{\sigma^{\omega}} P(\beta - 1, \sigma^{\omega}, \sigma^1)B_2(\omega C C).
    \label{B_2^{(2)}}
\end{equation}
\par
Substitution of \eqref{B_2^{(2)}} into \eqref{d_x_C} yields
\begin{equation}
\label{dxCJ}
    \dd_x C + [\omega, C]_* \cong J\,,
\end{equation}
where
\begin{multline}
    J = \dfrac{i\eta}{4}\fint{\tau \rho \beta \sigma^1 \sigma^2 \sigma^\omega} \D(\tau) \D(\sigma^1+\rho) \D (\sigma^2-\sigma^1-\overrightarrow{\sigma})\mu(\beta) \dd \Omega^2 \E(\Omega)\\  \big[
    P(-\overrightarrow{\sigma}, \sigma^{\omega}, \sigma^1, \sigma^2,\overrightarrow{\sigma})\omega(r_{\omega}, \bar{y}, K) \bar{*} C(r_1, \bar{y}, K) \bar{*}   C(r_2, \bar{y}, K) + \\ + P(-\overrightarrow{\sigma}, \sigma^1, \sigma^2, \sigma^{\omega},\overrightarrow{\sigma})C(r_1, \bar{y}, K) \bar{*}   C(r_2, \bar{y}, K) \bar{*} \omega(r_{\omega}, \bar{y}, K) + \\+ P(-\overrightarrow{\sigma}, \sigma^1, \sigma^{\omega}, \sigma^2,\overrightarrow{\sigma})C(r_1, \bar{y}, K) \bar{*} \omega(r_{\omega}, \bar{y}, K) \bar{*}   C(r_2, \bar{y}, K)
    \big]  \big|_{r_i = 0}\,.
    \label{J}
\end{multline}\par
Due to the choice of the solution \eqref{B2} $J$ is projectively compact spin-local \cite{Vasiliev:2023yzx} since, as is not hard to see, $\dd \Omega^2$ in \eqref{J} is proportional to $y^\a$.

\subsection{\texorpdfstring{$S_2$}{S2}}
$S_2$ is determined by the equation
\begin{equation}
    2i \dd S_2 \cong S_1 * S_1 - i \eta B_2 * \gamma\,.
\end{equation}\par

Calculation of  $S_1*S_1$ by virtue of \eqref{anz_mult} yields
\begin{multline}
    S_1 * S_1  = \dfrac{\eta^2}{4}\fint{\tau_1 \rho_1 \beta_1 \sigma_1^1 \sigma_1^2  \tau_2 \rho_2 \beta_2 \sigma_2^1 \sigma_2^2 \tau_{12}} \D(\tau_{12}-1) l(\tau_1) l(\tau_2) \D (\sigma_1^1)\D (\sigma_1^2-\overrightarrow{\sigma_1}) \D (\sigma_2^1-\overrightarrow{\sigma_2})\D (\sigma_2^2){\mu}(\rho_1, \beta_1) {\mu}(\rho_2, \beta_2) \times \\ \times \dd \Omega_1^2 \dd \Omega_2^2 \E(\Omega_1, \Omega_2) C(r_1, \bar{y}, K) \bar{*}   C(r_2, \bar{y}, K) \big|_{r_i = 0}
    \label{S_1*S_1}
\end{multline}
with $\Omega_1$ and $\Omega_2$ \eqref{omegas}.

To calculate  $-i\eta B_2 * \gamma$ we use that $\gamma$ can be represented in the form
\begin{equation}
    -i\eta \gamma = i\eta \fint{\tau \rho \beta}\D(1-\tau) \mu(\rho, \beta) \dd \Omega^2 \E(\Omega)  k.
\end{equation}
Then \eqref{anz_mult} yields
\begin{multline}
    -i\eta B_2 * \gamma = \dfrac{\eta^2}{4}\fint{\tau_1 \rho_1 \beta_1 \sigma_1^1 \sigma_1^2  \tau_2 \rho_2 \beta_2 \sigma_2^1 \sigma_2^2\tau_{12}} \D(\tau_{12}-1) l(\tau_1) \D(1-\tau_2) P(-\overrightarrow{\sigma_1}, \sigma_1^1, \sigma_1^2, \overrightarrow{\sigma_1})\D (\sigma_1^2-\sigma_1^1-\overrightarrow{\sigma_1}) \D (\sigma_2^1)\D(\sigma_2^2) \D(\rho_1+\sigma_1^1) \times \\ \times {\mu}(\beta_1) {\mu}(\rho_2, \beta_2) (\dd \Omega_1^2 + 2i(1-\beta_1)\D(\sigma_1^2)\dd \tau_1)\dd \Omega_2^2 \E(\Omega_1, \Omega_2) C(r_1, \bar{y}, K) \bar{*}   C(r_2, \bar{y}, K) \big|_{r_i = 0}.
    \label{B_2*gamma}
\end{multline} \par
$S_2$ can be found in the form
\begin{multline}
     S_2 = \dfrac{i\eta^2}{8} \fint{\tau_1 \rho_1 \beta_1 \sigma_1^1 \sigma_1^2  \tau_2 \rho_2 \beta_2 \sigma_2^1 \sigma_2^2 \tau_{12} \sigma_{12}^1 \sigma_{12}^2\beta_{12}}\mu'(\beta_{12}) l(\tau_{12}) l(\tau_1) l(\tau_2) \D (\sigma_1^1)\D (\sigma_1^2-\overrightarrow{\sigma_1}) \D (\sigma_2^1-\overrightarrow{\sigma_2})\D (\sigma_2^2)\D(\sigma_{12}^1)\D(\sigma_{12}^2)\times \\ \times{\mu}(\rho_1, \beta_1){\mu}(\rho_2, \beta_2)
    \dd {\Omega}_1^2 \dd {\Omega}_2^2 \E(\Omega_1, \Omega_2)  C(r_1, \bar{y}, K) \bar{*}   C(r_2, \bar{y}, K) \big|_{r_i = 0} + \\ + \dfrac{i\eta^2}{8} \fint{\tau_1 \rho_1 \beta_1 \sigma_1^1 \sigma_1^2  \tau_2 \rho_2 \beta_2 \sigma_2^1 \sigma_2^2 \tau_{12} \sigma_{12}^1 \sigma_{12}^2 \beta_{12}}\mu'(\beta_{12}) l(\tau_{12}) l(\tau_1) \D(1-\tau_2) P(-\overrightarrow{\sigma_1}, \sigma_1^1, \sigma_1^2, \overrightarrow{\sigma_1})\D (\sigma_1^2-\sigma_1^1-\overrightarrow{\sigma_1}) \D (\sigma_2^1)\D(\sigma_2^2)\D(\sigma_{12}^1)\D(\sigma_{12}^2)\times \\ \times\D(\rho_1+\sigma_1^1) {\mu}(\beta_1) {\mu}(\rho_2, \beta_2) (\dd \Omega_1^2 + 2i(1-\beta_1)\D(\sigma_1^2)\dd \tau_1) \dd {\Omega}_2^2 \E(\Omega_1, \Omega_2)  C(r_1, \bar{y}, K) \bar{*}   C(r_2, \bar{y}, K) \big|_{r_i = 0}.
    \label{S2}
\end{multline}
Equations \eqref{S_1*S_1}, \eqref{B_2*gamma},\eqref{S2} lack the $a_i$ parameters present in \eqref{fundament} due to the replacements $a_1 = 1-\beta_1$, $a_2 = 1-\beta_2$ and their subsequent omission. Additionally, the measure $\mu'(\beta_{12})$ satisfies the normalization condition $\int_{\beta_{12}} \mu'(\beta_{12}) = 1$. Equation \eqref{S2}
can be checked as follows.
\par
When $\dd$ acts on (\ref{S2}), the following terms arise. The terms containing $\D(\tau_1)$, $\D(\tau_2)$, and $\D(\tau_{12})$ are weak because the degree of the form to be integrated exceeds the dimension of $M$. When $\dd$ acts on the first term in \eqref{S2}, the terms with $\D(1-\tau_i)$  are zero for the following reason. Consider for definiteness $\D(1-\tau_2)$. In that case
\begin{equation}
    \begin{gathered}
        \Omega_1^{\alpha} =\tau_1 \Omega_{12}^{\alpha} - (1-\tau_1)(p_2^{\alpha}(1-\beta_1) + s^{\alpha}(1-\beta_1) - \beta_1 v_1^{\alpha} + \rho_1(y^{\alpha}+p_+^{\alpha}+s^{\alpha}+u_1^{\alpha})), \\
        \Omega_2^{\alpha} = \Omega_{12}^{\alpha}.
    \end{gathered}
\end{equation}
Consequently
\begin{equation}
    \dd \Omega_1^2 \dd \Omega_2^2 \propto \dd \tau_1 (\dots)^{\alpha} \dd \Omega_{12 \alpha} \dd \Omega_{12}^2 = 0
    \label{Schouten-S2}
\end{equation}
since $\dd\Omega_\a  \dd \Omega_{\b} \dd \Omega_\gamma = 0$. This highlights the benefits of introducing $\Omega_{12}$. The action of $\dd$ on $P(-\overrightarrow{\sigma_1}, \sigma_1^1, \sigma_1^2, \overrightarrow{\sigma_1}) = \theta(\sigma_1^1+\overrightarrow{\sigma_1})\theta(\sigma_1^2-\sigma_1^1)\theta(\overrightarrow{\sigma_1}-\sigma_1^2)$ in the second term in \eqref{S2} yields \eqref{Schouten-S2} again because $\dd\sigma_1^1$ and $\dd \sigma_1^2$ are present in the measure and only $\dd\tau_1$ and $\dd\Omega_{12}$ contribute from $\dd\Omega_1^2$.  When acting on the exponent in the term where $\dd\Omega_1^2$ is absent, one gets a form degree higher than the dimension of $M$. The term with $\D(1-\tau_{12})$ gives the sum of \eqref{S_1*S_1} and \eqref{B_2*gamma}.

Choosing a solution of the form  \eqref{S2} allows us to obtain spin-local vertex. The novelty compared to the lower-order case of \eqref{B2}
(without $\tau_{12}$) is that the physical field equations now retain a contribution from the $\tau_{12} = 0$ condition, which does not enforce $\tau_1 = 0$ and/or $\tau_2 = 0$. As a result, the coefficient in front of $p_{1\alpha}p_2^{\alpha}$ in the exponent responsible for locality may be different from zero. However, as shown below, it still can be eliminated  in the appropriate limit.

\subsection{\texorpdfstring{$W_2$}{W2}}

$W_2$ can be obtained  from the equation
\begin{equation}
    2i\dd W_2 \cong \dd _x S_1 + \dd _x S_2 + \omega * S_2 + S_2 * \omega + W_1 * S_1 + S_1 * W_1 - i \eta B_{2}^{(2)}*\gamma.
    \label{W2_eqution}
\end{equation}\par
The last term here is weakly zero since $B_{2}^{(2)} \cong 0$. On the other hand,  $\dd _x S_1$ is accounted since the action of  $\dd_x$  on $C$ contained in $S_1$\eqref{S1} yields (\ref{dxCJ}) with $J$ contributing to the second order,
\begin{equation}
    \dd_x S_1|_{CC\omega, C\omega C, \omega CC} = \dfrac{\eta}{2}\fint{\tau \rho \beta \sigma} l(\tau) \D(\sigma) \mu(\rho, \beta) \dd \Omega^2 \E(\Omega) J(r, \bar{y}, K)\big|_{r = 0} k\,.
    \label{d_xS_1}
\end{equation}\par
Consider the $C C\omega$ sector. Let us rewrite the expression \eqref{d_xS_1} in the form \eqref{fundament}. We use the convention that the lower index 1 denotes objects related to $J$, while 2 refers to $S_1$. We also redefine $p_{\alpha} \rightarrow t_{\alpha}$ and $r_{\alpha}\rightarrow s_{\alpha}$ in \eqref{d_xS_1}. As a result, the exponent in the integrand takes the form
\begin{equation}
    \exp i \big [\Omega_{\alpha}(y+t+u_1)^{\alpha} + \Omega_{J\alpha}(s+p_++u_1)^{\alpha}+s_{\alpha} t^{\alpha}+u_{1 {\alpha}}v_1^{\alpha}+u_{2 {\alpha}}v_2^{\alpha} - p_{1 {\alpha}} p_2^{\alpha} - p_{1 {\alpha}} p_{\omega}^{\alpha} - p_{2 {\alpha}} p_{\omega}^{\alpha} - \sum_{i=1}^2 p_{i{\alpha}} r_i^{\alpha}  \big].
\end{equation}
Shifting $s_{\alpha} \rightarrow s_{\alpha}+y_{\alpha}$ and $t_{\alpha} \rightarrow t_{\alpha}-p_{+\alpha}$ we obtain
\begin{multline}
    \exp i \big[\Omega_{\alpha}(y+t-p_++u_2)^{\alpha}  + (\Omega_{J\alpha}-s_{\alpha})(y+s+p_++u_1)^{\alpha}+s_{\alpha} t^{\alpha}+(s_{\alpha}-t_{\alpha}) y^{\alpha}+p_{+\alpha} y^{\alpha} +u_{1 {\alpha}}(v_1-s)^{\alpha}+ \\ +u_{2 {\alpha}}v_2^{\alpha} - p_{1 {\alpha}} p_2^{\alpha} - p_{1 {\alpha}} p_{\omega}^{\alpha} - p_{2 {\alpha}} p_{\omega}^{\alpha} - \sum_{i=1}^2 p_{i{\alpha}} r_i^{\alpha}  \big].
\end{multline}
Changing $v_{1\alpha}$ to $v_{1\alpha}+s_{\alpha}$ and taking into account the factor of $\D(\tau_1)$ in the expression for $J$ \eqref{J}, we find:
\begin{equation}
    \Omega_{J}^{\alpha}-s^{\alpha} = -(p^{\alpha} (\sigma_1) - \beta_1 (v_1^{\alpha}+s^{\alpha}) + \rho_1(y^{\alpha}+p_+^{\alpha}+s^{\alpha}+u_1^{\alpha}))-s^{\alpha} = \Omega_1^{\alpha}(\tau_1 = 0).
\end{equation}
Next, replacing $\tau$ by $\tau_{2}$ and adding integration over $\tau_{12}$ with $\D(1-\tau_{12})$ we bring $\Omega$ to the form of $\Omega_2$. This yields
\begin{multline}
    \dd_x S_1|_{C C \omega} = \dfrac{i\eta^2}{8} \fint{\tau_1 \rho_1 \beta_1 \sigma_1^1 \sigma_1^2 \sigma_1^{\omega}  \tau_2 \rho_2 \beta_2 \sigma_2^1 \sigma_2^2 \sigma_2^{\omega} \tau_{12} \sigma_{12}^1 \sigma_{12}^2 \sigma_{12}^{\omega} a_1 a_2 \beta_{12}} \mu'(\beta_{12}) \D(1-\tau_{12}) \D(\tau_1) l(\tau_2)\D(1-\beta_1 - a_1)\D(a_2) P(-\overrightarrow{\sigma_1}, \sigma_1^1, \sigma_1^2, \sigma_1^{\omega}, \overrightarrow{\sigma_1}) \times \\ \times \D (\sigma_1^2-\sigma_1^1-\overrightarrow{\sigma_1})  \D (\sigma_2^1)\D(\sigma_2^2)\D(\sigma_2^{\omega})\D(\sigma_{12}^1)\D(\sigma_{12}^2)\D(\sigma_{12}^{\omega})\D(\rho_1+\sigma_1^1){\mu}(\beta_1){\mu}(\rho_2, \beta_2)\times \\ \times
    \dd {\Omega}_1^2 \dd {\Omega}_2^2 \E(\Omega_1, \Omega_2)  C(r_1, \bar{y}, K) \bar{*}   C(r_2, \bar{y}, K) \bar{*} \omega(r_{\omega}, \bar{y}, K) \big|_{r_i = 0}.
    \label{d_xS_1_1}
\end{multline}\par
$\dd_x S_2|_{C C \omega}$ is found using formula \eqref{star-property_in_right} while $S_2 * \omega$ is accomplished by \eqref{product_r}. $S_1*W_1$ can be calculated with the aid of \eqref{anz_mult}. As a result, we obtain
\begin{multline}
     2i\dd W_2|_{C C \omega} = \dfrac{i\eta^2}{8} \fint{\tau_1 \rho_1 \beta_1 \sigma_1^1 \sigma_1^2 \sigma_1^{\omega}  \tau_2 \rho_2 \beta_2 \sigma_2^1 \sigma_2^2 \sigma_2^{\omega} \tau_{12} \sigma_{12}^1 \sigma_{12}^2 \sigma_{12}^{\omega} a_1 a_2 \beta_{12}}\mu'(\beta_{12})\D(\overrightarrow{\sigma_1} - a_1)\D (\sigma_2^1-a_2)\D (\sigma_2^2)\D(\sigma_{12}^1)\D(\sigma_{12}^2)\big\{ l(\tau_{12}) l(\tau_1) l(\tau_2) \D (\sigma_1^1)\times \\ \times \D (\sigma_1^2-\overrightarrow{\sigma_1}) \D(\sigma_1^{\omega}-\sigma_1^{2}) \left[  \D(\sigma_2^{\omega}+\overrightarrow{\sigma_2})\D(\sigma_{12}^{\omega}+\overrightarrow{\sigma_{12}})- \D(\sigma_2^{\omega}-\sigma_2^{2})\D(\sigma_{12}^{\omega}) \right]\D(\overrightarrow{\sigma_2} - a_2)\mu(\rho_1, \beta_1)\dd\Omega_1^2 - \\ -  l(\tau_{12}) l(\tau_1) \D(a_2) \D(1-\tau_2) P(-\overrightarrow{\sigma_1}, \sigma_1^1, \sigma_1^2, \overrightarrow{\sigma_1})\D (\sigma_1^2-\sigma_1^1-\overrightarrow{\sigma_1}) \D(\sigma_2^{\omega})\times \\ \times \left[ \D(\sigma_1^{\omega}-\overrightarrow{\sigma_1})\D(\sigma_{12}^{\omega}+\overrightarrow{\sigma_{12}}) -\D(\sigma_1^{\omega}-\sigma_1^{2})\D(\sigma_{12}^{\omega}) \right] \D(\rho_1+\sigma_1^1)\mu(\beta_1)(\dd \Omega_1^2 + 2i(1-\beta_1)\D(\sigma_1^2)\dd \tau_1)  + \\ +\D(1-\tau_{12}) \D(\tau_1) l(\tau_2)\D(a_2) P(-\overrightarrow{\sigma_1}, \sigma_1^1, \sigma_1^2, \sigma_1^{\omega}, \overrightarrow{\sigma_1})\D (\sigma_1^2-\sigma_1^1-\overrightarrow{\sigma_1}) \D(\sigma_2^{\omega}) \D(\sigma_{12}^{\omega})\D(\rho_1+\sigma_1^1)\mu(\beta_1)\dd \Omega_1^2  - \\ - \D(1-\tau_{12}) l(\tau_1) l(\tau_2) P(-\overrightarrow{\sigma_2},\sigma_2^{\omega},\sigma_2^2)\D (\sigma_1^1) \D(\sigma_1^2-\overrightarrow{\sigma_1}) \D(\sigma_1^{\omega}-\overrightarrow{\sigma_1}) \D(\sigma_{12}^{\omega})\D(\overrightarrow{\sigma_2} - a_2) {\mu}(\rho_1, \beta_1)\dd {\Omega}_1^2  \big\}\times \\ \times{\mu}(\rho_2, \beta_2)
     \dd {\Omega}_2^2 \E(\Omega_1, \Omega_2)  C(r_1, \bar{y}, K) \bar{*}   C(r_2, \bar{y}, K)\bar{*} \omega(r_{\omega}, \bar{y}, K) \big|_{r_i = 0}.
    \label{dW2}
\end{multline} \par
To find $W_2$ one has to represent  the {\it r.h.s.} of this expression in the $\dd$-exact form. This is achieved in terms of the measure of $\sigma_i^k$ yielding the following result
\begin{equation}
    \begin{gathered}
     W_2|_{C C \omega}  = \dfrac{\eta^2}{16} \fint{\tau_1 \rho_1 \beta_1 \sigma_1^1 \sigma_1^2 \sigma_1^{\omega}  \tau_2 \rho_2 \beta_2 \sigma_2^1 \sigma_2^2 \sigma_2^{\omega} \tau_{12} \sigma_{12}^1 \sigma_{12}^2 \sigma_{12}^{\omega} a_1 a_2 \beta_{12}}\mu'(\beta_{12}) \D(a_2-\overrightarrow{\sigma_2})\D(a_1-\overrightarrow{\sigma_1})l(\tau_{12})P(-\overrightarrow{\sigma_{12}},\sigma_{12}^{\omega},\sigma_{12}^2)\D(\sigma_{12}^1)\D(\sigma_{12}^2) \D (\sigma_2^1-\overrightarrow{\sigma_2})\D (\sigma_2^2) l(\tau_1)\times \\ \times  \big \{ l(\tau_2) \D (\sigma_1^1)\D (\sigma_1^2-\overrightarrow{\sigma_1})\D(\sigma_1^{\omega}-\overrightarrow{\sigma_1})  P(-\overrightarrow{\sigma_2},\sigma_2^{\omega},\sigma_2^2) \left[\D (\sigma_2^{\omega}+\overrightarrow{\sigma_2})-\D(\sigma_{12}^{\omega}-\sigma_{12}^2) \right]\mu(\rho_1, \beta_1)\dd\Omega_1^2 -\\-  \D(1-\tau_2) P(-\overrightarrow{\sigma_1}, \sigma_1^1, \sigma_1^2, \sigma_1^{\omega}, \overrightarrow{\sigma_1})\D (\sigma_1^2-\sigma_1^1-\overrightarrow{\sigma_1}) \D(\sigma_2^{\omega}) \left[\D(\sigma_1^{\omega}-\overrightarrow{\sigma_1})+\D(\sigma_{12}^{\omega}-\sigma_{12}^2)\right]   \D(\rho_1+\sigma_1^1)\times \\ \times\mu(\beta_1)(\dd \Omega_1^2 + 2i(1-\beta_1)\D(\sigma_1^2)\dd \tau_1)\big \}{\mu}(\rho_2, \beta_2) \dd {\Omega}_2^2 {\E}  C(r_1, \bar{y}, K) \bar{*}   C(r_2, \bar{y}, K)\bar{*} \omega(r_{\omega}, \bar{y}, K) \big|_{r_i = 0} +\\+\dfrac{\eta^2}{16} \fint{\tau_1 \rho_1 \beta_1 \sigma_1^1 \sigma_1^2 \sigma_1^{\omega}  \tau_2 \rho_2 \beta_2 \sigma_2^1 \sigma_2^2 \sigma_2^{\omega} \tau_{12} \sigma_{12}^1 \sigma_{12}^2 \sigma_{12}^{\omega} a_1 a_2\beta_{12}}\mu'(\beta_{12})\D(\overrightarrow{\sigma_1}-a_1)\D(a_2)\D(\tau_1) l(\tau_2) l(\tau_{12}) P(-\overrightarrow{\sigma_1}, \sigma_1^1, \sigma_1^2, \sigma_1^{\omega}, \overrightarrow{\sigma_1})\D (\sigma_1^2-\sigma_1^1-\overrightarrow{\sigma_1}) \times \\ \times \D (\sigma_2^1)\D(\sigma_2^2)\D(\sigma_2^{\omega})  \D(\sigma_{12}^1)\D(\sigma_{12}^2)\D(\sigma_{12}^{\omega}) \D(\rho_1+\sigma_1^1) {\mu}(\beta_1){\mu}(\rho_2, \beta_2) \times \\ \times
    \dd {\Omega}_1^2 \dd {\Omega}_2^2 {\E}  C(r_1, \bar{y}, K) \bar{*}   C(r_2, \bar{y}, K)\bar{*} \omega(r_{\omega}, \bar{y}, K) \big|_{r_i = 0} .
    \end{gathered}
    \label{W2}
\end{equation}
Note that the last term in \eqref{W2} is weak because integrating over $s^{\alpha}, t^{\alpha}, u_{12}^{\alpha}$ gives $v_{12}^{\alpha} \propto z^{\alpha}$ and hence $\dd \Omega_2^2\cong  2\dd\tau_2 \dd \tau_{12} \tau_2 (z-\beta_{12}v_{12})_{\alpha}\beta_{12}v_{12}^{\alpha} \cong 0$. One can verify that this expression (second integral in \eqref{W2}) coincides with $\tr_\l \dd_x S_1|_{ C C  \omega}$, with $\l= \mu'(\b_{12}) \D(\s_{12}^\w) \D(\s_{12}^1) \D(\s_{12}^2)$ and $\tr$ defined in \eqref{eq:shift_diff_op}. Further, the term under the first integral in \eqref{W2}  with factors $\D(1-\tau_2)$ and $\D(\sigma_{12}^{\omega}-\sigma_{12}^2)$
\begin{equation}
\begin{gathered}
    \dfrac{\eta^2}{16} \fint{\tau_1 \rho_1 \beta_1 \sigma_1^1 \sigma_1^2 \sigma_1^{\omega}  \tau_2 \rho_2 \beta_2 \sigma_2^1 \sigma_2^2 \sigma_2^{\omega} \tau_{12} \sigma_{12}^1 \sigma_{12}^2 \sigma_{12}^{\omega} a_1 a_2 \beta_{12}}\mu'(\beta_{12}) \D(a_2-\overrightarrow{\sigma_2})\D(a_1-\overrightarrow{\sigma_1})l(\tau_{12})P(-\overrightarrow{\sigma_{12}},\sigma_{12}^{\omega},\sigma_{12}^2)\D(\sigma_{12}^1)\D(\sigma_{12}^2) \D (\sigma_2^1-\overrightarrow{\sigma_2})\D (\sigma_2^2) l(\tau_1)\times \\ \times   \D(1-\tau_2) P(-\overrightarrow{\sigma_1}, \sigma_1^1, \sigma_1^2, \sigma_1^{\omega}, \overrightarrow{\sigma_1})\D (\sigma_1^2-\sigma_1^1-\overrightarrow{\sigma_1}) \D(\sigma_2^{\omega}) \D(\sigma_{12}^{\omega}-\sigma_{12}^2)   \D(\rho_1+\sigma_1^1)\times \\ \times\mu(\beta_1)(\dd \Omega_1^2 + 2i(1-\beta_1)\D(\sigma_1^2)\dd \tau_1) {\mu}(\rho_2, \beta_2) \dd {\Omega}_2^2 {\E}  C(r_1, \bar{y}, K) \bar{*}   C(r_2, \bar{y}, K)\bar{*} \omega(r_{\omega}, \bar{y}, K) \big|_{r_i = 0}
\end{gathered}
\end{equation}
is weakly zero since $\dd\tau_1$, $\dd\sigma_1^{1}$, $\dd\sigma_1^{2}$, $\dd\sigma_1^{\omega}$ can only occur from the three-form $\D (\sigma_1^2-\sigma_1^1-\overrightarrow{\sigma_1}) (\dd \Omega_1^2 + 2i(1-\beta_1)\D(\sigma_1^2)\dd \tau_1)$.\par
To explain how the measure in \eqref{dW2} is integrated over $\sigma_i^k$, let us consider the  action of $\dd$  on the first term under the first integral
in \eqref{W2}
\begin{multline}
    \dd P(-\overrightarrow{\sigma_{12}},\sigma_{2}^{\omega},\sigma_{12}^2) P(-\overrightarrow{\sigma_2},\sigma_2^{\omega},\sigma_2^2)\left[\D (\sigma_2^{\omega}+\overrightarrow{\sigma_2})-\D(\sigma_{12}^{\omega}-\sigma_{12}^2) \right] = \D(\sigma_{12}^{\omega}+\overrightarrow{\sigma_{12}})\D(\sigma_2^{\omega}+\overrightarrow{\sigma_2})+\\+\D(\sigma_{12}^2-\sigma_{12}^{\omega}) \D(\sigma_2^{\omega}+\overrightarrow{\sigma_2})- \D(\sigma_2^{\omega}+\overrightarrow{\sigma_2})\D(\sigma_{12}^{\omega}-\sigma_{12}^2) - \D(\sigma_2^2-\sigma_2^{\omega})\D(\sigma_{12}^{\omega}-\sigma_{12}^2) =\\ = \D(\sigma_2^{\omega} - \sigma_2^2) \D(\sigma_{12}^{\omega}-\sigma_{12}^2) -  \D(\sigma_2^{\omega}+\overrightarrow{\sigma_2}) \D(\sigma_{12}^{\omega}+\overrightarrow{\sigma_{12}})
    \label{W2_check}
\end{multline}
Thus, from here
 we obtain the term with the first square bracket in \eqref{dW2}. The second term of the first integral \eqref{W2} is obtained analogously. The measure of $\sigma_i^k$ allows us to integrate the part of the equation \eqref{W2_eqution} containing $S_2$. The action of $\dd$ on the $\tau_i$-dependent factors in the measure reproduces the rest terms in \eqref{W2_eqution}.\par
More in detail, the action of $\dd$ on \eqref{W2} modulo weak terms yields (the factors  arising from the action of $\dd$ are highlighted)
\begin{multline}
     2i\dd W_2|_{C C \omega}  = -\dfrac{i\eta^2}{8} \fint{\tau_1 \rho_1 \beta_1 \sigma_1^1 \sigma_1^2 \sigma_1^{\omega}  \tau_2 \rho_2 \beta_2 \sigma_2^1 \sigma_2^2 \sigma_2^{\omega} \tau_{12} \sigma_{12}^1 \sigma_{12}^2 \sigma_{12}^{\omega} a_1 a_2 \beta_{12}} \mu'(\beta_{12}) {\mu}(\rho_2, \beta_2) \dd {\Omega}_2^2 {\E}  C(r_1, \bar{y}, K) \bar{*}   C(r_2, \bar{y}, K)\bar{*} \omega(r_{\omega}, \bar{y}, K) \big|_{r_i = 0} 
     \\ 
     \bigg\{  \D(a_2-\overrightarrow{\sigma_2})\D(a_1-\overrightarrow{\sigma_1}) \bigg[ l(\tau_{12}) \D(\sigma_{12}^1)\D(\sigma_{12}^2) \D (\sigma_2^1-\overrightarrow{\sigma_2})\D (\sigma_2^2) l(\tau_1) \big \{ l(\tau_2) \D (\sigma_1^1)\D (\sigma_1^2-\overrightarrow{\sigma_1}) \times 
     \\ 
     \times \D(\sigma_1^{\omega}-\overrightarrow{\sigma_1}) \big [ \D(\sigma_2^{\omega}+\overrightarrow{\sigma_2}) \bm{\D(\sigma_{12}^{\omega}+\overrightarrow{\sigma_{12}})} \bm{\D(\sigma_2^{\omega} - \sigma_2^2)} \D(\sigma_{12}^{\omega}-\sigma_{12}^2) \big ] \mu(\rho_1, \beta_1)\dd\Omega_1^2- 
     \\
     -  \D(1-\tau_2) P(-\overrightarrow{\sigma_1}, \sigma_1^1, \sigma_1^2, \sigma_1^{\omega}, \overrightarrow{\sigma_1})\D (\sigma_1^2-\sigma_1^1-\overrightarrow{\sigma_1}) \D(\sigma_2^{\omega}) \times
     \\
     \times \big[\D(\sigma_1^{\omega}-\overrightarrow{\sigma_1})\bm{\D(\sigma_{12}^{\omega}+\overrightarrow{\sigma_{12}})}-\bm{\D(\sigma_1^{\omega}-\sigma_1^2)}\D(\sigma_{12}^{\omega}-\sigma_{12}^2) \big]   \D(\rho_1+\sigma_1^1) \mu(\beta_1)(\dd \Omega_1^2 + 2i(1-\beta_1)\D(\sigma_1^2)\dd \tau_1)\big\}+
    \label{dW2_1}
\end{multline}
\begin{multline}
     +l(\tau_{12})P(-\overrightarrow{\sigma_{12}},\sigma_{12}^{\omega},\sigma_{12}^2)\D(\sigma_{12}^1)\D(\sigma_{12}^2) \D (\sigma_2^1-\overrightarrow{\sigma_2})\D (\sigma_2^2) l(\tau_1)  \D(1-\tau_2) P(-\overrightarrow{\sigma_1}, \sigma_1^1, \sigma_1^2, \sigma_1^{\omega}, \overrightarrow{\sigma_1})\times \\ \times \D (\sigma_1^2-\sigma_1^1-\overrightarrow{\sigma_1}) \D(\sigma_2^{\omega}) \bm{\D(\sigma_1^1+\overrightarrow{\sigma_1})} \D(\sigma_{12}^{\omega}) \D(\rho_1+\sigma_1^1) \mu(\beta_1)\dd \Omega_1^2 +
    \label{dW2_2}
\end{multline}
\begin{multline}
     +l(\tau_{12})P(-\overrightarrow{\sigma_{12}},\sigma_{12}^{\omega},\sigma_{12}^2)\D(\sigma_{12}^1)\D(\sigma_{12}^2) \D (\sigma_2^1-\overrightarrow{\sigma_2})\D (\sigma_2^2) l(\tau_1)  \D(1-\tau_2) P(-\overrightarrow{\sigma_1}, \sigma_1^1, \sigma_1^2, \sigma_1^{\omega}, \overrightarrow{\sigma_1})\times \\ \times \D (\sigma_1^2-\sigma_1^1-\overrightarrow{\sigma_1}) \D(\sigma_2^{\omega}) \D(\sigma_{12}^{\omega}) \D(\rho_1+\sigma_1^1) \mu(\beta_1)(2i(1-\beta_1)\D(\sigma_1^2)\dd \tau_1) \E^{-1}\bm{\dd \E}-
    \label{dW2_2'}
\end{multline}
\begin{multline}
     - l(\tau_{12}) \D(\sigma_{12}^1)\D(\sigma_{12}^2) \D (\sigma_2^1-\overrightarrow{\sigma_2})\D (\sigma_2^2) \bm{\D(1-\tau_1)} l(\tau_2) \D (\sigma_1^1)\D (\sigma_1^2-\overrightarrow{\sigma_1})\D(\sigma_1^{\omega}-\overrightarrow{\sigma_1})\times \\ \times P(-\overrightarrow{\sigma_2},\sigma_2^{\omega},\sigma_2^2) \D(\sigma_{12}^{\omega})\mu(\rho_1, \beta_1)\dd\Omega_1^2 -
    \label{dW2_3}
\end{multline}
\begin{multline}
     - \bm{\D(1-\tau_{12})}\D(\sigma_{12}^1)\D(\sigma_{12}^2)  \D (\sigma_2^1-\overrightarrow{\sigma_2})\D (\sigma_2^2) l(\tau_1)l(\tau_2) \D (\sigma_1^1)\D (\sigma_1^2-\overrightarrow{\sigma_1})\D(\sigma_1^{\omega}-\overrightarrow{\sigma_1})  \times \\ \times P(-\overrightarrow{\sigma_2},\sigma_2^{\omega},\sigma_2^2) \D(\sigma_{12}^{\omega})\mu(\rho_1, \beta_1)\dd\Omega_1^2 -
    \label{dW2_4}
\end{multline}
\begin{multline}
     -l(\tau_{12})P(-\overrightarrow{\sigma_{12}},\sigma_{12}^{\omega},\sigma_{12}^2)\D(\sigma_{12}^1)\D(\sigma_{12}^2) \D (\sigma_2^1-\overrightarrow{\sigma_2})\D (\sigma_2^2) \bm{\D(\tau_1)} \D(1-\tau_2) P(-\overrightarrow{\sigma_1}, \sigma_1^1, \sigma_1^2, \sigma_1^{\omega}, \overrightarrow{\sigma_1})\times \\ \times\D (\sigma_1^2-\sigma_1^1-\overrightarrow{\sigma_1})  \D(\sigma_2^{\omega}) \D(\sigma_{12}^{\omega})  \D(\rho_1+\sigma_1^1) \mu(\beta_1)(\dd \Omega_1^2 + 2i(1-\beta_1)\D(\sigma_1^2)\dd \tau_1) \Bigg] -
    \label{dW2_5}
\end{multline}
\begin{multline}
     -\D(\overrightarrow{\sigma_1}-a_1)\D(a_2)\D(\tau_1) \{\bm{\D(1-\tau_2)} l(\tau_{12}) + l(\tau_2) \bm{\D(1-\tau_{12})} \} P(-\overrightarrow{\sigma_1}, \sigma_1^1, \sigma_1^2, \sigma_1^{\omega}, \overrightarrow{\sigma_1})\times \\ \times \D (\sigma_1^2-\sigma_1^1-\overrightarrow{\sigma_1})  \D (\sigma_2^1)\D(\sigma_2^2)\D(\sigma_2^{\omega})  \D(\sigma_{12}^1)\D(\sigma_{12}^2)\D(\sigma_{12}^{\omega}) \D(\rho_1+\sigma_1^1) {\mu}(\beta_1)
    \dd {\Omega}_1^2 \bigg \}.
    \label{dW2_7}
\end{multline}
\par As a result, \eqref{dW2_1} reproduces terms without $\D(1-\tau_{12})$ in \eqref{dW2},  \eqref{dW2_4}  equals to the last term of \eqref{dW2}, and
 the term with $\D(1-\tau_2)$ in \eqref{dW2_7} cancels \eqref{dW2_5}. The term with $\D(1-\tau_{12})$ equals to that  with $\D(\tau_1)$ in \eqref{dW2}.

\par Consider \eqref{dW2_2}, \eqref{dW2_2'}  and \eqref{dW2_3}, rewriting the last one using the symmetry property \eqref{symmetry}.
 \begin{multline}
     \dfrac{\eta^2}{16} \fint{\tau_1 \rho_1 \beta_1 \sigma_1^1 \sigma_1^2 \sigma_1^{\omega}  \tau_2 \rho_2 \beta_2 \sigma_2^1 \sigma_2^2 \sigma_2^{\omega} \tau_{12} \sigma_{12}^1 \sigma_{12}^2 \sigma_{12}^{\omega} a_1 a_2 \beta_{12}}\mu'(\beta_{12}) l(\tau_1) \D(1-\tau_2) l(\tau_{12}) P(-\overrightarrow{\sigma_1}, \sigma_1^1, \sigma_1^2, \sigma_1^{\omega}, \overrightarrow{\sigma_1})\D (\sigma_1^1+\overrightarrow{\sigma_1})\D(\sigma_1^2) \D (\sigma_2^1)\D(\sigma_2^2)\D(\sigma_2^{\omega}) \times \\ \times \D(\sigma_{12}^{\omega}) \D(\sigma_{12}^1)\D(\sigma_{12}^2)\big[-{\mu}(\rho_1, \beta_1)\dd {\Omega}_1^2 {\E} + \D(\rho_1 + \sigma_1^1) {\mu}(\beta_1)\dd {\Omega}_1^2 {\E} -2i(1-\beta_1)\D(\rho_1 + \sigma_1^1){\mu}(\beta_1)\dd \tau_1 \dd {\E}  \big] \times \\ \times {\mu}(\rho_2, \beta_2) \D(a_2-\overrightarrow{\sigma_2})\D(a_1-\overrightarrow{\sigma_1})
     \dd {\Omega}_2^2   C(r_1, \bar{y}, K) \bar{*}   C(r_2, \bar{y}, K)\bar{*} \omega(r_{\omega}, \bar{y}, K) \big|_{r_i = 0}.
    \label{W2_anal}
 \end{multline}
 \par
 Expansions of $\dd \Omega_1^2$ in the first and the second terms have the form
 \begin{equation}
     2\dd \tau_1 (\Omega_{12}-\beta_1 v_1 - (1-\beta_1)p_1 + (1-\beta_1)s)_{\alpha} p_{\omega}^{\alpha} \dd \sigma_1^{\omega}(1-\tau_1),
 \end{equation}
 \begin{equation}
     2\dd \tau_1 (\Omega_{12}-\beta_1 v_1 - (1-\beta_1)p_1 + (1-\beta_1)s -(1-\beta_1)(y+s+p_++u_1))_{\alpha} p_{\omega}^{\alpha} \dd \sigma_1^{\omega}(1-\tau_1).
 \end{equation}
 Their difference is $2\dd \tau_1 (1-\beta_1)(y+s+p_++u_1)_{\alpha} p_{\omega}^{\alpha} \dd \sigma_1^{\omega}(1-\tau_1)$. It is compensated by the third term. Thus, \eqref{W2_anal} vanishes.\par
 Let us show only two of weak terms  ignored in \eqref{dW2_1}.
\begin{equation}
    \begin{gathered}
     \dfrac{i\eta^2}{8} \fint{\tau_1 \rho_1 \beta_1 \sigma_1^1 \sigma_1^2 \sigma_1^{\omega}  \tau_2 \rho_2 \beta_2 \sigma_2^1 \sigma_2^2 \sigma_2^{\omega} \tau_{12} \sigma_{12}^1 \sigma_{12}^2 \sigma_{12}^{\omega} a_1 a_2 \beta_{12}} \mu'(\beta_{12}) {\mu}(\rho_2, \beta_2) \dd {\Omega}_2^2 {\E}  C(r_1, \bar{y}, K) \bar{*}   C(r_2, \bar{y}, K)\bar{*} \omega(r_{\omega}, \bar{y}, K) \big|_{r_i = 0} \D(a_2-\overrightarrow{\sigma_2})\D(a_1-\overrightarrow{\sigma_1})  \times 
     \\ 
     \times \Bigg[ \bm{\D(\tau_{12})}P(-\overrightarrow{\sigma_{12}},\sigma_{12}^{\omega},\sigma_{12}^2)\D(\sigma_{12}^1)\D(\sigma_{12}^2) \D (\sigma_2^1-\overrightarrow{\sigma_2})\D (\sigma_2^2) l(\tau_1) \big \{ l(\tau_2) \D (\sigma_1^1)\D (\sigma_1^2-\overrightarrow{\sigma_1}) \times 
     \\
     \times \D(\sigma_1^{\omega}-\overrightarrow{\sigma_1})  P(-\overrightarrow{\sigma_2},\sigma_2^{\omega},\sigma_2^2) \left[\D (\sigma_2^{\omega}+\overrightarrow{\sigma_2})-\D(\sigma_{12}^{\omega}) \right]\mu(\rho_1, \beta_1)\dd\Omega_1^2-  \D(1-\tau_2) P(-\overrightarrow{\sigma_1}, \sigma_1^1, \sigma_1^2, \sigma_1^{\omega}, \overrightarrow{\sigma_1}) \times 
     \\ 
     \times \D (\sigma_1^2-\sigma_1^1-\overrightarrow{\sigma_1}) \D(\sigma_2^{\omega}) \left[\D(\sigma_1^{\omega}-\overrightarrow{\sigma_1})+\D(\sigma_{12}^{\omega})\right]   \D(\rho_1+\sigma_1^1) \mu(\beta_1)(\dd \Omega_1^2 + 2i(1-\beta_1)\D(\sigma_1^2)\dd \tau_1)\big \}
    \end{gathered}
    \label{dW2_weak_1}
\end{equation}
\begin{equation}
    \begin{gathered}
     - \bm{\D(1-\tau_{12})}P(-\overrightarrow{\sigma_{12}},\sigma_{12}^{\omega},\sigma_{12}^2)\D(\sigma_{12}^1)\D(\sigma_{12}^2) \D (\sigma_2^1-\overrightarrow{\sigma_2})\D (\sigma_2^2) l(\tau_1)  \D(1-\tau_2) P(-\overrightarrow{\sigma_1}, \sigma_1^1, \sigma_1^2, \sigma_1^{\omega}, \overrightarrow{\sigma_1})\times \\ \times \D (\sigma_1^2-\sigma_1^1-\overrightarrow{\sigma_1}) \D(\sigma_2^{\omega}) \D(\sigma_{12}^{\omega}) \D(\rho_1+\sigma_1^1) \mu(\beta_1)(\dd \Omega_1^2 + 2i(1-\beta_1)\D(\sigma_1^2)\dd \tau_1) \Bigg ].
    \end{gathered}
    \label{dW2_weak_2}
\end{equation}
\par
\eqref{dW2_weak_1} is weak, as the form is integrated over a space dimension of which is smaller than the form degree ($\dd z_{\alpha} = \theta_{\alpha}$, which contributes via $\Omega_{12}$, is absent).
The expression \eqref{dW2_weak_2} is weak because the form is integrated over a space dimension of which is greater than its degree.
Other discarded weak terms can be analyzed analogously.
\par
 The resulting $W_2$ again respects the property necessary for spin-locality, discussed after \eqref{Schouten-S2}.\par
 To find $W_2$  in other sectors one has to consider $\dd_x S_2 + \omega*S_2 + S_2 * \omega$ and to integrate the measure of $\sigma_i^k$. In the ${C \omega C}$ sector we obtain
 \begin{multline}
    (\D(\sigma_1^{\omega}-\sigma_1^1)\D(\sigma_2^{\omega}-\sigma_2^1)-\D(\sigma_1^{\omega}-\sigma_1^2)\D(\sigma_2^{\omega}-\sigma_2^2))\D(\sigma_{12}^{\omega}) =\\= \dd P( \sigma_1^1, \sigma_1^{\omega}, \sigma_1^2) P(\sigma_2^2,\sigma_2^{\omega},\sigma_2^1) \left[\D (\sigma_2^{\omega}-\sigma_2^1)+\D(\sigma_{1}^{\omega}-\sigma_1^2) \right]\D(\sigma_{12}^{\omega}),
\end{multline}
that leads to the first term in the following expression:
\begin{equation}
    \begin{gathered}
     W_2|_{C \omega C}  = -\dfrac{\eta^2}{16} \fint{\tau_1 \rho_1 \beta_1 \sigma_1^1 \sigma_1^2 \sigma_1^{\omega}  \tau_2 \rho_2 \beta_2 \sigma_2^1 \sigma_2^2 \sigma_2^{\omega} \tau_{12} \sigma_{12}^1 \sigma_{12}^2 \sigma_{12}^{\omega} a_1 a_2  \beta_{12}}\mu'(\beta_{12})\D(a_2-\overrightarrow{\sigma_2})\D(a_1-\overrightarrow{\sigma_1}) l(\tau_{12}) l(\tau_1) l(\tau_2) P( \sigma_1^1, \sigma_1^{\omega}, \sigma_1^2)\D (\sigma_1^1)\D (\sigma_1^2-\overrightarrow{\sigma_1})\times \\ \times  \D (\sigma_2^1-\overrightarrow{\sigma_2})\D (\sigma_2^2)  P(\sigma_2^2,\sigma_2^{\omega},\sigma_2^1) \left[\D (\sigma_2^{\omega}-\sigma_2^1)+\D(\sigma_{1}^{\omega}-\sigma_1^2) \right]\D(\sigma_{12}^1)\D(\sigma_{12}^2)\D(\sigma_{12}^{\omega}) \times \\ \times {\mu}(\rho_1, \beta_1){\mu}(\rho_2, \beta_2)
    \dd {\Omega}_1^2 \dd {\Omega}_2^2 {\E}  C(r_1, \bar{y}, K)\bar{*} \omega(r_{\omega}, \bar{y}, K) \bar{*}   C(r_2, \bar{y}, K) \big|_{r_i = 0} +\\+\dfrac{\eta^2}{16} \fint{\tau_1 \rho_1 \beta_1 \sigma_1^1 \sigma_1^2 \sigma_1^{\omega}  \tau_2 \rho_2 \beta_2 \sigma_2^1 \sigma_2^2 \sigma_2^{\omega} \tau_{12} \sigma_{12}^1 \sigma_{12}^2 \sigma_{12}^{\omega} a_1 a_2\beta_{12}}\mu'(\beta_{12})\D(\overrightarrow{\sigma_1}-a_1)\D(a_2)\D(\tau_1) l(\tau_2) l(\tau_{12}) P(-\overrightarrow{\sigma_1}, \sigma_1^1, \sigma_1^{\omega}, \sigma_1^2, \overrightarrow{\sigma_1})\D (\sigma_1^2-\sigma_1^1-\overrightarrow{\sigma_1}) \times \\ \times \D (\sigma_2^1)\D(\sigma_2^2)\D(\sigma_2^{\omega})  \D(\sigma_{12}^1)\D(\sigma_{12}^2)\D(\sigma_{12}^{\omega}) \D(\rho_1+\sigma_1^1) {\mu}(\beta_1){\mu}(\rho_2, \beta_2) \times \\ \times
    \dd {\Omega}_1^2 \dd {\Omega}_2^2 {\E}  C(r_1, \bar{y}, K)\bar{*} \omega(r_{\omega}, \bar{y}, K) \bar{*}   C(r_2, \bar{y}, K) \big|_{r_i = 0} ,
    \end{gathered}
    \label{W2CwC}
\end{equation}
where the last term can be obtained as $\tr_\l \dd_x S_1|_{C \omega C}$, with $\l= \mu'(\b_{12}) \D(\s_{12}^1) \D(\s_{12}^\w) \D(\s_{12}^2)$, and is  weak.
\par
The ${\omega C C}$ sector yields
\begin{multline}
    (\D(\sigma_1^{\omega}+\overrightarrow{\sigma_1})\D(\sigma_{12}^{\omega}+\overrightarrow{\sigma_{12}})+\D(\sigma_1^{\omega}-\sigma_1^1)\D(\sigma_{12}^{\omega}-\sigma_{12}^1))\D(\sigma_{2}^{\omega}-\overrightarrow{\sigma_2}) =\\= -\dd P(-\overrightarrow{\sigma_{12}},\sigma_{12}^{\omega}, \sigma_{12}^1) P(-\overrightarrow{\sigma_1},\sigma_1^{\omega},\sigma_1^1) \left[\D (\sigma_1^{\omega}+\overrightarrow{\sigma_1})-\D(\sigma_{12}^{\omega}-\sigma_{12}^1) \right] \D(\sigma_2^{\omega}-\overrightarrow{\sigma_2}),
\end{multline}
and we can obtain
\begin{multline}
    W_2|_{\omega C C} =
     \\
    \begin{gathered}
     = \dfrac{\eta^2}{16} \fint{\tau_1 \rho_1 \beta_1 \sigma_1^1 \sigma_1^2 \sigma_1^{\omega}  \tau_2 \rho_2 \beta_2 \sigma_2^1 \sigma_2^2 \sigma_2^{\omega} \tau_{12} \sigma_{12}^1 \sigma_{12}^2 \sigma_{12}^{\omega}a_1 a_2 \beta_{12}}\mu'(\beta_{12})\D(a_2-\overrightarrow{\sigma_2})\D(a_1-\overrightarrow{\sigma_1}) l(\tau_{12})l(\tau_1) P(-\overrightarrow{\sigma_{12}},\sigma_{12}^{\omega}, \sigma_{12}^1)\D(\sigma_{12}^1)\D(\sigma_{12}^2) \D (\sigma_2^1-\overrightarrow{\sigma_2})\D (\sigma_2^2)\times 
     \\
     \times \D(\sigma_2^{\omega}-\overrightarrow{\sigma_2})   \big \{ l(\tau_2) \D (\sigma_1^1)\D (\sigma_1^2-\overrightarrow{\sigma_1})  P(-\overrightarrow{\sigma_1},\sigma_1^{\omega},\sigma_1^1) \left[\D (\sigma_1^{\omega}+\overrightarrow{\sigma_1})-\D(\sigma_{12}^{\omega}-\sigma_{12}^1) \right]{\mu}(\rho_1, \beta_1)
    \dd {\Omega}_1^2 -
    \\
    -  \D(1-\tau_2) P(-\overrightarrow{\sigma_1}, \sigma_1^{\omega}, \sigma_1^1, \sigma_1^2, \overrightarrow{\sigma_1})\D (\sigma_1^2-\sigma_1^1-\overrightarrow{\sigma_1}) \left[\D(\sigma_1^{\omega}+\overrightarrow{\sigma_1})-\D(\sigma_{12}^{\omega}-\sigma_{12}^1)\right]  \D(\rho_1+\sigma_1^1)\mu(\beta_1)\times \\ \times(\dd \Omega_1^2 + 2i(1-\beta_1)\D(\sigma_1^2)\dd \tau_1)\big \} {\mu}(\rho_2, \beta_2) \dd {\Omega}_2^2 {\E}  \omega(r_{\omega}, \bar{y}, K)\bar{*} C(r_1, \bar{y}, K) \bar{*}   C(r_2, \bar{y}, K) \big|_{r_i = 0} +\\+\dfrac{\eta^2}{16} \fint{\tau_1 \rho_1 \beta_1 \sigma_1^1 \sigma_1^2 \sigma_1^{\omega}  \tau_2 \rho_2 \beta_2 \sigma_2^1 \sigma_2^2 \sigma_2^{\omega} \tau_{12} \sigma_{12}^1 \sigma_{12}^2 \sigma_{12}^{\omega} a_1 a_2\beta_{12}}\mu'(\beta_{12})\D(\overrightarrow{\sigma_1}-a_1)\D(a_2)\D(\tau_1) l(\tau_2) l(\tau_{12}) P(-\overrightarrow{\sigma_1}, \sigma_1^{\omega}, \sigma_1^1, \sigma_1^2, \overrightarrow{\sigma_1})\D (\sigma_1^2-\sigma_1^1-\overrightarrow{\sigma_1}) \times 
    \end{gathered}
    \\
    \times \D (\sigma_2^1)\D(\sigma_2^2)\D(\sigma_2^{\omega})  \D(\sigma_{12}^1)\D(\sigma_{12}^2)\D(\sigma_{12}^{\omega}) \D(\rho_1+\sigma_1^1) {\mu}(\beta_1){\mu}(\rho_2, \beta_2) \times 
    \\
    \times
    \dd {\Omega}_1^2 \dd {\Omega}_2^2 {\E}  \omega(r_{\omega}, \bar{y}, K)\bar{*}C(r_1, \bar{y}, K) \bar{*}   C(r_2, \bar{y}, K) \big|_{r_i = 0},
    \label{W2wCC}
\end{multline}
where the last term is  weak again.\par
Modulo weak terms, $W_2$ can be represented in the following concise form:
\begin{multline}
    W_2 = \frac{1}{2i} \fint{\sigma_1^{\omega} \sigma_2^{\omega} \sigma_{12}^{\omega}} \Big \{ P(-\overrightarrow{\sigma_{12}},\sigma_{12}^{\omega},\sigma_{12}^2) P(-\overrightarrow{\sigma_2},\sigma_2^{\omega},\sigma_2^2) \D(\sigma_1^{\omega}-\overrightarrow{\sigma_1})  \left[\D (\sigma_2^{\omega}+\overrightarrow{\sigma_2})-\D(\sigma_{12}^{\omega}-\sigma_{12}^2) \right] S_2(CC\omega) - \\ - P( \sigma_1^1, \sigma_1^{\omega}, \sigma_1^2) P(\sigma_2^2,\sigma_2^{\omega},\sigma_2^1) \left[\D (\sigma_2^{\omega}-\sigma_2^1)+\D(\sigma_{1}^{\omega}-\sigma_1^2) \right]\D(\sigma_{12}^{\omega})S_2(C\omega C) +\\ +P(-\overrightarrow{\sigma_{12}},\sigma_{12}^{\omega}, \sigma_{12}^1)P(-\overrightarrow{\sigma_1},\sigma_1^{\omega},\sigma_1^1)\D(\sigma_2^{\omega}-\overrightarrow{\sigma_2}) \left[\D (\sigma_1^{\omega}+\overrightarrow{\sigma_1})-\D(\sigma_{12}^{\omega}-\sigma_{12}^1) \right] S_2(\omega CC) \Big \}.
\end{multline}
Here $S_2(\omega CC)$ denotes the expression for $S_2$ \eqref{S2} with $C \bar{*} C$ replaced by $\omega \bar{*}C \bar{*} C$ and analogously for $S_2(C\omega C)$ and $S_2(CC\omega)$. Analogous remarkable representation for $W_1$ via $S_1$ was found in \cite{Vasiliev:2023yzx}. As shown in \ref{solving_for_Wn}, this property can be proven without direct calculation because the r.h.s. of the equation for $S_2$ contains only a single term of the form $S_i * S_{n-i}$. However, in higher orders, the equation for $S_n$ does not share this structure. Consequently, proving that analogous relations hold beyond the quadratic order remains an open problem.

\subsection{Field equation on the physical field}
The second order interacting vertex in the one-form field sector is determined from the following equation:
\begin{equation}
    \dd_x W_1 + \dd_x W_2 +\omega * W_2 + W_2 * \omega +W_1*W_1 - 2i \dd W_2^{(2)} \cong \Upsilon\,.
 \end{equation}
Here $\dd_x W_1$ contributes as a result of the action of $\dd_x$ on $C$ and $\omega$ in $W_1$.\par
Let us consider the $C C\omega\omega$ sector. The following two terms come from $\dd_x W_1$:
 \begin{equation}
    \dd_x W_1|_{C C \omega \omega} = W_1(C \dd_x \omega)|_{C C \omega \omega} + W_1((\dd_x C) \omega)|_{C C \omega \omega}\,.
    \label{d_xW_1}
\end{equation}
The first one has the form
\begin{equation}
    W_1(C \dd_x \omega)|_{C C \omega \omega} = \dfrac{i\eta}{4}\fint{\tau \rho \beta \sigma \sigma^{\omega} } l(\tau) \D(\sigma) P(\sigma, \sigma^{\omega}, 1-\beta) \mu(\rho, \beta) \dd \Omega^2 \E(\Omega) C(r_1) \bar{*} (\dd_x \omega(r_{\omega}))|_{C \omega \omega} \big|_{r = 0} k\,.
    \label{d_xW_1_1}
\end{equation}\par
Here $\dd_x \omega$ has to be substituted by the contribution to the sector
$C \omega \omega$ from the {\it r.h.s.} of \eqref{D_omega}. Then \eqref{d_xW_1_1}
is transformed to the form \eqref{fundament}  analogously to how it has been proceeded  with \eqref{d_xS_1_1}. Let  the objects arising from $W_1$  and $\dd_x{\omega}$ be endowed by the labels 1 and  2, respectively. Then
we rename $p_{\alpha}^{\omega} \rightarrow s_{\alpha}$ and $r_{\alpha}^{\omega} \rightarrow -t_{\alpha}$. Note that  the coefficient in front of $s_{\alpha}$ in the resulting $\Omega_1$ is not constant since it arises from $\sigma^{\omega}$ in \eqref{d_xW_1_1}. The exponent in the integrand takes the form
\begin{multline}
    \exp i \big[\Omega_{\alpha}(y+s+p_1+u_1)^{\alpha} - (p(\sigma_2)-\beta_2 v_2)_{\alpha}(-t+p_2+p_{\omega1}+p_{\omega2}+u_2)^{\alpha}+s_{\alpha} t^{\alpha}+s_{\alpha} p_1^{\alpha}+ \\  +u_{1 {\alpha}}v_1^{\alpha}+u_{2 {\alpha}}v_2^{\alpha} - p_{2 {\alpha}} p_{\omega1}^{\alpha} - p_{2 {\alpha}} p_{\omega2}^{\alpha} - p_{\omega1 {\alpha}} p_{\omega2}^{\alpha} - \sum_{i=1}^4 p_{i{\alpha}} r_i^{\alpha}  \big] = \exp i \big[\Omega_{\alpha}(y+s+p_1+u_1)^{\alpha} - \\  - (-p(\sigma_2)-t+\beta_2 v_2)_{\alpha}(t-p_2-p_{\omega1}-p_{\omega2}-u_2)^{\alpha}+s_{\alpha} t^{\alpha}+s_{\alpha} p_1^{\alpha} +u_{1 {\alpha}}v_1^{\alpha}+u_{2 {\alpha}}v_2^{\alpha} - p_{2 {\alpha}} p_{\omega1}^{\alpha} - p_{2 {\alpha}} p_{\omega2}^{\alpha}+ \\   - p_{\omega1 {\alpha}} p_{\omega2}^{\alpha} - \sum_{i=1}^4 p_{i{\alpha}} r_i^{\alpha} - t_{\alpha}(-p_2-p_{\omega1}-p_{\omega2}-u_2)^{\alpha} \big].
\end{multline}
To bring this expression to the form \eqref{fundament} we change the signs of $u_{2\alpha}$ and $v_{2\alpha}$ and then make the substitutions $s_{\alpha} \rightarrow s_{\alpha}+p_{2\alpha}+p_{\omega1 {\alpha}}+p_{\omega2 {\alpha}}$, $t_{\alpha} \rightarrow t_{\alpha}+y_{\alpha}-p_{1\alpha}$, $v_{2\alpha} \rightarrow v_{2\alpha}-t_{\alpha}+p_{1\alpha}$. This yields
\begin{multline}
    W_1(C \dd_x \omega)|_{C C \omega \omega} = \dfrac{\eta^2}{16} \fint{\tau_1 \rho_1 \beta_1 \sigma_1^1 \sigma_1^2 \sigma_1^{\omega1} \sigma_1^{\omega2}  \tau_2 \rho_2 \beta_2 \sigma_2^1 \sigma_2^2 \sigma_2^{\omega 1} \sigma_2^{\omega 2} \tau_{12} \sigma_{12}^1 \sigma_{12}^2 \sigma_{12}^{\omega1} \sigma_{12}^{\omega2} a_1 a_2 \beta_{12}} \mu'(\beta_{12})  \D(1-\tau_{12}) \D(\overrightarrow{\sigma_2}-a_2) l(\tau_1) \D(\tau_2) P(0, a_1, \overrightarrow{\sigma_1}) \D (\sigma_1^1)\D (\sigma_1^2-a_1) \times \\ \times\D (\sigma_2^1-\overrightarrow{\sigma_2})\D (\sigma_2^2)\D(\sigma_1^{\omega1}-a_1)\D(\sigma_1^{\omega2}-a_1) P(-\overrightarrow{\sigma_2},\sigma_2^{\omega2}, \sigma_2^{\omega1},\sigma_2^2) \D(\sigma_{12}^{\omega1}) \D(\sigma_{12}^{\omega2}) \D(\sigma_{12}^1)\D(\sigma_{12}^2) \times \\ \times{\mu}(\rho_1, \beta_1){\mu}(\rho_2, \beta_2)
     \dd {\Omega}_1^2 \dd {\Omega}_2^2 {\E}(\Omega_1, \Omega_2)   C(r_1, \bar{y}, K) \bar{*}   C(r_2, \bar{y}, K) \bar{*} \omega(r_{\omega 1}, \bar{y}, K) \bar{*} \omega(r_{\omega 2}, \bar{y}, K) \big|_{r_i = 0}\,.
    \label{d_x_W2_dw}
\end{multline}\par
The second term in \eqref{d_xW_1} can be found analogously to \eqref{d_xS_1_1},
\begin{multline}
    W_1((\dd_x C) \omega)|_{C C \omega \omega} = 
    \\
    \dfrac{\eta^2}{16} \fint{\tau_1 \rho_1 \beta_1 \sigma_1^1 \sigma_1^2 \sigma_1^{\omega1} \sigma_1^{\omega2}  \tau_2 \rho_2 \beta_2 \sigma_2^1 \sigma_2^2 \sigma_2^{\omega 1} \sigma_2^{\omega 2} \tau_{12} \sigma_{12}^1 \sigma_{12}^2 \sigma_{12}^{\omega1} \sigma_{12}^{\omega2} a_1 a_2 \beta_{12}} \mu'(\beta_{12})\D(\overrightarrow{\sigma_1}-a_1)\D(a_2) \D(\tau_1) \D(1-\tau_{12}) l(\tau_{2}) P(-\overrightarrow{\sigma_1}, \sigma_1^1, \sigma_1^2, \sigma_1^{\omega1}, \sigma_1^{\omega2}, \overrightarrow{\sigma_1})\times 
    \\
    \times \D (\sigma_1^2-\sigma_1^1-\overrightarrow{\sigma_1})  P(-\overrightarrow{\sigma_{2}},\sigma_{2}^{\omega2}, \sigma_{2}^{\omega1}) \D (\sigma_2^1)\D(\sigma_2^2)  \D(\sigma_1^{\omega2}-\overrightarrow{\sigma_1})\D(\sigma_{12}^{1})\D(\sigma_{12}^2)\D(\sigma_{12}^{\omega1})\times 
    \\
    \times \D(\sigma_{12}^{\omega2}) \D(\sigma_{12}^1)\D(\sigma_{12}^2) \D(\sigma_{2}^{\omega1}) \D(\rho_1 + \sigma_1^1) {\mu}(\beta_1){\mu}(\rho_2, \beta_2) \times
    \\
    \times \dd {\Omega}_1^2 \dd {\Omega}_2^2 {\E}   C(r_1, \bar{y}, K) \bar{*}   C(r_2, \bar{y}, K) \bar{*} \omega(r_{\omega 1}, \bar{y}, K) \bar{*} \omega(r_{\omega 2}, \bar{y}, K) \big|_{r_i = 0}\,.
    \label{d_x_W2_dC}
\end{multline}\par

Let us consider the $\sigma$-dependent part of the measure in  $\dd_x W_2 + W_2*\omega + \omega * W_2$  discarding factors that are common for all these terms,
\begin{multline}
    \D(\sigma_{12}^{\omega2}+\overrightarrow{\sigma_{12}})\D (\sigma_2^{\omega1}+\overrightarrow{\sigma_2})\D (\sigma_2^{\omega2}+\overrightarrow{\sigma_2}) + \D(\sigma_{12}^{\omega1}-\sigma_{12}^{\omega2})\D (\sigma_2^{\omega1}+\overrightarrow{\sigma_2})\D (\sigma_2^{\omega2}+\overrightarrow{\sigma_2})+
    \\
    +\D (\sigma_2^2-\sigma_2^{\omega1})\D (\sigma_2^{\omega2}+\overrightarrow{\sigma_2})\D(\sigma_{12}^{\omega1}-\sigma_{12}^2)+\D (\sigma_2^{\omega2}+\overrightarrow{\sigma_2})\D(\sigma_{12}^{\omega1})\D(\sigma_{12}^{\omega2}+\overrightarrow{\sigma_{12}})+\D (\sigma_2^{\omega1}-\sigma_2^{\omega2})\D(\sigma_{12}^{\omega1}) \D(\sigma_{12}^{\omega2})+
    \\
    +\D (\sigma_2^2-\sigma_2^{\omega1})\D(\sigma_{12}^{\omega1}) \D(\sigma_{12}^{\omega2}) = \dd P(-\overrightarrow{\sigma_{12}},\sigma_{12}^{\omega2}, \sigma_{12}^{\omega1},\sigma_{12}^2) P(-\overrightarrow{\sigma_2},\sigma_2^{\omega2}, \sigma_2^{\omega1},\sigma_2^2) \times
    \\
    \times \left[ \D (\sigma_2^{\omega1}+\overrightarrow{\sigma_2})\D (\sigma_2^{\omega2}+\overrightarrow{\sigma_2}) +\D (\sigma_2^{\omega2}+\overrightarrow{\sigma_2})\D(\sigma_{12}^{\omega1}-\sigma_{12}^2) +\D(\sigma_{12}^{\omega1}) \D(\sigma_{12}^{\omega2}) \right].
    \label{DW2_sigma}
\end{multline}\par
Leaving details for \ref{App_D} we obtain $W_2^{(2)}$,
\begin{multline}
    W_2^{(2)}|_{C C \omega \omega} = \dfrac{i\eta^2}{32} \fint{{\tau_1 \rho_1 \beta_1 \sigma_1^1 \sigma_1^2 \sigma_1^{\omega1} \sigma_1^{\omega2}  \tau_2 \rho_2 \beta_2 \sigma_2^1 \sigma_2^2 \sigma_2^{\omega 1} \sigma_2^{\omega 2} \tau_{12} \sigma_{12}^1 \sigma_{12}^2 \sigma_{12}^{\omega1} \sigma_{12}^{\omega2}a_1 a_2\beta_{12}}} \mu'(\beta_{12})  l(\tau_{12}) l(\tau_1) P(-\overrightarrow{\sigma_{12}},\sigma_{12}^{\omega2}, \sigma_{12}^{\omega1},\sigma_{12}^2)\D(\sigma_{12}^1)\D(\sigma_{12}^2) \D (\sigma_2^1-\overrightarrow{\sigma_2})\D (\sigma_2^2)\times \\ \times\big \{ l(\tau_2) \D (\sigma_1^1)\D (\sigma_1^2-\overrightarrow{\sigma_1}) \D(\sigma_1^{\omega1}-\overrightarrow{\sigma_1})\D(\sigma_1^{\omega2}-\overrightarrow{\sigma_1}) P(-\overrightarrow{\sigma_2},\sigma_2^{\omega2}, \sigma_2^{\omega1},\sigma_2^2) \left[ \D (\sigma_2^{\omega1}+\overrightarrow{\sigma_2})\D (\sigma_2^{\omega2}+\overrightarrow{\sigma_2}) +\right.\\\left.+ \D (\sigma_2^{\omega2}+\overrightarrow{\sigma_2})\D(\sigma_{12}^{\omega1}-\sigma_{12}^2) +\D(\sigma_{12}^{\omega1}) \D(\sigma_{12}^{\omega2}) \right]{\mu}(\rho_1, \beta_1)
    \dd {\Omega}_1^2 + \D(1-\tau_2)  P(-\overrightarrow{\sigma_1}, \sigma_1^1, \sigma_1^2, \sigma_1^{\omega1}, \sigma_1^{\omega2}, \overrightarrow{\sigma_1})\times \\ \times\D (\sigma_1^2-\sigma_1^1-\overrightarrow{\sigma_1}) \D(\sigma_{2}^{\omega1})\D(\sigma_{2}^{\omega2})  \left[\D(\sigma_1^{\omega1}-\overrightarrow{\sigma_1})\D(\sigma_1^{\omega2}-\overrightarrow{\sigma_1}) -  \D(\sigma_1^{\omega2}-\overrightarrow{\sigma_1})\D(\sigma_{12}^{\omega1}) + \D(\sigma_{12}^{\omega1})\D(\sigma_{12}^{\omega2}) \right] \times \\ \times \D(\rho_1+\sigma_1^1){\mu}(\beta_1)(\dd \Omega_1^2 + 2i(1-\beta_1)\D(\sigma_1^2)\dd \tau_1) \big \}  {\mu}(\rho_2, \beta_2)
     \D(a_2-\overrightarrow{\sigma_2})\D(a_1-\overrightarrow{\sigma_1})  \dd {\Omega}_2^2 \E(\Omega_1, \Omega_2) \times \\ \times C(r_1, \bar{y}, K) \bar{*}   C(r_2, \bar{y}, K) \bar{*} \omega(r_{\omega 1}, \bar{y}, K) \bar{*} \omega(r_{\omega 2}, \bar{y}, K) \big|_{r_i = 0} + \\+ \dfrac{i\eta^2}{32} \fint{\tau_1 \rho_1 \beta_1 \sigma_1^1 \sigma_1^2 \sigma_1^{\omega1} \sigma_1^{\omega2}  \tau_2 \rho_2 \beta_2 \sigma_2^1 \sigma_2^2 \sigma_2^{\omega 1} \sigma_2^{\omega 2} \tau_{12} \sigma_{12}^1 \sigma_{12}^2 \sigma_{12}^{\omega1} \sigma_{12}^{\omega2} a_1 a_2\beta_{12}} \mu'(\beta_{12}) l(\tau_{12})P(0,a_1,\overrightarrow{\sigma_1})l(\tau_1)\D(\overrightarrow{\sigma_2}-a_2) \D(\tau_2) \D (\sigma_1^1)\D (\sigma_1^2-a_1) \D (\sigma_2^1-\overrightarrow{\sigma_2})\D (\sigma_2^2)\times \\ \times \D(\sigma_1^{\omega1}-a_1)\D(\sigma_1^{\omega2}-a_1)  P(-\overrightarrow{\sigma_2},\sigma_2^{\omega2}, \sigma_2^{\omega1},\sigma_2^2) \D(\sigma_{12}^{\omega1}) \D(\sigma_{12}^{\omega2})\D(\sigma_{12}^1)\D(\sigma_{12}^2) {\mu}(\rho_1, \beta_1){\mu}(\rho_2, \beta_2)\times \\ \times \dd {\Omega}_1^2 \dd {\Omega}_2^2 {\E}(\Omega_1, \Omega_2)   C(r_1, \bar{y}, K) \bar{*}   C(r_2, \bar{y}, K) \bar{*} \omega(r_{\omega 1}, \bar{y}, K) \bar{*} \omega(r_{\omega 2}, \bar{y}, K) \big|_{r_i = 0} + 
     \\
     + \dfrac{\eta^2}{32i} \fint{\tau_1 \rho_1 \beta_1 \sigma_1^1 \sigma_1^2 \sigma_1^{\omega1} \sigma_1^{\omega2}  \tau_2 \rho_2 \beta_2 \sigma_2^1 \sigma_2^2 \sigma_2^{\omega 1} \sigma_2^{\omega 2} \tau_{12} \sigma_{12}^1 \sigma_{12}^2 \sigma_{12}^{\omega1} \sigma_{12}^{\omega2} a_1 a_2\beta_{12}} \mu'(\beta_{12}) \D(\overrightarrow{\sigma_1}-a_1)\D(a_2) \D(\tau_1) l(\tau_2) l(\tau_{12})P(-\overrightarrow{\sigma_1}, \sigma_1^1, \sigma_1^2, \sigma_1^{\omega1}, \sigma_1^{\omega2}, \overrightarrow{\sigma_1})\D (\sigma_1^2-\sigma_1^1-\overrightarrow{\sigma_1}) \times
     \\
     \times \D (\sigma_2^1)\D(\sigma_2^2)  \D(\sigma_{12}^{\omega1}) \D(\sigma_{12}^1)\D(\sigma_{12}^2) \D(\sigma_2^{\omega1}) \big[ P(-\overrightarrow{\sigma_2},\sigma_{2}^{\omega2}, 0)\D(\sigma_{12}^{\omega_2}-\sigma_{12}^{\omega_1})\D(\sigma_1^{\omega_2}-\overrightarrow{\sigma_1}) +
     \\
     + \D(\sigma_2^{\omega_2}+\overrightarrow{\sigma_2})P(-\overrightarrow{\sigma_{12}}, \sigma_{12}^{\omega_2}, \sigma_{12}^{\omega_1})\D(\sigma_1^{\omega_2}-\overrightarrow{\sigma_1})+\D(\sigma_{12}^{\omega2})\D(\sigma_2^{\omega}) \big ] \times
     \\ 
     \times \D(\rho_1 + \sigma_1^1) {\mu}(\beta_1){\mu}(\rho_2, \beta_2)
    \dd {\Omega}_1^2 \dd {\Omega}_2^2 {\E}   C(r_1, \bar{y}, K) \bar{*}   C(r_2, \bar{y}, K) \bar{*} \omega(r_{\omega 1}, \bar{y}, K) \bar{*} \omega(r_{\omega 2}, \bar{y}, K) \big|_{r_i = 0}\,.
    \label{W2^2}
\end{multline}
\par
The contribution to the equation on the physical field \eqref{eq_omega} results from the terms proportional to $\D(\tau_{12})$ upon the action of $\dd$ on $W_2^{(2)}$, being therefore independent of $z_{\alpha}$ and $\theta_{\alpha}$. Note that the term with $P(0,a_1,\overrightarrow{\sigma_1})$ does not contribute  which can be proven by expanding $\dd\Omega_1^2$ and integrating over $u_i, v_i, s, t$, that entails $\dd \Omega_1^2 \cong 0$. The same happens with the contribution from the term with $\D(\tau_1)$.\par
The vertex in the sector in question takes the form
\begin{multline}
    \Upsilon(CC\omega\omega) = \dfrac{\eta^2}{16} \fint{\tau_1 \rho_1 \beta_1 \sigma_1^1 \sigma_1^2 \sigma_1^{\omega1} \sigma_1^{\omega2}  \tau_2 \rho_2 \beta_2 \sigma_2^1 \sigma_2^2 \sigma_2^{\omega 1} \sigma_2^{\omega 2} \tau_{12} \sigma_{12}^1 \sigma_{12}^2 \sigma_{12}^{\omega1} \sigma_{12}^{\omega2}\beta_{12}} \mu'(\beta_{12}) \D(\tau_{12}) l(\tau_1)P(-\overrightarrow{\sigma_{12}},\sigma_{12}^{\omega2}, \sigma_{12}^{\omega1},\sigma_{12}^2)\D(\sigma_{12}^1)\D(\sigma_{12}^2)\D(\sigma_1^{\omega1}-\overrightarrow{\sigma_1})\D(\sigma_1^{\omega2}-\overrightarrow{\sigma_1}) \times \\ \times \D (\sigma_2^1-\overrightarrow{\sigma_2})\D (\sigma_2^2)\D (\sigma_2^{\omega2}+\overrightarrow{\sigma_2}) \big \{ l(\tau_2) \D (\sigma_1^1)\D (\sigma_1^2-\overrightarrow{\sigma_1})   P(-\overrightarrow{\sigma_2},\sigma_2^{\omega2}, \sigma_2^{\omega1},\sigma_2^2) \times \\ \times \big[ \D (\sigma_2^{\omega1}+\overrightarrow{\sigma_2}) - \D(\sigma_{12}^{\omega1}) \big ] {\mu}(\rho_1, \beta_1)
    \dd {\Omega}_1^2 + \D(1-\tau_2) P(-\overrightarrow{\sigma_1}, \sigma_1^1, \sigma_1^2, \sigma_1^{\omega1}, \sigma_1^{\omega2}, \overrightarrow{\sigma_1})\times \\ \times \D (\sigma_1^2-\sigma_1^1-\overrightarrow{\sigma_1}) \D(\sigma_{2}^{\omega1})  \D(\rho_1+\sigma_1^1){\mu}(\beta_1)(\dd \Omega_1^2 + 2i(1-\beta_1)\D(\sigma_1^2)\dd \tau_1) \big \} \times \\ \times {\mu}(\rho_2, \beta_2) \dd {\Omega}_2^2 \E(\Omega_1, \Omega_2)  C(r_1, \bar{y}, K) \bar{*}   C(r_2, \bar{y}, K) \bar{*} \omega(r_{\omega 1}, \bar{y}, K) \bar{*} \omega(r_{\omega 2}, \bar{y}, K) \big|_{r_i = 0}
    \label{vertexCCww}
\end{multline}
\par
Now let $\beta_1 = \beta_2 = 0$, $\beta_{12} = \beta$. A local result is obtained in the limit $\beta \rightarrow -\infty$. As shown in \ref{App_E}, only
\eqref{vertex-4} contributes in the limit taken analogously to \cite{Didenko:2019xzz}. As a result we obtain
\begin{multline}
    \Upsilon(CC\omega\omega) = \dfrac{\eta^2}{4} \fint{\tau_1  \tau_2 \sigma_{2}^{\omega1} \sigma_{12}^{\omega2}}  l(\tau_1) l(\tau_2) P(-1,\sigma_{12}^{\omega2}, 0)P(-1,\sigma_{2}^{\omega1}, 0) \dd \tau_1 \dd \tau_2  \dd \sigma_{12}^{\omega2} \dd \sigma_2^{\omega1} \tau_1 (1+\sigma_{12}^{\omega2}) \sigma_{2}^{\omega1} (p_{\omega1\alpha}p_{\omega2}^{\alpha})^2 \times \\ \times \exp i \Big[ \tau_1 A + \tau_2 B + (1-\tau_1-\tau_2)C +D \Big]  C(r_1, \bar{y}, K) \bar{*}   C(r_2, \bar{y}, K) \bar{*} \omega(r_{\omega 1}, \bar{y}, K) \bar{*} \omega(r_{\omega 2}, \bar{y}, K) \big|_{r_i = 0}
    \label{vertex-lim}\,,
\end{multline}
where
\begin{equation}
    \begin{gathered}
        A = \sigma_2^{\omega1}(\sigma_{12}^{\omega2}+1)p_{\omega1\alpha}p_{\omega2}^{\alpha} +(\sigma_{12}^{\omega2}+1)p_{1\alpha}p_{\omega2}^{\alpha}-\sigma_2^{\omega1}p_{1\alpha}p_{\omega1}^{\alpha}+(\sigma_{12}^{\omega2}+1)y_{\alpha}p_{\omega2}^{\alpha}\,, \\
        B = (\sigma_{12}^{\omega2}+1)p_{\omega1\alpha}p_{\omega2}^{\alpha} +(\sigma_{12}^{\omega2}+1)p_{1\alpha}p_{\omega2}^{\alpha}-(\sigma_{12}^{\omega2}+1)p_{2\alpha}p_{\omega2}^{\alpha}+(\sigma_{12}^{\omega2}+1)y_{\alpha}p_{\omega2}^{\alpha}\,,\\
        C = \sigma_2^{\omega1}p_{\omega1\alpha}p_{\omega2}^{\alpha}+\sigma_2^{\omega1} y_{\alpha}p_{\omega1}^{\alpha}, \quad
        D = -y_{\alpha}p_{\omega2}^{\alpha} - (\sigma_2^{\omega1}+1)p_{2\alpha}p_{\omega1}^{\alpha}\,.
    \end{gathered}
\end{equation}
This expression reproduces the ultra-local result obtained in \cite{Didenko:2019xzz}.
\par
The final results for vertices in other sectors obtained
in the form of differential homotopy analogously to the sector $CC\omega\omega$ read as
\begin{multline}
    \Upsilon(C \omega C \omega) = 
    \\
    = -\dfrac{\eta^2}{16} \fint{\tau_1 \rho_1 \beta_1 \sigma_1^1 \sigma_1^2 \sigma_1^{\omega1} \sigma_1^{\omega2}  \tau_2 \rho_2 \beta_2 \sigma_2^1 \sigma_2^2 \sigma_2^{\omega 1} \sigma_2^{\omega 2} \tau_{12} \sigma_{12}^1 \sigma_{12}^2 \sigma_{12}^{\omega1} \sigma_{12}^{\omega2}\beta_{12}} \mu'(\beta_{12}) \D(\tau_{12}) l(\tau_1) \big \{  l(\tau_2) \D (\sigma_1^1)\D (\sigma_1^2-\overrightarrow{\sigma_1}) \D (\sigma_2^1-\overrightarrow{\sigma_2})\D (\sigma_2^2)\D(\sigma_1^{\omega2}-\overrightarrow{\sigma_1})\times \\ \times  P(-\overrightarrow{\sigma_{12}},\sigma_{12}^{\omega2},\sigma_{12}^2) P(\sigma_1^1,\sigma_1^{\omega1}, \sigma_1^2)P(-\overrightarrow{\sigma_2},\sigma_2^{\omega2}, \sigma_2^2, \sigma_2^{\omega1},\sigma_2^1) \left[ \D (\sigma_2^{\omega1}-\sigma_2^1) \D (\sigma_2^{\omega2}+\overrightarrow{\sigma_2}) +\right.\\\left.+ \D (\sigma_1^{\omega1}-\sigma_1^2) \D (\sigma_2^{\omega2}+\overrightarrow{\sigma_2}) + \D(\sigma_{12}^{\omega2})\D (\sigma_2^{\omega1}-\sigma_2^1)  \right]\D(\sigma_{12}^1)\D(\sigma_{12}^2)\D(\sigma_{12}^{\omega1}) \times \\ \times {\mu}(\rho_1, \beta_1){\mu}(\rho_2, \beta_2)
    \dd {\Omega}_1^2 \dd {\Omega}_2^2 \E(\Omega_1, \Omega_2)  C(r_1, \bar{y}, K) \bar{*} \omega(r_{\omega 1}, \bar{y}, K) \bar{*}   C(r_2, \bar{y}, K) \bar{*} \omega(r_{\omega 2}, \bar{y}, K) \big|_{r_i = 0}\,,
\end{multline}
\begin{multline}
    \Upsilon(C \omega \omega C) = 
    \\
    = \dfrac{\eta^2}{16} \fint{\tau_1 \rho_1 \beta_1 \sigma_1^1 \sigma_1^2 \sigma_1^{\omega1} \sigma_1^{\omega2}  \tau_2 \rho_2 \beta_2 \sigma_2^1 \sigma_2^2 \sigma_2^{\omega 1} \sigma_2^{\omega 2} \tau_{12} \sigma_{12}^1 \sigma_{12}^2 \sigma_{12}^{\omega1} \sigma_{12}^{\omega2}\beta_{12}} \mu'(\beta_{12}) \D(\tau_{12}) l(\tau_1) l(\tau_2) \D (\sigma_1^1)\D (\sigma_1^2-\overrightarrow{\sigma_1}) \D (\sigma_2^1-\overrightarrow{\sigma_2})\D (\sigma_2^2)\times \\ \times  P(\sigma_1^1,\sigma_1^{\omega1}, \sigma_1^{\omega2}, \sigma_1^2)P(\sigma_2^2,\sigma_2^{\omega2}, \sigma_2^{\omega1}, \sigma_2^1)  \D (\sigma_1^{\omega2}-\sigma_1^2)\D (\sigma_2^{\omega1}-\sigma_2^1) \D(\sigma_{12}^1)\D(\sigma_{12}^2)\D(\sigma_{12}^{\omega1})\D(\sigma_{12}^{\omega2}) \times \\ \times {\mu}(\rho_1, \beta_1){\mu}(\rho_2, \beta_2)
    \dd {\Omega}_1^2 \dd {\Omega}_2^2 \E(\Omega_1, \Omega_2)  C(r_1, \bar{y}, K)  \bar{*} \omega(r_{\omega 1}, \bar{y}, K) \bar{*} \omega(r_{\omega 2}, \bar{y}, K) \bar{*}  C(r_2, \bar{y}, K) \big|_{r_i = 0}\,,
\end{multline}
\begin{multline}
    \Upsilon(\omega C C \omega) =
    \\
    = -\dfrac{\eta^2}{16} \int\limits
    _\mathclap{{\qquad \qquad  \qquad  \qquad \qquad  \qquad  \qquad \qquad \tau_1 \rho_1 \beta_1 \sigma_1^1 \sigma_1^2 \sigma_1^{\omega1} \sigma_1^{\omega2}  \tau_2 \rho_2 \beta_2 \sigma_2^1 \sigma_2^2 \sigma_2^{\omega 1} \sigma_2^{\omega 2} \tau_{12} \sigma_{12}^1 \sigma_{12}^2 \sigma_{12}^{\omega1} \sigma_{12}^{\omega2}\beta_{12}}} \mu'(\beta_{12}) \D(\tau_{12}) l(\tau_1) P(-\overrightarrow{\sigma_1}, \sigma_1^{\omega1}, \sigma_1^1, \sigma_1^2, \sigma_1^{\omega2}, \overrightarrow{\sigma_1})P(-\overrightarrow{\sigma_2}, \sigma_2^{\omega2}, \sigma_2^2, \sigma_2^1, \sigma_2^{\omega1}, \overrightarrow{\sigma_2})\times 
    \\
    \times P(-\overrightarrow{\sigma_{12}}, \sigma_{12}^{\omega2}, \sigma_{12}^2)P(-\overrightarrow{\sigma_{12}}, \sigma_{12}^{\omega1},\sigma_{12}^1)\D(\sigma_{12}^1)\D(\sigma_{12}^2)\D(\sigma_{1}^{\omega2}-\overrightarrow{\sigma_1})\D(\sigma_{2}^{\omega1}-\overrightarrow{\sigma_2})\D (\sigma_2^1-\overrightarrow{\sigma_2})\D (\sigma_2^2) \times
    \\
    \times \big \{ l(\tau_2)  \D (\sigma_1^1)\D (\sigma_1^2-\overrightarrow{\sigma_1})  \big[ \D(\sigma_{12}^{\omega1})\D(\sigma_{12}^{\omega2}) +\D(\sigma_2^{\omega2}+\overrightarrow{\sigma_2})\D(\sigma_{12}^{\omega1}) + \D(\sigma_1^{\omega1}+\overrightarrow{\sigma_1})\D(\sigma_{12}^{\omega2}) -
    \\
    - \D(\sigma_1^{\omega1}+\overrightarrow{\sigma_1})\D(\sigma_2^{\omega2}+\overrightarrow{\sigma_2}) \big] {\mu}(\rho_1, \beta_1)
    \dd {\Omega}_1^2 - 
    \\
    -  \D(1-\tau_2) \D (\sigma_1^2-\sigma_1^1-\overrightarrow{\sigma_1})\D(\sigma_1^{\omega1}+\overrightarrow{\sigma_1}) \D(\sigma_{2}^{\omega2})  \D(\rho_1+\sigma_1^1){\mu}(\beta_1)(\dd \Omega_1^2 + 2i(1-\beta_1)\D(\sigma_1^2)\dd \tau_1)\big \}\times 
    \\
    \times {\mu}(\rho_2, \beta_2)
     \dd {\Omega}_2^2 \E(\Omega_1, \Omega_2)  \omega(r_{\omega 1}, \bar{y}, K) \bar{*}C(r_1, \bar{y}, K) \bar{*}   C(r_2, \bar{y}, K)  \bar{*} \omega(r_{\omega 2}, \bar{y}, K) \big|_{r_i = 0}\,,
\end{multline}
\begin{multline}
    \Upsilon(\omega \omega C C)=
    \\
    = -\dfrac{\eta^2}{16} \fint{\tau_1 \rho_1 \beta_1 \sigma_1^1 \sigma_1^2 \sigma_1^{\omega1} \sigma_1^{\omega2}  \tau_2 \rho_2 \beta_2 \sigma_2^1 \sigma_2^2 \sigma_2^{\omega 1} \sigma_2^{\omega 2} \tau_{12} \sigma_{12}^1 \sigma_{12}^2 \sigma_{12}^{\omega1} \sigma_{12}^{\omega2}} \mu'(\beta_{12}) \D(\tau_{12}) l(\tau_1) P(-\overrightarrow{\sigma_{12}}, \sigma_{12}^{\omega1}, \sigma_{12}^{\omega2}, \sigma_{12}^2)P(-\overrightarrow{\sigma_1}, \sigma_1^{\omega1}, \sigma_1^{\omega2}, \sigma_1^1, \sigma_1^2, \overrightarrow{\sigma_1})  \D(\sigma_{12}^1)\D(\sigma_{12}^2) \times \\ \times \D (\sigma_2^1-\overrightarrow{\sigma_2})\D (\sigma_2^2)\D(\sigma_2^{\omega1}-\overrightarrow{\sigma_2})\D(\sigma_2^{\omega2}-\overrightarrow{\sigma_2})\D (\sigma_1^{\omega1}+\overrightarrow{\sigma_1}) \big \{ l(\tau_2) \D (\sigma_1^1)\D (\sigma_1^2-\overrightarrow{\sigma_1})\   \big[ \D (\sigma_1^{\omega2}+\overrightarrow{\sigma_1})- \D(\sigma_{12}^{\omega2}) \big]\times \\ \times \mu(\rho_1,\beta_1)\dd {\Omega}_1^2  -  \D(1-\tau_2)   \D (\sigma_1^2-\sigma_1^1-\overrightarrow{\sigma_1})   \D(\sigma_1^{\omega2}+\overrightarrow{\sigma_1}) \D(\rho_1+\sigma_1^1){\mu}(\beta_1)(\dd \Omega_1^2 + 2i(1-\beta_1)\D(\sigma_1^2)\dd \tau_1) \big \}\times \\ \times  {\mu}(\rho_2, \beta_2)
    \dd {\Omega}_2^2 \E(\Omega_1, \Omega_2)  \omega(r_{\omega 1}, \bar{y}, K) \bar{*} \omega(r_{\omega 2}, \bar{y}, K) \bar{*} C(r_1, \bar{y}, K) \bar{*}   C(r_2, \bar{y}, K) \big|_{r_i = 0}\,,
\end{multline}
\begin{multline}
    \Upsilon(\omega C \omega C) = 
    \\ 
    = \dfrac{\eta^2}{16} \fint{\tau_1 \rho_1 \beta_1 \sigma_1^1 \sigma_1^2 \sigma_1^{\omega1} \sigma_1^{\omega2}  \tau_2 \rho_2 \beta_2 \sigma_2^1 \sigma_2^2 \sigma_2^{\omega 1} \sigma_2^{\omega 2} \tau_{12} \sigma_{12}^1 \sigma_{12}^2 \sigma_{12}^{\omega1} \sigma_{12}^{\omega2}\beta_{12}} \mu'(\beta_{12}) \D(\tau_{12}) l(\tau_1) \big \{  l(\tau_2) \D (\sigma_1^1)\D (\sigma_1^2-\overrightarrow{\sigma_1}) \D (\sigma_2^1-\overrightarrow{\sigma_2})\D (\sigma_2^2)\D(\sigma_2^{\omega1}-\overrightarrow{\sigma_2})\times \\ \times  P(-\overrightarrow{\sigma_{12}}, \sigma_{12}^{\omega1}, \sigma_{12}^1) P(-\overrightarrow{\sigma_1}, \sigma_1^{\omega1}, \sigma_1^1,\sigma_1^{\omega2}, \sigma_1^2)P(\sigma_2^2, \sigma_2^{\omega2}, \sigma_2^1, \sigma_2^{\omega1}, \overrightarrow{\sigma_2}) \left[ \D (\sigma_1^{\omega2}-\sigma_1^2) \D (\sigma_1^{\omega1}+\overrightarrow{\sigma_1}) +\right.\\\left.+ \D (\sigma_2^{\omega2}-\sigma_2^1) \D (\sigma_1^{\omega1}+\overrightarrow{\sigma_1}) + \D(\sigma_{12}^{\omega1})\D (\sigma_1^{\omega2}-\sigma_1^2)  \right]\D(\sigma_{12}^1)\D(\sigma_{12}^2)\D(\sigma_{12}^{\omega2}) \times \\ \times {\mu}(\rho_1, \beta_1){\mu}(\rho_2, \beta_2)
    \dd {\Omega}_1^2 \dd {\Omega}_2^2 \E(\Omega_1, \Omega_2)  \omega(r_{\omega 1}, \bar{y}, K) \bar{*} C(r_1, \bar{y}, K)\bar{*}  \omega(r_{\omega 2}, \bar{y}, K) \bar{*}   C(r_2, \bar{y}, K)  \big|_{r_i = 0}\,.
\end{multline}

This list exhausts all quadratic holomorphic  ({\it i.e.,} proportional to $\eta^2$) vertices. Antiholomorphic vertices can be obtained from the holomorphic ones by
complex conjugation. The construction of mixed vertices involving both $\eta$ and $\bar \eta$ is an important open problem for the future.

\section{Conclusion}
\label{con}

In this paper, the differential homotopy approach of \cite{Vasiliev:2023yzx},
originally developed for the lower order analysis of HS equations, is
further extended  in a way appropriate for the higher-order computations. We have modified the Fundamental Ansatz of \cite{Vasiliev:2023yzx} in a way appropriate for the next order of perturbation theory. Useful properties of this modification have been presented.
In particular, we have generalized the star-multiplication formulae of \cite{Vasiliev:2023yzx} and derived symmetry properties \eqref{symmetry}, \eqref{symmetry-2}. The form of the extended Ansatz enriches the area of applicability of differential homotopy. The variable $\Omega_{12}$ in \eqref{Omega12} packages new homotopy parameters, making it easier to analyze their contribution. The introduction of the parameter $\b_{12}$ makes it possible to achieve spin-local vertices.  A particularly interesting new result is the star-multiplication  formula \eqref{star_mult_any}  involving an
arbitrary number of Ansatzes from \cite{Vasiliev:2023yzx} (canonical functions \eqref{anz}).

One of  significant results of \cite{Vasiliev:2023yzx} was the appropriate homotopy procedure for finding the second-order field $B_{2pc}$, that leads to projectively compact vertices in the zero-form sector. In our paper, we provide  a comparison between the differential homotopy and shifted homotopy approaches. It is  shown how the shifted homotopy approach can be represented in terms of the differential one. Thus, the differential homotopy generalizes and extends the shifted one. In addition, we have shown in \ref{App_C} that the difference between $B_{2pc}$ and $B_{2sh}$ of \cite{Vasiliev:2023yzx} coincides with the shift $\delta B$ of \cite{Vasiliev:2016xui}, that leads to a minimal number of derivatives in the HS equations.

The developed formalism has made it possible to obtain quadratic vertices in the
one-form equation of the holomorphic sector. These results match those of \cite{Didenko:2019xzz}, reproducing in particular such their properties as spinor spin-locality and projective-compactness, which in turn imply space-time spin-locality \cite{Vasiliev:2022med}.
\par
Our results were derived with an arbitrary dependence of the measure on the parameter
 $\beta$  playing  instrumental role in taking the spin-local limit $\beta\to-\infty$. The approach developed in this paper provides a basis
 for extending the obtained results to higher orders in a reasonably simple
 and systematic way.
In particular, it allows one to make computations directly without using the sophisticated analysis of the type of \cite{Gelfond:2021two} relying on numerous auxiliary theorems about properties of various classes of HS field functions.

An interesting direction for future research would be to extend the developed formalism to HS equations in the mixed sector (\ie\!\!, neither purely holomorphic nor purely antiholomorphic) of the theory.
Some progress in this direction was achieved  in \cite{Gelfond:2023fwe}, where the so-called moderately nonlocal cubic vertices in the zero-form sector were obtained. An interesting task is to construct local vertices in both the zero-form sector and the one-form sector within the differential homotopy formalism.

\vspace{-0.2cm}

 \section*{Acknowledgments}

\vspace{-0.1cm}
 
We are  grateful to Olga Gelfond for many
useful discussions and comments and to Ruslan Metsaev for the correspondence. Also we thank Anatoly Korybut and Aleksander Tarusov for careful reading the draft and useful comments. We are most grateful to Viacheslav Vereitin for pointing out some incorrect sign factors. PK is  grateful to the teachers of MIPT for fruitful lectures and seminars, which made it possible to carry out this work. MV wishes to thank
Ofer Aharony, Theoretical High Energy Physics Group of Weizmann Institute of Science
for hospitality and creating favourable conditions for this work.
This work was supported by Theoretical Physics and
Mathematics Advancement Foundation “BASIS” Grant No 24-1-1-9-5.

\begin{appendices}

    \addtocontents{toc}{\protect\setcounter{tocdepth}{1}}
    \section{Ansatz multiplication computations}
    \label{App_A}

    \subsection{The proof by induction}
    \label{App_A:sec_1}
    Formula \eqref{star_mult_any} can be proven by induction. At $J = 1$ it reproduces \eqref{anz} because $s_1$
    and $t_1$ are absent in that case. To proceed, consider
    \begin{multline}
        \label{F}
        F_J * f_{J+1} = \fint{p r u v s t \tau \rho \beta \s} \delta^2(t_1) \delta^2(s_{J+1})  \E_{J+1} \prod_{j=1}^{J+1}  \dd^2 u_j \dd^2 v_j \dd^2 p_j \dd^2 r_j \dd^2 s_j \dd^2 t_j \mu_j(\tau_j, \rho_j, \beta_j, \s_j) \dd \tilde{\Omega}_j^2 \times \\
        \times
        g_{j,1}(r_{j,1})\overline{*}\dots\overline{*}g_{j,n_j}(r_{j,n_j}) k^{J+1},
    \end{multline}
    where
    \begin{multline}
        \E_{J+1} = \exp i \sum_{j=1}^J \big [ (s_{j\alpha
        } + s_{j-1 \alpha}) t_j^{\alpha} + (s_{j\alpha
        }-t_{j\alpha
        })(q_j^{\alpha} + s_{J}^\alpha+u_j^{\alpha}) + u_{j\alpha
        }v_j^{\alpha
        } +\\+ (\tilde{\Omega}_{j\alpha}+s_{J \alpha})(q_j^{\alpha} + s_{J}^\alpha+u_j^{\alpha} + s_{j}^\alpha
        +t_{j}^\alpha
        ) - p_{j\alpha} r^{\alpha} + (\tilde{\Omega}_{J+1\alpha}-t_{J+1 \alpha})(q_{J+1}^{\alpha} + t_{J+1 }^{\alpha}+u_{J+1}^{\alpha} ) +\\
        + s_{J \alpha} t_{J+1}^{\alpha} + u_{J+1\alpha}v_{J+1}^{\alpha} -\sum_i p_{j,i\alpha} r_{j,i}^{\alpha} -\sum_{i<i'}p_{j,i\a}p_{j,i'}^\a\big],
    \end{multline}
    \begin{multline}
        \tilde{\Omega}_{j\alpha} = \tau_j z_{\alpha}-(1-\tau_j)((-1)^{j+1}p_{j\alpha}(\sigma_j)- \beta_j v_{j\alpha} + s_{j\alpha}+ s_{J\alpha} - t_{j\alpha} + \rho_j(q_{j\alpha}+u_{j\alpha}+s_{j\alpha}+t_{j\alpha})) \\ \text{for } j \leq J,
        \label{omega_j'}
    \end{multline}
    \begin{equation}
        \tilde{\Omega}_{J+1\alpha} = \tau_{J+1} z_{\alpha}-(1-\tau_{J+1})((-1)^{J}p_{J+1\alpha}(\sigma_{J+1})- \beta_{J+1} v_{J+1\alpha} - t_{J+1\alpha} + \rho_{J+1}(q_{J+1\alpha}+u_{J+1\alpha}+t_{J+1\alpha})).
        \label{omega_J+1}
    \end{equation}
    Here $s_{J}^\a$ and $t_{J+1}^\a$ denote, respectively, the integration variables $u^{\a}$ and $v^{\a}$ in the
    star product \eqref{star_kernel}.\par
    The change of variables
    \begin{equation}
        s_{j\a} \rightarrow s_{j\a}-s_{J\a}, \quad j<J
    \end{equation}
     brings $\Omega_{i \a}$ \eqref{omega_j'}, \eqref{omega_J+1} to the canonical form \eqref{omega_j}
    with $1\leq j\leq J+1$
    (recall that we use the convention that $t_{1\a}=s_{j_{max}\a}=0$ with
    $j_{max}=J+1$) and
    $\E_{J+1}$ of the same form as $\E_{J}$ \eqref{Ej} with
    $J\to J+1$.

    \subsection{Reduced formula}
    \label{App_A:sec_2}
    Shifting $u_i^\a \rightarrow u_i^\a - q_i^\a$, formulae \eqref{Ej} and \eqref{omega_j} can be rewritten in the following way
    \begin{multline}
        \E_J = \exp i \sum_{j=1}^J \left[ (s_{j\a} + s_{j-1 \a}) t_j^\a +(s_{j\a} - t_{j\a})u_j^\a + (u_{j\a} - q_{j\a})v_j^\a + \tO_{j\a}(u_j^\a + s_j^\a + t_j^\a) -\right.\\\left.\vphantom{\sum_i} -   \sum_i p_{j,i\alpha} r_{j,i}^{\alpha} -\sum_{i<i'}p_{j,i\a}p_{j,i'}^\a  \right],
        \label{Ej11}
    \end{multline}
    \begin{equation}
        \tO_j^\a = \tau_j z^\a - (1-\tau_j)\left((-1)^{j+1}p_{j}^\a(\s_j) - \b_j v_j^\a + s_j^\a -t_j^\a + \r_j(u_j^\a + s_j^\a + t_j^\a)\right).
        \label{oi1}
    \end{equation}
    These formulae can be further elaborated on at $J \geq 2$ by observing that the integration variables $u_i^\a$ and $v_i^\a$ can be excluded from $\tO_i^\a$ and then integrated away in the Gaussian integral. Indeed, changing the variables
    \begin{equation}
        s_i^\a \to s_i^\a - c_i u_i^\a, \qquad t_i^\a \to t_i^\a - b_i u_i^\a
    \end{equation}
    with some coefficients $c_i$ and $b_i$ such that
    \begin{equation}
        c_i + b_i = 1
        \label{apb}
    \end{equation}
    we rewrite \eqref{Ej11} and $\eqref{oi1}$ in the form
    \begin{multline}
        \E_J = \exp i \sum_{j=1}^J \left[(s_{j\a} + s_{j-1 \a})t_j^\a + (c_j(s_{j\a} + t_{j+1 \a}) - b_j(t_{j\a} + s_{j-1 \a}))u_j^\a  \vphantom{\sum_i}\right. + \\
        + \left. c_{j-1} b_j u_{j-1 \a} u_j^\a + (u_{j\a} - q_{j\a})v_j^\a + \tO_{j\a}(s_j^\a + t_j^\a)  -   \sum_i p_{j,i\alpha} r_{j,i}^{\alpha} -\sum_{i<i'}p_{j,i\a}p_{j,i'}^\a\right],
        \label{Ej12}
    \end{multline}
    \begin{equation}
        \tO_i^\a = \tau_i z^\a - (1-\tau_i)\left((-1)^{i+1}p_i^\a(\s_i) - \b_i v_i^\a + s_i^\a - t_i^\a - (c_i - b_i)u_i^\a + \r_i(s_i^\a + t_i^\a)\right).
        \label{oi2}
    \end{equation}
    Shifting the variables $v_j^\a$
    \begin{equation}
        v_i^\a \to v_i^\a - \b_i^{-1}(c_i - b_i)(u_i^\a - q_i^\a)
    \end{equation}
    we observe that this does not affect the exponent in \eqref{Ej12} but replaces $u_i^\a$ by $q_i^\a$ in \eqref{oi2},
    \begin{equation}
         \tO_i^\a = \tau_i z^\a - (1-\tau_i)\left((-1)^{i+1}p_i^\a(\s_i) - \b_i v_i^\a + s_i^\a - t_i^\a - (c_i - b_i)q_i^\a + \r_i(s_i^\a + t_i^\a)\right).
        \label{oi3}
    \end{equation}
    The seeming singularity at $\b_i = 0$ is fictitious since in that case the integration over $v_i^\a$ in the exponent gives the delta-function that replaces $u_i^\a$ by $q_i^\a$ leading again to $\tO_i^\a$ \eqref{oi2}.

     Now shifting $v_i^\a$
    \begin{equation}
         v_i^\a \to v_i^\a + c_i(s_i^\a + t_{i+1}^\a) - b_i(t_i^\a + s_{i-1}^\a)
         \label{v}
    \end{equation}
    we obtain
    \begin{multline}
        \E_J = \exp i \sum_{j=1}^J \left[(s_{j\a} + s_{j-1 \a})t_j^\a + (c_j(s_{j\a} + t_{j+1 \a}) - b_j(t_{j\a} + s_{j-1 \a}))q_j^\a \vphantom{\sum_i} \right. + \\
        + \left. c_{j-1} b_j u_{j-1 \a} u_j^\a + (u_{j\a} - q_{j\a})v_j^\a + \tO_{j\a}(s_j^\a + t_j^\a)-   \sum_i p_{j,i\alpha} r_{j,i}^{\alpha} -\sum_{i<i'}p_{j,i\a}p_{j,i'}^\a \right],
        \label{Ej14}
    \end{multline}
    \begin{multline}
        \tO_i^\a = \tau_i z^\a - (1-\tau_i)\left((-1)^{i+1}p_i^\a(\s_i) - \b_i( v_i^\a + c_i(s_i^\a + t_{i+1}^\a) - b_i(t_i^\a + s_{i-1}^\a))  + s_i^\a - t_i^\a -\right.\\\left.- (c_i - b_i)q_i^\a + \r_i(s_i^\a + t_i^\a)\right).
        \label{oi4}
    \end{multline}
    Let us now specify $c_i$ and $b_i$ obeying \eqref{apb} in the form \eqref{ab-def}. For that choice
    \begin{equation}
        c_{i-1}b_i = \delta_{i-k-1, 0}\,.
        \label{abd}
    \end{equation}
    From \eqref{abd} it follow that the term bilinear in $u_i^\a$ in \eqref{Ej14} takes the form
    \begin{equation}
        c_{i-1}b_i u_{i-1 \a}u_i^\a = \delta_{i-k-1, 0}u_{k\a}u_{k+1}^\a.
    \end{equation}
    As a result, integration over $u_l^\a$ with $l < k$ or $l > k+1$ yields
    \begin{equation}
        v_l^\a = 0, \quad \text{for } l \neq k, k+1,
    \end{equation}
    \ie the only leftover components of $v_i^\a$ and $u_i^\a$ are $v_k^\a, v_{k+1}^\a$ and $u_k^\a, u_{k+1}^\a$. Integration over $u_k^\a$ yields $\delta^{2}(u_{k+1} + v_k)$ bringing \eqref{Ej14} to the form
    \begin{multline}
        \E_J = \exp i \Bigg( \sum_{j=1}^J \left[(s_{j\a} + s_{j-1 \a})t_j^\a + (c_j(s_{j\a} + t_{j+1 \a}) - b_j(t_{j\a} + s_{j-1 \a}))q_j^\a + \tO_{j\a}(s_j^\a + t_j^\a) \vphantom{\sum_i}\right. - \\
        \left.-   \sum_i p_{j,i\alpha} r_{j,i}^{\alpha} -\sum_{i<i'}p_{j,i\a}p_{j,i'}^\a \right]-v_{k\a} v_{k+1}^\a - q_{k\a} v_k^\a -q_{k+1\a}v_{k+1}^\a \Bigg ) ,
        \label{Ej15}
    \end{multline}
    where $\tO_{i\a}$ still has the form \eqref{oi4} with the convention that $v_j^\a = 0$ at $j \neq k, k+1$.

    To proceed, we represent $q_i^\a$ in the form
    \begin{equation}
        q_i^\a = (-1)^i(n_i^\a - n_{i-1}^\a)
        \label{qn}
    \end{equation}
    with $n_i$ defined in \eqref{n_i}.

    To compensate for the terms linear in $q_i^\a$ in \eqref{Ej15}, we make the following shifts of the integration variables
    \begin{equation}
        s_i^\a \to s_i^\a +(-1)^i (n_J^\a - c_i n_{i-1}^\a - b_i n_i^\a), \qquad t_i^\a \to t_i^\a + (-1)^i (c_i n_{i-1}^\a + b_i n_i^\a),
        \label{stn}
    \end{equation}
    and
    \begin{equation}
        v_{k+1}^\a \to v_{k+1}^\a + q_k^\a, \qquad v_k^\a \to v_k^\a - q_{k+1}^\a.
        \label{vq}
    \end{equation}

    The shift \eqref{vq} cancels the $qv$-type terms. One can also see that the shift \eqref{stn} compensates the $sq$- and $tq$-type terms. Note that this shift is consistent with $s_J^\a = t_1^\a = 0$ since we consider $c_J = b_1 = 0, c_1 = b_J = 1$:
    \begin{equation}
        s_J^\a \to s_J^\a - 1 \cdot n_J^\a + n_J^\a = s_J^\a, \qquad t_1^\a \to t_1^\a.
    \end{equation}
    Using \eqref{stn} from here it follows that
    \begin{equation}
        s_i^\a + t_i^\a \to s_i^\a +t_i^\a + (-1)^i n_J^\a.
    \end{equation}

    The leftover terms bilinear in $n_i^\a$ can be seen to acquire a simple form $-\sum_{j=1}^J n_{j-1 \a}n_j^\a$.
    To prove this, it is useful to employ equation \eqref{abd} and some other identities involving $c_i$ and $b_i$:
    \begin{equation}
        c_i b_i = 0, \qquad c_i + b_{i-1} = 1 - \delta_{i - k -1, 0},
    \end{equation}
    \begin{equation}
        \quad b_i b_{i+i'} = b_i, \quad c_i c_{i-i'} = c_i \quad \text{for }i'\geq0\,,
    \end{equation}
    that are valid for $c_i$ and $b_i$ defined as in \eqref{ab-def}. This brings \eqref{Ej15} to the form
    \begin{multline}
        \E_J = \exp i \Bigg( \sum_{j=1}^J \bigg[(s_{j\a} + s_{j-1 \a})t_j^\a + \tO_{j\a}(s_j^\a + t_j^\a) - n_{j-1 \a} n_j^\a -  \\  -   \sum_i p_{j,i\alpha} r_{j,i}^{\alpha} -\sum_{i<i'}p_{j,i\a}p_{j,i'}^\a\bigg]
        -v_{k\a} v_{k+1}^\a \Bigg ).
        \label{Ej16}
    \end{multline}
    To evaluate $\sum_{j=1}^Jn_{j-1 \a}n_j^\a$ we use \eqref{qn}
    \begin{equation}
        -\sum_{j=1}^J n_{j-1 \a}n_j^\a = \sum_{j=1}^J (-1)^jq_{j\a}n_j^\a =  \sum_{J\geq j > j'\geq1}(-1)^{j+j'} q_{j\a}q_{j'}^\a.
    \end{equation}
    From \eqref{qj}, it follows that this expression contains the terms bilinear in $p$ and those linear in $y$. The former have the form $\sum_{j>j'} p_{(j)+\a} p_{(j')+}^\a$. The form of $y$-dependent terms depends on whether $J$ is odd or even. The final result is
    \begin{equation}
        -\sum_{j=1}^J n_{j-1 \a}n_j^\a = -\sum_{j<j'} p_{(j)+\a} p_{(j')+}^\a - |J+1|_2\sum_{j=1}^J y_\a p_{(j)+}^\a,
    \end{equation}
    with the convention \eqref{modulo}.

    This yields \eqref{Ej17}, \eqref{oi5}.

    \section{Relation to the \texorpdfstring{$\beta$}{beta}-shifted contracting homotopy method}
    \label{App_B}

    \subsection{Homotopy operator}
    \label{subsec:homotopy_op}

    We prove the statement \eqref{eq:hom_ops_conection_in text}, showing that the operator $\tr_\lambda$ \eqref{eq:shift_diff_op} with the specific choice of measure $\l$ in the form \eqref{eq:hom_op_measure} reproduces the action of the standard $\b$-shifted contracting homotopy operator \eqref{eq:b-shifted_hom_op}.
    To proceed, we decompose $F$ as follows:
    \begin{equation}
        F(z, y|\theta) = F_0(z, y) + F_1(z , y)_\a \theta^\a + F_2(z, y) \theta_\a \theta^\a\,,
    \end{equation}
     where it is used that
    \begin{equation}
        q_\a \theta^\a k_\b \theta^\b = \dfrac{1}{2} q_\a k^\a \theta_\b \theta^\b.
    \end{equation}
    Then, let us check that
    \begin{equation}
        \fint{u^2 v^2 \tau \s^i \b } \quad \tr_{\l_0} F \cong \Delta_{p(\s_0), \b_0} F \qquad \text{with } \l_0(\s_\z, \b) = \prod_{i} \D(\s^i - \s_0^i) \cdot \D(\beta - \beta_0)
        \label{eq:hom_ops_conection}
    \end{equation}
    for each  $\theta$-grading separately. Firstly, $\tr_{\l_0} F_0(z, y) \cong 0$, because it does not contain the differential $d\tau$, which agrees with $\Delta_{p(\s_0), \b_0} F_0(z, y) \sim \dfrac{\partial}{\partial \theta} F_0(z, y) = 0$. Further, the only
    potentially non-weak term in the expression $\tr_{\l_0} F_1(z,y)_\a \theta^\a$ is that proportional to $d\tau$. Namely,
    \begin{equation}
        \tr_{\l_0} F_1(z,y)_\a \theta^\a \cong  d^2 u d^2 v \l_0(\s, \beta) l(\tau) F_1(\Omega, y + u )_\a d\tau (z + p(\s) - \b v)^\a \exp i u_\gamma v^\gamma.
    \end{equation}
    On the other hand,
    \begin{multline}
        \Delta_{p(\s_0), \b_0} \left(F_{1\a}\theta^\a\right) = \fint{u^2 v^2} d^2 u d^2 v \fint{\tau} \dfrac{d\tau}{\tau} l(\tau) (z + p(\s_0) -\b_0 v)^\a \dfrac{\partial}{\partial \theta^\a} F_1(\Omega_0, y + u)_\b \tau \theta^\b \exp i u_\gamma v^\gamma = \\
        = \fint{u^2 v^2} d^2 u d^2 v \fint{\tau}d\tau l(\tau) F_1(\Omega_0, y + u)_\a  (z + p(\s_0) -\b_0 v)^\a  \exp i u_\gamma v^\gamma.
    \end{multline}
    Thus, the correctness of  \eqref{eq:hom_ops_conection} is checked by direct comparison. For a two-form in $\theta$ this is checked analogously.
    The only term in $\tr_{\l_0} F_2(z,y) \theta_\a \theta^\a \sim F_2(z, y) \dd \Omega_\a \dd \Omega^\a$ that can be non-weak is that  proportional to $2 \tau d\tau (z + p(\s) - \b v)_\a \theta^\a$.

    This leads to the following interesting properties. Firstly, if $F \cong 0$, then $\tr_{\l_0} F \cong 0$. This is true because
    \begin{equation}
         \fint{u^2 v^2 \tau \s^i \b \{h\}} \quad  \tr_{\l_0} F \cong \fint{\{h\}} \Delta_{p(\s_0), \b_0} F = \Delta_{p(\s_0), \b_0} \left(\fint{\{h\}} F \right) = 0\,,
    \end{equation}
    where the symbol $\{h\}$ denotes all homotopy parameters in $F$ as before. In particular, this can be applied to the associator \eqref{eq:assoc}: $\tr_\l \left[(F * G) * H - F * (G * H) \right] \cong 0$. We recall that the product $*$ defined in \eqref{star_kernel} involves no actual integration. From this definition of $*$ it follows that the associator itself is non-vanishing. This distinction is fundamental: the associator vanishes only in the weak sense due to the absence of integration in the star-product kernel. The more consequential manifestation for our work --- and crucially extendable to symmetry properties --- appears in the commutator with $\gamma$ \eqref{eq:gamma_com}:
    \begin{equation}
    \tr_\l \left[F, \gamma \right]_* \cong 0.
    \label{delta_Fg}
    \end{equation}
    One can verify by straightforward computation that when $F$ is canonical \eqref{anz},
    the relation \eqref{delta_Fg} becomes a special case of the more general one \eqref{symmetry}. This is because the measure $\l$ is chosen in the particular form \eqref{eq:hom_op_measure}. Specifically, \eqref{delta_Fg} coincides with \eqref{symmetry} if in the latter  the measure for variables $\tau_{12}, \s_{12}^i, \b_{12}$ is given by
    \begin{equation}
        l(\tau_{12}) \prod_{i} \D(\s_{12}^i-\s_{0}^i) \D(\b_{12}-\b_0)\,.
    \end{equation}

    \subsection{Fields derivation}

    The equations that we are to solve at each step of the perturbation procedure can be written as follows
    \begin{align}
        2i \dd B_n = & \mathcal{R}_{B_n} + \mathcal{V}_{B_n} = \sum_{i = 1}^{n-1} \left[S_i, B_{n-i} \right]_* + \mathcal{V}_{B_n}\,,
        \label{eq:equation_Bn}\\
        2i \dd S_n = & \mathcal{R}_{S_n} + \mathcal{V}_{S_n} = \sum_{i = 1}^{n-1} S_i * S_{n-1} - i\eta B_n * \gamma + \mathcal{V}_{S_n}\,,
        \label{eq:equation_Sn}\\
        2i \dd W_n = & \mathcal{R}_{W_n} + \mathcal{V}_{W_n} = \sum_{i = 1}^{n-1} \left\{S_i, W_{n-i} \right\}_* + \sum_{i = 1}^{n} \dd_x S_i - i \eta B_{n\omega} * \gamma + \mathcal{V}_{W_n}\,,
        \label{eq:equation_Wn}
    \end{align}
    where $\mathcal{V}_i$ denote some weak terms while $\mathcal{R}_i$ are the non-weak {\it r.h.s.}  of the corresponding equations.

    \subsubsection{Solving  for \texorpdfstring{$S_n$}{Sn}}
    One may wish to reproduce the shifted homotopy scheme choosing the solution for $S_n$ in the following form
    \begin{equation}
        S_n = \dfrac{1}{2i} \tr_{\l_0} \left(\sum_{i = 1}^{n-1} S_i * S_{n-1} - i\eta B_n * \gamma\right).
    \end{equation}
    It solves equations since \eqref{eq:d_tF=0}, \eqref{eq:d_weak_term} and \eqref{eq:hom_ops_conection} take place. So,
    \begin{equation}
        S_n \cong \dfrac{1}{2i} \Delta_{p(\s_0), \b_0} \left(\mathcal{R}_{S_n} + \mathcal{V}_{S_n}\right) \Rightarrow \dd S_n = \dd_z S_n + \dd_t S_n \cong \dd_z S_n \cong \dfrac{1}{2i} \dd_z  \Delta_{p(\s_0), \b_0} \mathcal{R}_{S_n} = \dfrac{1}{2i} \mathcal{R}_{S_n}\,.
    \end{equation}

    \subsubsection{Solving for \texorpdfstring{$W_n$}{Wn}}
    \label{solving_for_Wn}
    Solution to this field can also be written in terms of the homotopy operator,
    \begin{equation}
        W_n = \dfrac{1}{2i} \tr_{\l_0} \mathcal{R}_{W_n} \cong  \dfrac{1}{2i} \tr_{\l_0} \left( \sum_{k=0}^{n} \left\{W_k, S_{n-k} \right\}_* - i \eta B_{n\omega} * \gamma \right).
    \end{equation}
    The last weak equality is fair because,
     just as in the shifted homotopy approach,
        \begin{equation}
            \tr_{\l_0} \dd_x S_k \sim \dd_x \tr_{\l_0} S_k \sim \dd_x \tr_{\l_0}\tr_{\l_0} \mathcal{R}_{S_k} \cong \dd_x \Delta_{p(\s_0), \b_0} \Delta_{p(\s_0), \b_0} \mathcal{R}_{S_k} = 0
        \end{equation}
        due to $\{ \Delta_{q_1, \b_1}, \Delta_{q_2, \b_2}\} = 0$.
    However, differential homotopy approach and its fundamental Ansatz allows us to control the form of this expression in the second order. First of all, $B_{2\w}$ is weak. Further, we should note that according to \eqref{product_l}, taking into account  that $S_2$ cointans $\delta (a_i -( 1-\b_i))$,
    \begin{equation}
        \omega(y)*S_2 = \D(\sigma_{12}^{\w}+(1-\beta_{12}))\D(\sigma_2^{\w}-(1-\beta_2))\D(\sigma_{1}^{\w}+(1-\beta_{1})) S_2 (w C C) ,
    \end{equation}
    and, consequently, the expression $\tr_{\l_0} \w * S_2$ takes the form
    \begin{equation}
            \tr_{\l_0} \omega(y)*S_2 = P(-1+\beta_{12} , \sigma_{12}^{\w}, 0)\D(\sigma_2^{\w}-(1-\beta_2))\D(\sigma_{1}^{\w}+(1-\beta_{1})) S_2 (w C C).
            \label{eq:tr_w*S_2}
    \end{equation}
    To prove this one has to use that
    \begin{equation}
        S_2 = \dfrac{1}{2i} \tr_{\l_0} \mathcal{R}_{S_2} = d^2 u_{12} d^2 v_{12} \l_0(\s_{12}, \beta_{12}) l(\tau_{12})  \mathcal{R}_{S_2}(\Omega_{12}, y + u_{12} | \dd \Omega_{12} ) \exp i u_{12\a} v_{12}^\a
    \end{equation}
    so that
    \begin{multline}
        \tr_{\l_0} \left(\w * \tr_{\l_0} \mathcal{R}_{S_2}\right) = \tr_{\l_0} \D(\sigma_{12}^{\w}+(1-\beta_{12}))\D(\sigma_2^{\w}-(1-\beta_2))\D(\sigma_{1}^{\w}+(1-\beta_{1})) \\
        d^2 u_{12} d^2 v_{12} \l_0(\s_{12}^1, \s_{12}^2, \beta_{12}) l(\tau_{12}) \mathcal{R}_{S_2}(\Omega_{12}, y + u_{12} | \dd \Omega_{12} ) \exp i u_{12\a} v_{12}^\a = \\
        =d^2 u_\z d^2 v_\z \l_0(\s_\z^1, \s_\z^2,  \b_\z)l(\tau_\z) \D(\sigma_{12}^{\w}+(1-\beta_{12}))\D(\sigma_2^{\w}-(1-\beta_2))\D(\sigma_{1}^{\w}+(1-\beta_{1})) \\
        d^2 u_{12} d^2 v_{12} \l_0(\s_{12}^1, \s_{12}^2, \beta_{12}) l(\tau_{12}) \mathcal{R}_{S_2}(\Omega_{12}', y + u_{12} + u_\z | \dd \Omega_{12}' ) \exp i\left( u_{12\a} v_{12}^\a + u_{\z \b} v_\z^\b \right),
    \end{multline}
    where now $\Omega_{12}' = \tau_{12} \z - (1-\tau_{12})\left[p(\s_{12}) - \b_{12} v_{12}\right], \quad \z = \tau_\z z - (1-\tau_\z)\left[p(\s_{\z}) - \b_{\z} v_{\z}\right]$. Since $\l_0(\s, \b) = \prod_{i} \D(\s^i - \s_0^i) \cdot \D(\beta - \beta_0)$ is composed from delta functions, the product $p(\s_{12})$ and $p(\s_\z)$ can be replaced with $p(\s_0)$ keeping the expression unchanged in a weak sense. For simplicity, $\s_0^i$ can be set to zero. Then, shifting the variables  $u_{12} \rightarrow u_{12} - u_\z$ we arrive at the expression
    \begin{multline}
        d^2 u_{12} d^2 v_{12} d^2 u_\z d^2 v_\z \mathcal{R}_{S_2}(\Omega_{12}', y + u_{12} | \dd \Omega_{12}') \exp i \left(u_{12 \a} v_{12}^\a + u_{\z \b} (v_\z - v_{12})^\b \right) \cong \\
        \cong d^2 u_{12} d^2 v_{12} d^2 v_\z \delta^{(2)}(v_\z - v_{12}) \mathcal{R}_{S_2}(\Omega_{12}', y + u_{12} | \dd \Omega_{12}') \exp i \left(u_{12 \a} v_{12}^\a \right)
        \cong \\
        \cong
        d^2 u_{12} d^2 v_{12} \mathcal{R}_{S_2}(\Omega_{12}'', y + u_{12} | \dd \Omega_{12}'') \exp i \left(u_{12 \a} v_{12}^\a \right),
    \end{multline}
    where
    \begin{multline}
        \Omega_{12}'' = \tau_{12}\left[\tau_\z z - (1-\tau_\z)\left( p(\s_0) - \b_0 v_{12} \right) \right] - (1-\tau_{12}) \left[p(\s_0) - p_\w (1-\b_{0}) - \b_{0} v_{12} \right] =
        \\
        = \tau_{12} \tau_\z z - (1 - \tau_{12} \tau_\z) \left[p(\s_{12}') + p_\w \s_{12}' - \b_{0}' v_{12} \right].
    \end{multline}
    The new coefficients in front of $p_i$ and $v_{12}$ can be easily computed
    \begin{align}
        {\s_{12}'}^{i} = & \left( \dfrac{\tau_{12}(1-\tau_\z)}{1 - \tau_{12}  \tau_\z} + \dfrac{1-\tau_{12}}{1 - \tau_{12} \tau_\z} \right) \s_0^i = \s_0^i, \quad \text{for } i = 1,2 \\
        {\s_{12}'}^{\w} = & - \dfrac{1-\tau_{12}}{1 - \tau_{12} \tau_\z} (1 - \b_{0}) \\
        \b_{12}' = & \left( \dfrac{\tau_{12}(1-\tau_\z)}{1 - \tau_{12} \tau_\z} + \dfrac{1-\tau_{12}}{1 - \tau_{12} \tau_\z} \right) \b_0 = \b_0.
    \end{align}
    As a result of change of the integration variables from $\tau_{12}, \tau_\z$ to $\tau = \tau_{12} \tau_\z, {\s_{12}'}^{\w}$, taking into account that for $0 < \tau_{12} < 1$ and $0< \tau_\z < 1$
    \begin{equation}
        \tau_{12} \tau_\z \in (0, 1), \qquad - \dfrac{1-\tau_{12}}{1 - \tau_{12} \tau_\z} (1 - \b_{0}) \in (-(1-\b_0), 0),
    \end{equation}
    the measure transforms as follows
    \begin{multline}
        l(\tau_{12}) l(\tau_\z) \l_0(\s_{12}^1, \s_{12}^2,  \b_{12})\D(\sigma_{12}^{\w}+(1-\beta_{12}))\D(\sigma_2^{\w}-(1-\beta_2))\D(\sigma_{1}^{\w}+(1-\beta_{1}))
        \rightarrow \\  \rightarrow l(\tau)  \l_0(\s_{12}^1, \s_{12}^2,  \b_{12})P(-1+\beta_{12} , \sigma_{12}^{\w}, 0)\D(\sigma_2^{\w}-(1-\beta_2))\D(\sigma_{1}^{\w}+(1-\beta_{1})),
    \end{multline}
    which proves \eqref{eq:tr_w*S_2}. Analogously it can be proven that
    \begin{equation}
            \tr_{\l_0} S_2*\omega(y) = P(-1+\beta_{12} , \sigma_{12}^{\w}, 0)\D(\sigma_2^{\w}+(1-\beta_2))\D(\sigma_{1}^{\w}-(1-\beta_{1})) S_2 (w C C).
    \end{equation}
    In \cite{Vasiliev:2023yzx} it was shown that $W_1$ admits the following representation via $S_1$:
    \begin{align}
        W_1(\w C) =\hphantom{-} & \dfrac{1}{2i} P(\b - 1, \s^\w, \s^1) S_1(\w C),\\
        W_1(C \w) =-& \dfrac{1}{2i} P(\s^1, \s^\w, 1-\b) S_1(C \w)\,,
    \end{align}
    that can be used to analyze terms $\tr_{\l_0} \left\{S_1, W_1\right\}_*$. For instance, for the $\w C C$ ordering only $\tr_{\l_0} \w*S_2(C C)$ and $\tr_{\l_0} W_1(\w C)*S_1(C)$ contributes. And $\frac{1}{2i} \tr_{\l_0} W_1 * S_1$ can be written as
    \begin{multline}
        \frac{1}{2i} \tr_{\l_0} W_1(\w C) * S_1(C) = \frac{1}{2i} \tr_{\l_0} \dfrac{1}{2i} P(\b_1 - 1, \s_{1}^\w, \s_1^1) S_1(\w C) * S_1( C) \cong \\
        \cong \dfrac{1}{2i} P(\b_1 - 1, \s_{1}^\w, \s_1^1) \D(\s_{12}^\w) \D(\s_2^\w - (1-\b_2)) \tr_{\l_0} S_1*S_1(\w C C) \cong
        \\
        \cong \dfrac{1}{2i} P(\b_1 - 1, \s_{1}^\w,0) \D(\s_{12}^\w) \D(\s_2^\w - (1-\b_2) S_2(\w C C).
        \label{eq:making_I_Sn}
    \end{multline}
    In the last weak equality, since $S_2 = \frac{1}{2i} \tr_{\l_0} \left(S_1*S_1 - i \eta B_2 * \gamma \right)$, it was asserted that the added expression  $P(\b_1 - 1, \s_{1}^\w,0) \D(\s_{12}^\w) \D(\s_2^\w - (1-\b_2) \tr_{\l_0} B_2 * \gamma $ is obviously  weak because now $\Omega_1$ contains three parameters requiring differentials, but at most two of them can  be present in $\dd \Omega_1^2$. So, a solution to $W_2$ can be  written in a concise form
    \begin{multline}
        W_2(\w C C) = \dfrac{1}{2i}\left[ P(-1+\beta_{12} , \sigma_{12}^{\w}, 0)\D(\sigma_2^{\w}-(1-\beta_2))\D(\sigma_{1}^{\w}+(1-\beta_{1})) + \right.\\
        + \left. P(\b_1 - 1, \s_{1}^\w, \s_1^1) \D(\s_{12}^\w) \D(\s_2^\w - (1-\b_2) \right] S_2 (w C C)\,.
    \end{multline}
    Analogous representations for $W_2$ can be obtained in other orderings of $\w$ and $C$.

    One might attempt to generalize this result to higher orders in the form $W_n = P_n(\sigma^\w) S_n$ with an appropriate measure $P_n(\sigma^\w)$. While we cannot rule out the existence of some relation between $W_n$ and $S_n$, the naive approach of directly replicating the above calculations fails. Concretely,
    when generalizing \eqref{eq:making_I_Sn} to higher orders to obtain an expression proportional to $S_n$, we have to add not only $P_n(\s^\w) \tr_{\l_0} B_n *\gamma$, but also summands like $P_n(\s^\w) \tr S_i*S_{n-i}$, which are no longer weak. Here, $P_n(\s^\w)$ denotes the measure for the variables $\s_i^\w$ (e.g., $P(\b_1 - 1, \s_{1}^\w,0) \D(\s_{12}^\w) \D(\s_2^\w - (1-\b_2)$).

    \section{Comparison of shifted and differential homotopy \texorpdfstring{$\delta B_2$}{delta B2}}
    \label{App_C}

    \noindent
    To obtain a projectively compact spin-local vertex,
    in \cite{Vasiliev:2016xui} it was suggested to replace $B_{2sh}$ \eqref{B2sh}  by $  B_{2sh}' := B_{2sh} + \delta B_{2sh}$
    with
    \begin{equation}
        \delta B_{2sh} = \dfrac{\eta}{2}\int \limits_0^1 d\tau C(\tau y) \bar{*} C((\tau - 1)y) k\,.
    \end{equation}

    On the other hand, in \cite{Vasiliev:2023yzx} it was shown, that $B_{2pc}$ \eqref{B2}
    also leads to a projectively compact vertex.
    Naively, there is no good reason to expect $B_{2sh}' = B_{2pc}$, despite both generating the same vertex $\Upsilon(\w C C)$. Nevertheless, as we show in this Appendix, these two fields coincide.

    Since, as shown in \cite{Didenko:2019xzz}, the lowest-order corrections, including $B_2$, $S_1$ and $W_1$, are $\b$-independent,
    without loss of generality, we set $\b=0$ throughout this section.
    Let $\delta B_{2pc} := B_{2pc} - B_{2sh}$. Here, we prove that $\delta B_{ 2pc} = \delta B_{2sh}$ hence establishing $B_{2pc } = B_{2sh}'$.

    Let us first compute $\delta B_{2sh}$.
    \begin{multline}
         \delta B_{2sh} = \dfrac{\eta}{2}\int \limits_0^1 d\tau C(\tau y) \bar{*} C((\tau - 1)y) k = \\
         = \dfrac{\eta}{2}\int d\tau d^2 r_1 d^2 r_2 \delta^{2}(r_1 - \tau y) \delta^{2}(r_2 - (\tau - 1)y) C(r_1)\bar{*} C(r_2) k = \\
         =\dfrac{\eta}{2} \int d\tau d^2 r_i d^2 p_i e^{-i p_{1\a} (r_1 - \tau y)^{\a}} e^{-i p_{2\a} (r_2 - (\tau - 1)y)^{\a}}C(r_1) \bar{*}C(r_2) k = \\
         =\dfrac{\eta}{2} \int d\tau d^2 r_i d^2 p_i e^{i\left[\tau p_{+\a} y^{\a} - p_{2\a} y^\a - p_{1\a} r_1^\a - p_{2\a} r_2^\a \right]} C(r_1) \bar{*}C(r_2) k\,,
     \end{multline}
    where, as usual, $p_{+\a} = p_{1\a} + p_{2\a}$. Now it is easy to evaluate the integral over $\tau$:
    \begin{equation}
         \int \limits_0^1 d\tau e^{i \tau p_{+\a}y^\a} = \dfrac{e^{i p_{+\a} y^\a} - 1}{i p_{+\a} y^\a}.
     \end{equation}
    Therefore, denoting $\exp -i (p_{1\a} r_1^{\a} + p_{2\a} r_2^\a) \eqqcolon \E^{(pr)} $,
    \begin{equation}
        \delta B_{2sh} =\dfrac{\eta}{2i} \int d^2 r_i d^2 p_i \;  e^{-i p_{2\a} y^\a} \cdot  \dfrac{e^{i p_{+\a} y^\a} - 1}{ p_{+\a} y^\a} \E^{(pr)} C(r_1) \bar{*}C(r_2) k.
    \end{equation}
    Here and after, entire functions like $\frac{e^{i p_{+\a} y^\a} - 1}{p_{+\a} y^\a}$ are understood  as the power series $\sum_{n=1} \frac{i^n}{n!} (p_{+\a} y^\a)^{n-1}$.

    Now, let us compute $\delta B_{2pc}$ resulting from the differential homotopy method. Recall that (for $\b = 0$)
    \begin{equation}
        B_{2sh} = -\dfrac{\eta}{4i}\fint{\tau \rho \s^1 \s^2} l(\tau) \D(\s^2 - \s^1 - 1) l(-\s^1) \D(\r) \dd \Omega^2 \E(\Omega) C(r_1) \bar{*} C(r_2) k\,,
    \end{equation}
    \begin{equation}
        B_{2pc} = \dfrac{\eta}{4i}\fint{\tau \rho \s^1 \s^2} l(\tau) \D(\s^1 + \r) \D(\s^2 - \s^1 - 1) \left[ l(-\s^1) \dd \Omega^2 + 2i \D(\s^2) d\tau \right]\E(\Omega) C(r_1) \bar{*} C(r_2) k\,,
    \end{equation}
    \begin{equation}
    \begin{gathered}
        \E(\Omega) = \exp i \left[\Omega_\a(y + p_+)^\a -p_{1\a} p_2^\a - p_{1\a} r_1^\a - p_{2\a} r_2^\a \right], 
        \\
        \Omega_\a = \tau z_\a -(1-\tau)[p_{1\a} \s^1 + p_{2\a} \s^2 + \r(y+p_+)_\a]\,.
    \end{gathered}
    \end{equation}
    Then
    \begin{align}
        \delta B_{2pc} = \delta B_2^{[1]} + \delta B_2^{[2]} + \delta B_2^{[3]}=& \dfrac{\eta}{4i}\fint{\tau \rho \s^1 \s^2} l(\tau) \D(\s^1 + \r) \D(\s^2 -\s^1 - 1) l(-\s^1) \dd \Omega^2 \E(\Omega) C(r_1) \bar{*} C(r_2) k - \label{dB_1} \\
        -& \dfrac{\eta}{4i}\fint{\tau \rho \s^1 \s^2} l(\tau) \D(\r) \D(\s^2 -\s^1 - 1) l(-\s^1) \dd \Omega^2 \E(\Omega) C(r_1) \bar{*} C(r_2) k + \label{dB_2} \\
        +& \dfrac{\eta}{2}\fint{\tau \rho \s^1 \s^2} l(\tau) \D(\s^1 + \r) \D(\s^2 -\s^1 - 1) \D(\s^2) d\tau \E(\Omega) C(r_1) \bar{*} C(r_2) k. \label{dB_3}
    \end{align}
    Firstly, consider the last term \eqref{dB_3}. Evaluating most of integrals we obtain
    \begin{multline}
       \delta B_2^{[3]} = \dfrac{\eta}{2}\fint{\tau \rho \s^1 \s^2} l(\tau) \D(\s^1 + \r) \D(\s^2 -\s^1 - 1) \D(\s^2) d\tau \E(\Omega) C(r_1) \bar{*} C(r_2) k =  \\
       = \dfrac{\eta}{2}\fint{\tau \rho \s^1 \s^2} l(\tau) \D(\s^2) \D(\s^1 + 1) \D(\r - 1) d\tau \E(\Omega) C(r_1) \bar{*} C(r_2) k =  \dfrac{\eta}{2}\fint{\tau} l(\tau) d\tau \E_1(\Omega) C(r_1) \bar{*} C(r_2) k\nonumber\,,
    \end{multline}
    where $\E_1(\Omega) = \E(\Omega(\s^1 = -1, \s^2 = 0))$. Let us consider the exponent in more detail,
    \begin{multline}
        \E_1 = \exp i\left[ \left(\tau z + (1-\tau) p_1 \right)_\a (y+p_+)^\a - p_{1\a} p_2^\a - \sum p_{i\a} r_i^\a \right] = \\
        = \exp i \left[ \tau(z_\a y^\a + z_\a p_+^\a - p_{1\a} p_2^\a) - (1-\tau) y_\a p_+^\a + (1-\tau)y_\a p_2^\a - \sum p_{i\a} r_i^\a \right] = \E_0 \cdot e^{-i(1-\tau)y_\a p_+^\a},
    \end{multline}
    where $\E_0 = \exp i\left[ \tau(z_\a y^\a + z_\a p_+^\a - p_{1\a} p_2^\a) + (1-\tau)y_\a p_2^\a - \sum p_{i\a} r_i^\a \right]$.

    Now compute the sum of the first two terms, \eqref{dB_1} and \eqref{dB_2}. To this end, we make the shift $\s^2 \rightarrow \s^2 + \s^1 + 1$ in the both terms, $\r \rightarrow \r - \s^1$ in the first term, and then evaluate the integrals over all variables except for $\tau$.

    In the first term \eqref{dB_1}, $\Omega_\a = \tau z_\a - (1-\tau)\left[p_{1\a} \s^1 + p_{2\a} (\s^1 + 1) - \s^1(y + p_+)_\a \right] =$\\$= \tau z_\a - (1-\tau)(p_{2\a} - \s^1 y_\a)$
    \begin{equation}
        \delta B_2^{[1]} = \dfrac{\eta}{4i}\fint{\tau \s^1} l(\tau) l(-\s^1) 2 d\s^1 d\tau y_\a(z + p_2)^\a(\tau - 1) \E C(r_1) \bar{*} C(r_2) k.
    \end{equation}

    The exponent in this expression is
    \begin{multline}
        \E = \exp i \left[(\tau z - (1-\tau)(p_2 - \s^1 y))_\a(y + p_+)^\a - p_{1\a} p_2^\a - \sum p_{i\a} r_i^\a \right] = \\
        = \exp i \left[ \tau(z_\a y^\a + z_\a p_+^\a -p_{1\a} p_2^\a) + (1-\tau) y_\a p_2^\a + (1-\tau)\s^1 y_\a p_+^\a - \sum p_{i\a} r_i^\a\right] = \E_0 e^{i\s^1 (1-\tau)y_\a p_+^\a}.
    \end{multline}
    In the second term \eqref{dB_2}, the exponent has the same form, as it is $\r$-independent. Now we are in a position to compute the preexponential factor. Since in the second term \eqref{dB_2}  $\Omega = \tau z - (1-\tau)\left[ p_2 + \s^1 p_+ \right]$ we find
    \begin{equation}
        \delta B_2 ^{[2]} = \dfrac{\eta}{4i}\fint{\tau \s^1} l(\tau) l(-\s^1) 2 d\s^1 d\tau p_{+\a}(z + p_2)^\a(\tau - 1) \E C(r_1) \bar{*} C(r_2) k\,.
    \end{equation}
    Thus,
    \begin{multline}
        \delta B_2^{[1]}+ \delta B_2^{[2]} = \dfrac{\eta}{2i}\fint{\tau \s^1} d\s_1 d\tau l(\tau) l(-\s^1) e^{i\s^1 (1-\tau) y_\a p_+^\a} (\tau - 1) (y+p_+)_\a (z+p_2)^\a \E_0 C(r_1) \bar{*} C(r_2) k = \\
        = \dfrac{\eta}{2}\fint{\tau} d\tau l(\tau) \dfrac{1 - e^{-i(1-\tau) y_\a p_+^\a}}{y_\b p_+^\b}  (y+p_+)_\a (z+p_2)^\a \E_0 C(r_1) \bar{*} C(r_2) k\,.
    \end{multline}
    Since $y_\a p_+^\a - (y+ p_+)_\a (z+p_2)^\a = y_\a p_1^\a - y_\a z^\a - p_{+\a} z^\a + p_{2\a} p_1^\a = (y + p_+)_\a p_1^\a - (y + p_+)_\a z^\a = (y + p_+)_\a (p_1 - z)^\a\,,$
    the total correction is
    \begin{multline}
        \delta B_{2pc} = 
        \\
        = \dfrac{\eta}{2} \int_0^1 d\tau \dfrac{(y+p_+)_\a (z+p_2)^\a + \left[y_\a p_+^\a - (y+p_+)_\a (z+p_2)^\a e^{-i(1-\tau)y_\a p_+^\a} \right]}{y_\b p_+^\b} \E_0 C(r_1) \bar{*} C(r_2) k = \\
        = \dfrac{\eta}{2} \int_0^1 d\tau \dfrac{(y+p_+)_\a (z+p_2)^\a + (y+p_+)_\a (p_1-z)^\a e^{-i(1-\tau)y_\a p_+^\a}}{y_\b p_+^\b} \E_0 C(r_1) \bar{*} C(r_2) k.
    \end{multline}
    Let us evaluate the integral
    \begin{multline}
        \int_0^1 d\tau \E_0 = \int_0^1 d\tau \exp i\left[ \tau(z_\a y^\a + z_\a p_+^\a - p_{1\a} p_2^\a) + (1-\tau)y_\a p_2^\a \right] \E^{(pr)} = \\
        = \dfrac{1}{i(z_\a y^\a + z_\a p_+^\a - p_{1\a} p_2^\a - y_\a p_2^\a)} \cdot \left( e^{i(z_\a y^\a + z_\a p_+^\a - p_{1\a} p_2^\a)} - e^{iy_\a p_2^\a} \right) \E^{(pr)} = \\
        = \dfrac{1}{i(z+p_2)_\a (y+p_+)^\a} \cdot \left(e^{i(z_\a y^\a + z_\a p_+^\a - p_{1\a} p_2^\a)} - e^{iy_\a p_2^\a} \right) \E^{(pr)}.
    \end{multline}
    Analogous calculation yields
    \begin{multline}
        \int_0^1 d\tau \E_0 e^{-i(1-\tau)y_\a p_+^\a} = \int_0^1 d\tau \exp i\left[ \tau(z_\a y^\a + z_\a p_+^\a - p_{1\a} p_2^\a) - (1-\tau)y_\a p_1^\a \right] \E^{(pr)} = \\
        = \dfrac{1}{i(z_\a y^\a + z_\a p_+^\a - p_{1\a} p_2^\a + y_\a p_1^\a)} \cdot \left( e^{i(z_\a y^\a + z_\a p_+^\a - p_{1\a} p_2^\a)} - e^{-iy_\a p_1^\a} \right) \E^{(pr)} = \\
        = \dfrac{1}{i(z-p_1)_\a (y+p_+)^\a} \cdot \left(e^{i(z_\a y^\a + z_\a p_+^\a - p_{1\a} p_2^\a)} - e^{-iy_\a p_1^\a} \right) \E^{(pr)}.
    \end{multline}
    Thus,
    \begin{multline}
        \delta B_{2pc} =
        \\
        = \dfrac{\eta}{2i} \left[e^{iy_\a p_2^\a} - e^{i(z_\a y^\a + z_\a p_+^\a - p_{1\a} p_2^\a)} + e^{i(z_\a y^\a + z_\a p_+^\a - p_{1\a} p_2^\a)} - e^{-iy_\a p_1^\a}  \right] \dfrac{1}{y_\a p_+^\a} \E^{(pr)} C(r_1) \bar{*} C(r_2) k =\\
        = \dfrac{\eta}{2i} \cdot \dfrac{e^{ip_{1\a} y^\a} - e^{-ip_{2\a} y^\a}}{p_{+\a} y^\a} \E^{(pr)} C(r_1) \bar{*} C(r_2) k = \dfrac{\eta}{2i} e^{-ip_{2\a} y^\a} \dfrac{e^{ip_{+\a} y^\a} - 1}{p_{+\a} y^\a} \E^{(pr)} C(r_1) \bar{*} C(r_2) k = \delta B_{2sh}.
    \end{multline}
    $\Box$

    \newpage
    \section{\texorpdfstring{$\mathbf{W_2^{(2)}}$}{W2(2)}-computations}
    \label{App_D}

    \noindent
    Consider the action of $\dd$ on \eqref{W2^2} (the factors resulting from the action of $\dd$ are highlighted while the weak terms are discarded)
    \begin{multline}
    	2i\dd W_2^{(2)}|_{C C \omega \omega}  \cong \frac{\eta^2}{16} \fint{\tau_1 \rho_1 \beta_1 \sigma_1^1 \sigma_1^2 \sigma_1^{\omega1} \sigma_1^{\omega2}  \tau_2 \rho_2 \beta_2 \sigma_2^1 \sigma_2^2 \sigma_2^{\omega1} \sigma_2^{\omega2}  \tau_{12} \sigma_{12}^{1} \sigma_{12}^{2} \sigma_{12}^{\omega1} \sigma_{12}^{\omega2} a_1 a_2 \beta_{12}}   \dd \Omega_2^2  C(r_1, \bar{y}, K) \bar{*}   C(r_2, \bar{y}, K) \bar{*} \omega(r_{\omega 1}, \bar{y}, K) \bar{*} \omega(r_{\omega 2}, \bar{y}, K) \big|_{r_i = 0} \times \\ \times \mu'(\beta_{12})\mu(\rho_2, \beta_2) l(\tau_{12})  P(-\overrightarrow{\sigma_{12}}, \sigma_{12}^{\omega2}, \sigma_{12}^{\omega1}, 0)    \D(\sigma_{12}^1)\D(\sigma_{12}^2) \times \\ \times \Bigg[ \bm{\D(1-\tau_1)} \D (\sigma_2^1-\overrightarrow{\sigma_2})\D (\sigma_2^2) l(\tau_2) \D (\sigma_1^1)\D (\sigma_1^2-\overrightarrow{\sigma_1}) \D(\sigma_1^{\omega1}-\overrightarrow{\sigma_1})\D(\sigma_1^{\omega2}-\overrightarrow{\sigma_1})  P(-\overrightarrow{\sigma_2},\sigma_2^{\omega2}, \sigma_2^{\omega1},\sigma_2^2)\times \\ \times  \left[ \D (\sigma_2^{\omega1}+\overrightarrow{\sigma_2})\D (\sigma_2^{\omega2}+\overrightarrow{\sigma_2})+ \D (\sigma_2^{\omega2}+\overrightarrow{\sigma_2})\D(\sigma_{12}^{\omega1}) +\D(\sigma_{12}^{\omega1}) \D(\sigma_{12}^{\omega2}) \right]\times \\ \times {\mu}(\rho_1, \beta_1) \D(a_2-\overrightarrow{\sigma_2})\D(a_1-\overrightarrow{\sigma_1})
    	\dd {\Omega}_1^2 \E (\Omega_1, \Omega_2)
    	\label{dW2^2_1}
    \end{multline}
     \begin{multline}
    	+ l(\tau_1)  \D (\sigma_2^1-\overrightarrow{\sigma_2})\D (\sigma_2^2) \bm{\D(1-\tau_2)} \D (\sigma_1^1)\D (\sigma_1^2-\overrightarrow{\sigma_1}) \D(\sigma_1^{\omega1}-\overrightarrow{\sigma_1})\D(\sigma_1^{\omega2}-\overrightarrow{\sigma_1}) \times \\ \times  \D (\sigma_2^{\omega1}+\overrightarrow{\sigma_2})\D (\sigma_2^{\omega2}+\overrightarrow{\sigma_2}) {\mu}(\rho_1, \beta_1) \D(a_2-\overrightarrow{\sigma_2})\D(a_1-\overrightarrow{\sigma_1})
    	\dd {\Omega}_1^2 \E (\Omega_1, \Omega_2)
    	\label{dW2^2_2}
    \end{multline}
     \begin{multline}
    	+ l(\tau_1) P(-\overrightarrow{\sigma_{12}}, \sigma_{12}^{\omega2}, \sigma_{12}^{\omega1}, 0) \D (\sigma_2^1-\overrightarrow{\sigma_2})\D (\sigma_2^2)\bm{ \D(\tau_2) }\D (\sigma_1^1)\D (\sigma_1^2-\overrightarrow{\sigma_1}) \D(\sigma_1^{\omega1}-\overrightarrow{\sigma_1})\D(\sigma_1^{\omega2}-\overrightarrow{\sigma_1}) \times \\ \times P(-\overrightarrow{\sigma_2},\sigma_2^{\omega2}, \sigma_2^{\omega1},\sigma_2^2)  \D(\sigma_{12}^{\omega1}) \D(\sigma_{12}^{\omega2}){\mu}(\rho_1, \beta_1)
    	\D(a_2-\overrightarrow{\sigma_2})\D(a_1-\overrightarrow{\sigma_1}) \dd {\Omega}_1^2 \E (\Omega_1, \Omega_2)
    	\label{dW2^2_3}
    \end{multline}
    \begin{multline}
    	+ \bm{\D(\tau_1)}\D (\sigma_2^1-\overrightarrow{\sigma_2})\D (\sigma_2^2) \D(1-\tau_2)  P(-\overrightarrow{\sigma_1}, \sigma_1^1, \sigma_1^2, \sigma_1^{\omega1}, \sigma_1^{\omega2}, \overrightarrow{\sigma_1}) \D (\sigma_1^2-\sigma_1^1-\overrightarrow{\sigma_1})  \D(\sigma_1^{\omega2}-\overrightarrow{\sigma_1})\D(\sigma_{12}^{\omega1}) \times \\ \times \D(\sigma_{2}^{1})\D(\sigma_{2}^{2}) \D(\sigma_{2}^{\omega1})\D(\sigma_{2}^{\omega2}) \D(\rho_1+\sigma_1^1)(\dd \Omega_1^2 + 2i \D(\sigma_1^2)\dd \tau_1 ) {\mu}(\beta_1) \D(a_2-\overrightarrow{\sigma_2})\D(a_1-\overrightarrow{\sigma_1}) \E (\Omega_1, \Omega_2)
    	\label{dW2^2_5}
    \end{multline}
    \begin{multline}
    	- l(\tau_1)  \D (\sigma_2^1-\overrightarrow{\sigma_2})\D (\sigma_2^2) \D(1-\tau_2)  P(-\overrightarrow{\sigma_1}, \sigma_1^1, \sigma_1^2, \sigma_1^{\omega1}, \sigma_1^{\omega2}, \overrightarrow{\sigma_1}) \D (\sigma_1^2-\sigma_1^1-\overrightarrow{\sigma_1})  \times \\ \times \left[-\D(\sigma_1^{\omega1}-\overrightarrow{\sigma_1})\D(\sigma_1^{\omega2}-\overrightarrow{\sigma_1}) +  \D(\sigma_1^{\omega2}-\overrightarrow{\sigma_1})\D(\sigma_{12}^{\omega1}) - \D(\sigma_{12}^{\omega1})\D(\sigma_{12}^{\omega2}) \right] \D(\sigma_{2}^{\omega1})\D(\sigma_{2}^{\omega2}) \D(\rho_1+\sigma_1^1)\times \\ \times( \bm{\D(\sigma_1^1+\overrightarrow{\sigma_1})} \dd \Omega_1^2 \E (\Omega_1, \Omega_2) + 2i \D(\sigma_1^2)\dd \tau_1 \bm{\dd \E (\Omega_1, \Omega_2)} ) {\mu}(\beta_1) \D(a_2-\overrightarrow{\sigma_2})\D(a_1-\overrightarrow{\sigma_1})
    	\label{dW2^2_6}
    \end{multline}
    \begin{multline}
    	- l(\tau_1) \bm{\D(\sigma_1^1)} \D (\sigma_2^1-\overrightarrow{\sigma_2})\D (\sigma_2^2) \D(1-\tau_2)  P(-\overrightarrow{\sigma_1}, \sigma_1^1, \sigma_1^2, \sigma_1^{\omega1}, \sigma_1^{\omega2}, \overrightarrow{\sigma_1}) \D (\sigma_1^2-\sigma_1^1-\overrightarrow{\sigma_1})  \times \\ \times \D(\sigma_1^{\omega1}-\overrightarrow{\sigma_1})\D(\sigma_1^{\omega2}-\overrightarrow{\sigma_1}) \D(\sigma_{2}^{\omega1})\D(\sigma_{2}^{\omega2}) \D(\rho_1+\sigma_1^1){\mu}(\beta_1) \dd \Omega_1^2 \E (\Omega_1, \Omega_2)\D(a_2-\overrightarrow{\sigma_2})\D(a_1-\overrightarrow{\sigma_1})
    	\label{dW2^2_6_}
    \end{multline}
    \begin{multline}
    	+ \bm{\D(a_1)}l(\tau_1)\D(\overrightarrow{\sigma_2}-a_2) \D(\tau_2) \D (\sigma_1^1)\D (\sigma_1^2-\overrightarrow{\sigma_1}) \D (\sigma_2^1-\overrightarrow{\sigma_2})\D (\sigma_2^2)\times \\ \times \D(\sigma_1^{\omega1}-a_1)\D(\sigma_1^{\omega2}-a_1)  P(-\overrightarrow{\sigma_2},\sigma_2^{\omega2}, \sigma_2^{\omega1},\sigma_2^2) \D(\sigma_{12}^{\omega1}) \D(\sigma_{12}^{\omega2}) {\mu}(\rho_1,\beta_1) \dd {\Omega_1}^2 \E (\Omega_1, \Omega_2)
    	\label{dW2^2_7}
    \end{multline}
    \begin{multline}
    	+ \bm{\D(\overrightarrow{\sigma_1}-a_1)}l(\tau_1)\D(\overrightarrow{\sigma_2}-a_2) \D(\tau_2) \D (\sigma_1^1)\D (\sigma_1^2-a_1) \D (\sigma_2^1-\overrightarrow{\sigma_2})\D (\sigma_2^2)\times \\ \times \D(\sigma_1^{\omega1}-a_1)\D(\sigma_1^{\omega2}-a_1)  P(-\overrightarrow{\sigma_2},\sigma_2^{\omega2}, \sigma_2^{\omega1},\sigma_2^2) \D(\sigma_{12}^{\omega1}) \D(\sigma_{12}^{\omega2}) {\mu}(\rho_1,\beta_1) \dd {\Omega_1}^2 \E (\Omega_1, \Omega_2)
    	\label{dW2^2_8}
    \end{multline}
    \begin{multline}
    	+ \bm{\D(1-\tau_{12})}P(0, a_1, \overrightarrow{\sigma_1})l(\tau_1)\D(\overrightarrow{\sigma_2}-a_2) \D(\tau_2) \D (\sigma_1^1)\D (\sigma_1^2-a_1) \D (\sigma_2^1-\overrightarrow{\sigma_2})\D (\sigma_2^2)\times \\ \times \D(\sigma_1^{\omega1}-a_1)\D(\sigma_1^{\omega2}-a_1)  P(-\overrightarrow{\sigma_2},\sigma_2^{\omega2}, \sigma_2^{\omega1},\sigma_2^2) \D(\sigma_{12}^{\omega1}) \D(\sigma_{12}^{\omega2}) {\mu}(\rho_1,\beta_1)\dd {\Omega_1}^2 \E (\Omega_1, \Omega_2)
    	\label{dW2^2_9}
    \end{multline}
    \begin{multline}
    	+ \bm{\D(\tau_{12})}P(0, a_1, \overrightarrow{\sigma_1})l(\tau_1)\D(\overrightarrow{\sigma_2}-a_2) \D(\tau_2) \D (\sigma_1^1)\D (\sigma_1^2-a_1) \D (\sigma_2^1-\overrightarrow{\sigma_2})\D (\sigma_2^2)\times \\ \times \D(\sigma_1^{\omega1}-a_1)\D(\sigma_1^{\omega2}-a_1)  P(-\overrightarrow{\sigma_2},\sigma_2^{\omega2}, \sigma_2^{\omega1},\sigma_2^2) \D(\sigma_{12}^{\omega1}) \D(\sigma_{12}^{\omega2}) {\mu}(\rho_1,\beta_1)\dd {\Omega_1}^2 \E (\Omega_1, \Omega_2)
    	\label{dW2^2_9_1}
    \end{multline}
    \begin{multline}
    	- \bm{\D(1-\tau_2)}\D(\overrightarrow{\sigma_1}-a_1)\D(a_2) \D(\tau_1) P(-\overrightarrow{\sigma_1}, \sigma_1^1, \sigma_1^2, \sigma_1^{\omega1}, \sigma_1^{\omega2}, \overrightarrow{\sigma_1})\D (\sigma_1^2-\sigma_1^1-\overrightarrow{\sigma_1}) \D (\sigma_2^1)\D(\sigma_2^2) \times \\ \times\D(\sigma_{12}^{\omega1})  \D(\sigma_1^{\omega_2}-\overrightarrow{\sigma_1})\D(\sigma_2^{\omega1}) \D(\sigma_2^{\omega_2}+\overrightarrow{\sigma_2})P(-\overrightarrow{\sigma_{12}}, \sigma_{12}^{\omega_2}, \sigma_{12}^{\omega_1}) \D(\rho_1 + \sigma_1^1) {\mu}(\beta_1)
    	\dd {\Omega}_1^2 \E (\Omega_1, \Omega_2)
    	\label{dW2^2_10}
    \end{multline}
    \begin{multline}
    	- \bm{\D(1-\tau_{12})}\D(\overrightarrow{\sigma_1}-a_1)\D(a_2) \D(\tau_1) l(\tau_2) P(-\overrightarrow{\sigma_1}, \sigma_1^1, \sigma_1^2, \sigma_1^{\omega1}, \sigma_1^{\omega2}, \overrightarrow{\sigma_1})\D (\sigma_1^2-\sigma_1^1-\overrightarrow{\sigma_1}) \D (\sigma_2^1)\D(\sigma_2^2) \times \\ \times\D(\sigma_{12}^{\omega1})  \D(\sigma_1^{\omega_2}-\overrightarrow{\sigma_1})\D(\sigma_2^{\omega1}) P(-\overrightarrow{\sigma_2}, \sigma_{2}^{\omega2}, 0)\D(\sigma_{12}^{\omega_2}-\sigma_{12}^{\omega_1})  \D(\rho_1 + \sigma_1^1) {\mu}(\beta_1)
    	\dd {\Omega}_1^2 \E (\Omega_1, \Omega_2)
    	\label{dW2^2_11}
    \end{multline}
    \begin{multline}
    	- \bm{\D(\tau_{12})}\D(\overrightarrow{\sigma_1}-a_1)\D(a_2) \D(\tau_1) l(\tau_2) P(-\overrightarrow{\sigma_1}, \sigma_1^1, \sigma_1^2, \sigma_1^{\omega1}, \sigma_1^{\omega2}, \overrightarrow{\sigma_1})\D (\sigma_1^2-\sigma_1^1-\overrightarrow{\sigma_1}) \times \\ \times \D (\sigma_2^1)\D(\sigma_2^2) \D(\sigma_{12}^{\omega1})   \D(\sigma_1^{\omega_2}-\overrightarrow{\sigma_1})\D(\sigma_2^{\omega1}) \big[P(-\overrightarrow{\sigma_2},\sigma_{2}^{\omega2}, 0)\D(\sigma_{12}^{2}-\sigma_{12}^{\omega_1}) +\\+ \D(\sigma_2^{\omega_2}+\sigma_2)P(-\overrightarrow{\sigma_{12}}, \sigma_{12}^{\omega_2}, \sigma_{12}^{\omega_1}) \big]  \D(\rho_1 + \sigma_1^1) {\mu}(\beta_1)
    	\dd {\Omega}_1^2 \E (\Omega_1, \Omega_2)
    	\label{dW2^2_13}
    \end{multline}
    \begin{multline}
    	+ \D(\overrightarrow{\sigma_1}-a_1)\D(a_2) \D(\tau_1) l(\tau_2) \bm{\D(\sigma_1^1+\overrightarrow{\sigma_1})} P(-\overrightarrow{\sigma_1}, \sigma_1^1, \sigma_1^2, \sigma_1^{\omega1}, \sigma_1^{\omega2}, \overrightarrow{\sigma_1})\D (\sigma_1^2-\sigma_1^1-\overrightarrow{\sigma_1}) \D (\sigma_2^1)\D(\sigma_2^2) \times \\ \times\D(\sigma_{12}^{\omega1})  \D(\sigma_2^{\omega1}) \D(\sigma_{12}^{\omega_2}-\sigma_{12}^{\omega_1})\D(\sigma_2^{\omega_2})  \D(\rho_1 + \sigma_1^1) {\mu}(\beta_1)
    	\dd {\Omega}_1^2 \E (\Omega_1, \Omega_2)  \Bigg]
    	\label{dW2^2_12}
    \end{multline}
    \begin{equation}
        +(\dd_x W_2 + W_2*\omega - \Upsilon)|_{C C \omega\omega}.
        \label{dW2^2_14}
    \end{equation}

    \par
    Using \eqref{symmetry}, \eqref{symmetry-2} by computations analogous to \eqref{W2_anal}
    one finds that \eqref{dW2^2_1} cancels \eqref{dW2^2_6},
     \eqref{dW2^2_2} cancels \eqref{dW2^2_6_}, \eqref{dW2^2_3} cancels \eqref{dW2^2_8}, \eqref{dW2^2_5} cancels \eqref{dW2^2_10},
     \eqref{dW2^2_7} cancels \eqref{dW2^2_12},
     \eqref{dW2^2_9} is equal to \eqref{d_x_W2_dw} and \eqref{dW2^2_11} is equal to \eqref{d_x_W2_dC}.

    \par \eqref{dW2^2_13} is weak since $\Omega_{2\alpha} \propto (\dots)p_{\omega_2 \alpha}$ after integration over $s_{\alpha}, t_{\alpha}, u_{i\alpha}, v_{j\alpha}$ and hence $\dd \Omega_{2\alpha} \dd \Omega_{2}^{\alpha} \cong 0$. Analogously \eqref{dW2^2_9_1} is weak.
    \par

    The other sectors differ by the $\sigma_i^k$  dependence of the measure $\dd_x W_2 + W_2*\omega + \omega*W_2$. As a result we obtain the following expressions:
    \begin{multline}
        W_2^{(2)}|_{C \omega C \omega} = \dfrac{i\eta^2}{32} \fint{{\tau_1 \rho_1 \beta_1 \sigma_1^1 \sigma_1^2 \sigma_1^{\omega1} \sigma_1^{\omega2}  \tau_2 \rho_2 \beta_2 \sigma_2^1 \sigma_2^2 \sigma_2^{\omega 1} \sigma_2^{\omega 2} \tau_{12} \sigma_{12}^1 \sigma_{12}^2 \sigma_{12}^{\omega1} \sigma_{12}^{\omega2} a_1 a_2\beta_{12}}} \E(\Omega_1, \Omega_2)  C(r_1, \bar{y}, K) \bar{*}   C(r_2, \bar{y}, K) \bar{*} \omega(r_{\omega 1}, \bar{y}, K) \bar{*} \omega(r_{\omega 2}, \bar{y}, K) \big|_{r_i = 0} \times \\ \times \mu'(\beta_{12}) l(\tau_{12}) l(\tau_1){\mu}(\rho_2, \beta_2)
        \dd {\Omega}_2^2 \Big[ \D (\sigma_2^1-\overrightarrow{\sigma_2})\D (\sigma_2^2)\D(\sigma_{12}^1)\D(\sigma_{12}^2)P(-\overrightarrow{\sigma_{12}},\sigma_{12}^{\omega2},\sigma_{12}^2)\times \\ \times \big \{  l(\tau_2) \D (\sigma_1^1)\D (\sigma_1^2-\overrightarrow{\sigma_1}) \D(\sigma_1^{\omega2}-\overrightarrow{\sigma_1})  P(\sigma_1^1,\sigma_1^{\omega1}, \sigma_1^2)P(-\overrightarrow{\sigma_2},\sigma_2^{\omega2}, \sigma_2^2, \sigma_2^{\omega1},\sigma_2^1) \left[ \D (\sigma_2^{\omega1}-\sigma_2^1) \D (\sigma_2^{\omega2}+\overrightarrow{\sigma_2}) +\right.\\\left.+ \D (\sigma_1^{\omega1}-\sigma_1^2) \D (\sigma_2^{\omega2}+\overrightarrow{\sigma_2}) + \D(\sigma_{12}^{\omega2})\D (\sigma_2^{\omega1}-\sigma_2^1) + \D(\sigma_{12}^{\omega2})\D (\sigma_1^{\omega1}-\sigma_1^2) \right]\D(\sigma_{12}^{\omega1})\D(\rho_1)\dd {\Omega}_1^2 -\\- \D(1-\tau_2)    P(-\overrightarrow{\sigma_1}, \sigma_1^1, \sigma_1^{\omega1}, \sigma_1^2, \sigma_1^{\omega2}, \overrightarrow{\sigma_1})\D (\sigma_1^2-\sigma_1^1-\overrightarrow{\sigma_1}) \D(\sigma_{2}^{\omega1})\D(\sigma_{2}^{\omega2})  \D (\sigma_1^{\omega2}-\overrightarrow{\sigma_1})\D(\sigma_{12}^{\omega1}) \D(\rho_1+\sigma_1^1)\times \\ \times(\dd \Omega_1^2 + 2i(1-\beta_1)\D(\sigma_1^2)\dd \tau_1) \big \} \D(a_2-\overrightarrow{\sigma_2})\D(a_1-\overrightarrow{\sigma_1}) {\mu}(\beta_1) - \\- P(0,a_1,\overrightarrow{\sigma_1}) \D(\tau_2)\D(\overrightarrow{\sigma_2}-a_2)  \D (\sigma_1^1)\D (\sigma_1^2-a_1) \D (\sigma_2^1-\overrightarrow{\sigma_2})\D (\sigma_2^2) \D(\sigma_1^{\omega1}-a_1)\D(\sigma_1^{\omega2}-a_1) \times \\ \times P(-\overrightarrow{\sigma_2},\sigma_2^{\omega2}, \sigma_2^2, \sigma_2^{\omega1},\sigma_2^1) \D(\sigma_{12}^{\omega1}) \D(\sigma_{12}^{\omega2})\D(\sigma_{12}^1)\D(\sigma_{12}^2) {\mu}(\rho_1, \beta_1)\dd {\Omega}_1^2+ \\ + \D(\overrightarrow{\sigma_1}-a_1)\D(a_2) \D(\tau_1) l(\tau_2) P(-\overrightarrow{\sigma_1}, \sigma_1^1, \sigma_1^{\omega1}, \sigma_1^2, \sigma_1^{\omega2}, \overrightarrow{\sigma_1}) \D (\sigma_1^2-\sigma_1^1-\overrightarrow{\sigma_1})  \D (\sigma_2^1)\D(\sigma_2^2)\D(\sigma_2^{\omega1}) \times \\ \times \D(\sigma_{12}^{\omega1}) \D(\sigma_{12}^1)\D(\sigma_{12}^2) \big[P(-\overrightarrow{\sigma_2},\sigma_{2}^{\omega2}, \sigma_2^2)\D(\sigma_{12}^{\omega2})\D(\sigma_{1}^{\omega_2}-\overrightarrow{\sigma_1}) +\\+ P(-\overrightarrow{\sigma_{12}},\sigma_{12}^{\omega2},\sigma_{12}^2)\D(\sigma_{1}^{\omega_2}-\overrightarrow{\sigma_1})\D(\sigma_2^{\omega_2}+\overrightarrow{\sigma_2})+ \D(\sigma_{12}^{\omega2})\D(\sigma_{2}^{\omega2}-\sigma_2^2)\big] \D(\rho_1 + \sigma_1^1) {\mu}(\beta_1)
        \dd {\Omega}_1^2 \Big ],
    \end{multline}\par
    \begin{multline}
        W_2^{(2)}|_{C \omega \omega C} = \dfrac{i\eta^2}{32} \fint{{\tau_1 \rho_1 \beta_1 \sigma_1^1 \sigma_1^2 \sigma_1^{\omega1} \sigma_1^{\omega2}  \tau_2 \rho_2 \beta_2 \sigma_2^1 \sigma_2^2 \sigma_2^{\omega 1} \sigma_2^{\omega 2} \tau_{12} \sigma_{12}^1 \sigma_{12}^2 \sigma_{12}^{\omega1} \sigma_{12}^{\omega2}a_1 a_2\beta_{12}}} {\mu}(\rho_2, \beta_2)
        \dd {\Omega}_1^2 \dd {\Omega}_2^2 {\E}   C(r_1, \bar{y}, K)\bar{*} \omega(r_{\omega 1}, \bar{y}, K) \bar{*} \omega(r_{\omega 2}, \bar{y}, K) \bar{*}   C(r_2, \bar{y}, K)\big|_{r_i = 0} \times \\ \times \mu'(\beta_{12}) l(\tau_{12}) l(\tau_1) l(\tau_2) \D(\sigma_{12}^1)\D(\sigma_{12}^2) \Big[ \D (\sigma_1^1)\D (\sigma_1^2-\overrightarrow{\sigma_1}) \D (\sigma_2^1-\overrightarrow{\sigma_2})\D (\sigma_2^2) \D(a_2-\overrightarrow{\sigma_2})\D(a_1-\overrightarrow{\sigma_1}) \times \\ \times P(\sigma_1^1,\sigma_1^{\omega1}, \sigma_1^{\omega2}, \sigma_1^2)P(\sigma_2^2,\sigma_2^{\omega2}, \sigma_2^{\omega1}, \sigma_2^1) \big[ \D (\sigma_2^{\omega1}-\sigma_2^1) \D (\sigma_2^{\omega2}-\sigma_2^1) +\\+ \D (\sigma_1^{\omega1}-\sigma_1^2) \D (\sigma_1^{\omega2}-\sigma_1^2) - \D (\sigma_1^{\omega2}-\sigma_1^2)\D (\sigma_2^{\omega1}-\sigma_2^1) \big]\D(\sigma_{12}^{\omega1})\D(\sigma_{12}^{\omega2}){\mu}(\rho_1, \beta_1)+ \\+ P(0,a_1,\overrightarrow{\sigma_1})\D(\overrightarrow{\sigma_2}-a_2) \D(\tau_2) \D (\sigma_1^1)\D (\sigma_1^2-a_1) \D (\sigma_2^1-\overrightarrow{\sigma_2})\D (\sigma_2^2)\times \\ \times \D(\sigma_1^{\omega1}-a_1)\D(\sigma_1^{\omega2}-a_1)  P( \sigma_2^2, \sigma_2^{\omega2}, \sigma_2^{\omega1}, \sigma_2^1) \D(\sigma_{12}^{\omega1}) \D(\sigma_{12}^{\omega2}){\mu}(\rho_1, \beta_1)+ \\+ P(0,a_2,\overrightarrow{\sigma_2})\D(\overrightarrow{\sigma_1}-a_1) \D(\tau_1) \D (\sigma_1^1)\D (\sigma_1^2-\overrightarrow{\sigma_1}) \D (\sigma_2^1-a_2)\D (\sigma_2^2)\times \\ \times \D(\sigma_2^{\omega1}-a_2)\D(\sigma_2^{\omega2}-a_2)  P(\sigma_1^1,\sigma_1^{\omega1}, \sigma_1^{\omega2}, \sigma_1^2) \D(\sigma_{12}^{\omega1}) \D(\sigma_{12}^{\omega2}){\mu}(\rho_1, \beta_1)+ \\ +\D(\overrightarrow{\sigma_1}-a_1)\D(a_2) \D(\tau_1) P(-\overrightarrow{\sigma_1}, \sigma_1^1, \sigma_1^{\omega1}, \sigma_1^{\omega2}, \sigma_1^2, \overrightarrow{\sigma_1}) \D (\sigma_1^2-\sigma_1^1-\overrightarrow{\sigma_1}) \times \\ \times \D (\sigma_2^1)\D(\sigma_2^2) \D(\sigma_{12}^{\omega1})\D(\sigma_{12}^{\omega2}) \D(\sigma_2^{\omega_1})\D(\sigma_2^{\omega_2})\D(\rho_1 + \sigma_1^1) {\mu}(\beta_1) \Big],
    \end{multline}\par
    \begin{multline}
        W_2^{(2)}|_{\omega C C \omega} = \dfrac{i\eta^2}{32} \fint{{\tau_1 \rho_1 \beta_1 \sigma_1^1 \sigma_1^2 \sigma_1^{\omega1} \sigma_1^{\omega2}  \tau_2 \rho_2 \beta_2 \sigma_2^1 \sigma_2^2 \sigma_2^{\omega 1} \sigma_2^{\omega 2} \tau_{12} \sigma_{12}^1 \sigma_{12}^2 \sigma_{12}^{\omega1} \sigma_{12}^{\omega2}a_1 a_2 \beta_{12}}}    \omega(r_{\omega 1}, \bar{y}, K) \bar{*} C(r_1, \bar{y}, K) \bar{*}   C(r_2, \bar{y}, K) \bar{*} \omega(r_{\omega 2}, \bar{y}, K) \big|_{r_i = 0} \times \\ \times \mu'(\beta_{12}) l(\tau_{12}) l(\tau_1) l(\tau_2) {\mu}(\rho_2, \beta_2)
        \dd {\Omega}_2^2 {\E}(\Omega_1, \Omega_2) P(-\overrightarrow{\sigma_1}, \sigma_1^{\omega1}, \sigma_1^1, \sigma_1^2, \sigma_1^{\omega2}, \overrightarrow{\sigma_1}) \times \\ \times \Big[ P(-\overrightarrow{\sigma_2}, \sigma_2^{\omega2}, \sigma_2^2, \sigma_2^1, \sigma_2^{\omega1}, \overrightarrow{\sigma_2})P(-\overrightarrow{\sigma_{12}}, \sigma_{12}^{\omega2}, \sigma_{12}^2)P(-\overrightarrow{\sigma_{12}}, \sigma_{12}^{\omega1},\sigma_{12}^1) \D(\sigma_{12}^1)\D(\sigma_{12}^2) \times \\ \times \D(a_2-\overrightarrow{\sigma_2})\D(a_1-\overrightarrow{\sigma_1}) \big \{ l(\tau_2)  \D (\sigma_1^1)\D (\sigma_1^2-\overrightarrow{\sigma_1}) \D(\sigma_{1}^{\omega2}-\overrightarrow{\sigma_1})\D(\sigma_{2}^{\omega1}-\overrightarrow{\sigma_2}) \D (\sigma_2^1-\overrightarrow{\sigma_2})\D (\sigma_2^2)\times  \\ \times \big[ \D(\sigma_{12}^{\omega1})\D(\sigma_{12}^{\omega2}) + \D(\sigma_2^{\omega2}+\overrightarrow{\sigma_2})\D(\sigma_{12}^{\omega1}) + \D(\sigma_1^{\omega1}+\overrightarrow{\sigma_1})\D(\sigma_{12}^{\omega2})- \D(\sigma_1^{\omega1}+\overrightarrow{\sigma_1})\D(\sigma_2^{\omega2}+\overrightarrow{\sigma_2}) \big] \D(\rho_1)\dd {\Omega}_1^2 -\\-  \D(1-\tau_2) \D (\sigma_1^2-\sigma_1^1-\overrightarrow{\sigma_1}) \D (\sigma_2^1)\D(\sigma_2^2)\D(\sigma_{2}^{\omega1})\D(\sigma_{2}^{\omega2}) \big [ \D(\sigma_{12}^{\omega1})\D(\sigma_{12}^{\omega2}) +\D(\sigma_1^{\omega2}-\overrightarrow{\sigma_1})\D(\sigma_{12}^{\omega1}) +\\+ \D(\sigma_1^{\omega1}+\overrightarrow{\sigma_1})\D(\sigma_{12}^{\omega2})- \D(\sigma_1^{\omega1}+\overrightarrow{\sigma_1})\D(\sigma_1^{\omega2}-\overrightarrow{\sigma_1}) \big ]\D(\rho_1+\sigma_1^1)(\dd \Omega_1^2 + 2i(1-\beta_1)\D(\sigma_1^2)\dd \tau_1)\big\}{\mu}(\beta_1) +
        \\
        +  \D(\overrightarrow{\sigma_1}-a_1) \D(\tau_1) \D(a_2)l(\tau_2) P(-\overrightarrow{\sigma_1}, \sigma_1^{\omega1}, \sigma_1^1, \sigma_1^2, \sigma_1^{\omega2}, \overrightarrow{\sigma_1}) \D (\sigma_1^2-\sigma_1^1-\overrightarrow{\sigma_1})
        \times 
        \\
        \times P(-\overrightarrow{\sigma_{12}}, \sigma_{12}^{\omega2}, \sigma_{12}^2)P(-\overrightarrow{\sigma_{12}}, \sigma_{12}^{\omega1},\sigma_{12}^1) \times 
        \\
        \times  \big [ \D(\sigma_{12}^{\omega1})\D(\sigma_2^{\omega_1}) \big\{ P(-\overrightarrow{\sigma_2}, \sigma_2^{\omega2}, \sigma_2^2)\D(\sigma_1^{\omega2}-\overrightarrow{\sigma_1})\D(\sigma_{12}^{\omega2}+\overrightarrow{\sigma_{12}}) + \D(\sigma_2^{\omega_2})\D(\sigma_1^{\omega2}-\overrightarrow{\sigma_1}) + \D(\sigma_2^{\omega_2})\D(\sigma_{12}^{\omega2})  \big\} + 
        \\
        +\D(\sigma_{12}^{\omega2})\D(\sigma_2^{\omega_2}) \big\{ P(\sigma_2^1,\sigma_2^{\omega1},  \overrightarrow{\sigma_2})\D(\sigma_1^{\omega1}+\overrightarrow{\sigma_1})\D(\sigma_{12}^{\omega1}+\overrightarrow{\sigma_{12}}) + \D(\sigma_1^{\omega1}+\overrightarrow{\sigma_1})\D(\sigma_2^{\omega_1}) + \D(\sigma_2^{\omega_1})\D(\sigma_{12}^{\omega1})  \big\} \big ]\times 
        \\
        \times \D (\sigma_2^1)\D(\sigma_2^2)  \D(\sigma_{12}^1)\D(\sigma_{12}^2)  \D(\rho_1 + \sigma_1^1) {\mu}(\beta_1) \dd {\Omega}_1^2 \Big].
    \end{multline}\par
    These expressions allow us to obtain vertices of all orderings of $C$ and $\omega$, resulting from the action of $\dd$  on $W_2^{(2)}$. Straightforward computations show that terms with $\D(\tau_{12})$ resulting from such action produce the non-zero contributions to the vertices.

    \section{The \texorpdfstring{$\mathbf{\beta \rightarrow -\infty}$}{beta -> -oo} limit}
    \label{App_E}

    \noindent
    Consider the limit $\beta \rightarrow -\infty$. To this end let us decompose \eqref{vertexCCww} into four terms:
    \begin{equation}
        \Upsilon(CC\omega\omega) = \Upsilon_1 + \Upsilon_2 + \Upsilon_3 + \Upsilon_4\,,
        \label{vertex}
    \end{equation}
    \begin{multline}
        \Upsilon_1 = -\dfrac{\eta^2}{16} \fint{\tau_1 \rho_1 \beta_1 \sigma_1^1 \sigma_1^2 \sigma_1^{\omega1} \sigma_1^{\omega2}  \tau_2 \rho_2 \beta_2 \sigma_2^1 \sigma_2^2 \sigma_2^{\omega 1} \sigma_2^{\omega 2} \tau_{12} \sigma_{12}^1 \sigma_{12}^2 \sigma_{12}^{\omega1} \sigma_{12}^{\omega2} a_1 a_2 \beta_{12}} \mu'(\beta_{12}) \D(\overrightarrow{\sigma_2}-a_2) \D(\overrightarrow{\sigma_1}-a_1) \D(\tau_{12})l(\tau_1) l(\tau_2) \D (\sigma_1^1)\D (\sigma_1^2-\overrightarrow{\sigma_1}) \D(\sigma_1^{\omega1}-\overrightarrow{\sigma_1})\D(\sigma_1^{\omega2}-\overrightarrow{\sigma_1}) \times \\ \times P(-\overrightarrow{\sigma_{12}},\sigma_{12}^{\omega2}, \sigma_{12}^{\omega1},\sigma_{12}^2)  \D (\sigma_2^{\omega1}+\overrightarrow{\sigma_2})\D (\sigma_2^{\omega2}+\overrightarrow{\sigma_2})\D (\sigma_2^1-\overrightarrow{\sigma_2})\D(\sigma_2^2)\D(\sigma_{12}^1)\D(\sigma_{12}^2) {\mu}(\rho_1, \beta_1){\mu}(\rho_2, \beta_2)\times \\ \times
        \dd {\Omega}_1^2 \dd {\Omega}_2^2 {\E}   C(r_1, \bar{y}, K) \bar{*}   C(r_2, \bar{y}, K) \bar{*} \omega(r_{\omega 1}, \bar{y}, K) \bar{*} \omega(r_{\omega 2}, \bar{y}, K) \big|_{r_i = 0},
        \label{vertex-1}
    \end{multline}
    \begin{multline}
        \Upsilon_2 = -\dfrac{\eta^2}{16} \fint{\tau_1 \rho_1 \beta_1 \sigma_1^1 \sigma_1^2 \sigma_1^{\omega1} \sigma_1^{\omega2}  \tau_2 \rho_2 \beta_2 \sigma_2^1 \sigma_2^2 \sigma_2^{\omega 1} \sigma_2^{\omega 2} \tau_{12} \sigma_{12}^1 \sigma_{12}^2 \sigma_{12}^{\omega1} \sigma_{12}^{\omega2} a_1 a_2 \beta_{12}} \mu'(\beta_{12}) \D(\overrightarrow{\sigma_2}-a_2) \D(\overrightarrow{\sigma_1}-a_1) \D(\tau_{12})l(\tau_1) \D(1-\tau_2)   P(-\overrightarrow{\sigma_1}, \sigma_1^1, \sigma_1^2, \sigma_1^{\omega1}, \sigma_1^{\omega2}, \overrightarrow{\sigma_1})\times \\ \times \D (\sigma_1^2-\sigma_1^1-\overrightarrow{\sigma_1})\D(\sigma_1^{\omega1}-\overrightarrow{\sigma_1})\D(\sigma_1^{\omega2}-\overrightarrow{\sigma_1})  \D (\sigma_2^1)\D(\sigma_2^2) P(-\overrightarrow{\sigma_{12}},\sigma_{12}^{\omega2}, \sigma_{12}^{\omega1},\sigma_{12}^2)\D(\sigma_{12}^1)\D(\sigma_{12}^2) \D(\sigma_{2}^{\omega1})\times \\ 
         \times
          \D(\sigma_{2}^{\omega2}) \D(\rho_1+\sigma_1^1){\mu}(\beta_1){\mu}(\rho_2, \beta_2)
        \dd {\Omega}_1^2 \dd {\Omega}_2^2 {\E}   C(r_1, \bar{y}, K) \bar{*}   C(r_2, \bar{y}, K) \bar{*} \omega(r_{\omega 1}, \bar{y}, K) \bar{*} \omega(r_{\omega 2}, \bar{y}, K) \big|_{r_i = 0},
        \label{vertex-2}
    \end{multline}
    \begin{multline}
        \Upsilon_3 = \dfrac{\eta^2}{16} \fint{\tau_1 \rho_1 \beta_1 \sigma_1^1 \sigma_1^2 \sigma_1^{\omega1} \sigma_1^{\omega2}  \tau_2 \rho_2 \beta_2 \sigma_2^1 \sigma_2^2 \sigma_2^{\omega 1} \sigma_2^{\omega 2} \tau_{12} \sigma_{12}^1 \sigma_{12}^2 \sigma_{12}^{\omega1} \sigma_{12}^{\omega2} a_1 a_2 \beta_{12}} \mu'(\beta_{12}) \D(\overrightarrow{\sigma_2}-a_2) \D(\overrightarrow{\sigma_1}-a_1) l(\tau_1) \D(1-\tau_2) \D(\tau_{12}) P(-\overrightarrow{\sigma_{12}},\sigma_{12}^{\omega2}, \sigma_{12}^{\omega1},\sigma_{12}^2) \D (\sigma_1^1+\overrightarrow{\sigma_1})\D(\sigma_1^2)\times \\ \times \D (\sigma_2^1)\D(\sigma_2^2)  \D(\sigma_1^{\omega1}-\overrightarrow{\sigma_1})\D(\sigma_1^{\omega2}-\overrightarrow{\sigma_1})\D(\sigma_{12}^1)\D(\sigma_{12}^2) \D(\sigma_{2}^{\omega1})\D(\sigma_{2}^{\omega2})\D(\rho_1+\sigma_1^1)  {\mu}(\beta_1){\mu}(\rho_2, \beta_2)
        \times \\ \times (2i(1-\beta_1)\dd \tau_{1}) \dd {\Omega}_2^2 {\E}   C(r_1, \bar{y}, K) \bar{*}   C(r_2, \bar{y}, K) \bar{*} \omega(r_{\omega 1}, \bar{y}, K) \bar{*} \omega(r_{\omega 2}, \bar{y}, K) \big|_{r_i = 0},
        \label{vertex-3}
    \end{multline}
    \begin{multline}
        \Upsilon_4 = -\dfrac{\eta^2}{16} \fint{\tau_1 \rho_1 \beta_1 \sigma_1^1 \sigma_1^2 \sigma_1^{\omega1} \sigma_1^{\omega2}  \tau_2 \rho_2 \beta_2 \sigma_2^1 \sigma_2^2 \sigma_2^{\omega 1} \sigma_2^{\omega 2} \tau_{12} \sigma_{12}^1 \sigma_{12}^2 \sigma_{12}^{\omega1} \sigma_{12}^{\omega2} \beta_{12}} \mu'(\beta_{12}) \D(\tau_{12}) l(\tau_1) l(\tau_2) \D (\sigma_1^1)\D (\sigma_1^2-\overrightarrow{\sigma_1}) \D (\sigma_2^1-\overrightarrow{\sigma_2})\D (\sigma_2^2)P(-\overrightarrow{\sigma_2},\sigma_2^{\omega2}, \sigma_2^{\omega1},\sigma_2^2) \times \\ \times \D(\sigma_1^{\omega1}-\overrightarrow{\sigma_1})\D(\sigma_1^{\omega2}-\overrightarrow{\sigma_1}) \D (\sigma_2^{\omega2}+\overrightarrow{\sigma_2}) P(-\overrightarrow{\sigma_{12}},\sigma_{12}^{\omega2}, \sigma_{12}^{\omega1},\sigma_{12}^2)\D(\sigma_{12}^{\omega1}-\sigma_{12}^2) \D(\sigma_{12}^1)\D(\sigma_{12}^2)\times \\ \times {\mu}(\rho_1, \beta_1){\mu}(\rho_2, \beta_2)
        \dd {\Omega}_1^2 \dd {\Omega}_2^2 {\E}   C(r_1, \bar{y}, K) \bar{*}   C(r_2, \bar{y}, K) \bar{*} \omega(r_{\omega 1}, \bar{y}, K) \bar{*} \omega(r_{\omega 2}, \bar{y}, K) \big|_{r_i = 0}\,.
        \label{vertex-4}
    \end{multline}
    Upon integration over $s_{\alpha}, t_{\alpha}, u_{12\alpha}, v_{12\alpha}$ the expression in the exponent of the vertex takes the form
    \begin{multline}
        \frac{1}{1-\beta \tau_{\circ}}y_{\alpha}(\tau_{\circ}p(\sigma_{12})+(1-\tau_2)(1-\tau_1)(p(\sigma_1) + p(\sigma_2) - p_+) - \tau_1 \tau_2 p_+)^{\alpha}+\\+\frac{1}{1-\beta \tau_{\circ}}p(\sigma_{12})_{\alpha}(\tau_2 (1-\tau_1)p(\sigma_1) - \tau_1 (1-\tau_2)p(\sigma_2))^{\alpha}+\\+ \frac{1-\beta}{1-\beta \tau_{\circ}}\tau_1(1-\tau_2)p(\sigma_2)_{\alpha}p_+^{\alpha} - \frac{1-\beta}{1-\beta \tau_{\circ}}\tau_2 (1-\tau_1)p(\sigma_1)_{\alpha}p_{+}^{\alpha} +\\+ \frac{1}{1-\beta \tau_{\circ}} (1-\tau_1) (1-\tau_2) p(\sigma_2)_{\alpha} p(\sigma_1)^{\alpha} - p_{1\alpha} p_2^{\alpha} - (p_{1\alpha} + p_{2\alpha}) (p_{\omega1}^{\alpha}+p_{\omega2}^{\alpha}) - p_{\omega1 \alpha} p_{\omega2}^{\alpha}\,,
    \end{multline}
    where $$\tau_{\circ} = \tau_1(1-\tau_2)+\tau_2(1-\tau_1)\,.$$\par
    Upon integration, the pre-exponent in \eqref{vertex-1} amounts to
    \begin{multline}
        2\tau_1 \tau_2 \dd \tau_1 \dd \tau_2  (\dd \Omega_{12})^2 \Bigg [ \frac{1}{(1-\beta \tau_{\circ})^4}\bigg((1-\tau_2)(-p(\sigma_{12})+(1-\beta)(p(\sigma_2)+p(\sigma_1)- p_+))-
        \\
        - \tau_2 (y - p(\sigma_1)) \bigg)^{\alpha} \bigg((1-\tau_1)(-p(\sigma_{12})+(1-\beta)(p(\sigma_1)+p(\sigma_2)-p_+))+ \tau_1 (y + p(\sigma_2)) \bigg)_{\alpha} - 
        \\
        - \frac{2i}{(1-\beta \tau_{\circ})^3} \Bigg ] = \dd \tau_1 \dd \tau_2 \dd \sigma_{12}^{\omega_1} \dd \sigma_{12}^{\omega_2} p_{\omega_1 \alpha} p_{\omega_2}^{\alpha} \frac{4 \tau_1 \tau_2 }{(1-\beta \tau_{\circ})^4}\Big( \tau_1 \tau_2  (-y + p_2+p_{\omega1}+p_{\omega2})_{\alpha}(p_1+p_2)^{\alpha}+
        \\
        +\big[-p_{\omega_1}(\sigma_{12}^{\omega_1}+(1-\beta))-p_{\omega_2}(\sigma_{12}^{\omega_2}+(1-\beta)) \big]_{\alpha}\big[\tau_{\circ} (y-p_{\omega1}-p_{\omega2}) + (1-\tau_2)\tau_1 p_1 - (1-\tau_1)\tau_2 p_2 \big]^{\alpha} -
        \\
        -2i (1-\beta \tau_{\circ}) \Big)\,,
        \label{preexp1}
    \end{multline}
    where $-(1-\beta)\leq \sigma_{12}^{\omega2} \leq \sigma_{12}^{\omega1} \leq 0$.\par
    Each term in \eqref{preexp1} converges absolutely when integrated over $\tau_1$ and $\tau_2$ and the  limit $\beta \rightarrow -\infty$ of the integral is finite. It should be stressed that in this limit the integral is sensitive to the behavior of the integrand only in the vicinity of $\tau_{\circ} \rightarrow 0$. To find the limiting exponent, we decompose it into a series. Then we get a sum of terms of the form $ \frac{1}{(1-\beta \tau_{\circ})^k} \tau_1^{m_1} \tau_2^{m_2} (1-\tau_1)^{n_1}(1-\tau_2)^{n_2} $, that are integrated over $\tau_1$ and $\tau_2$.\par
    To find the limiting expression, we boil integration over $\tau_1$ and $\tau_2$ down to integration over $\tau = \tau_{\circ}$ and use for the function
    \begin{equation}
        J_{n_1, n_2}^{m_1, m_2} (\tau) = \int_0^1 \dd \tau_1 \int_0^1 \dd \tau_2 \quad \tau_1^{m_1} \tau_2^{m_2} (1-\tau_1)^{n_1}(1-\tau_2)^{n_2} \delta (\tau - \tau_{\circ})
        \label{J(n,m)}
    \end{equation}
    the formula obtained in \cite{Didenko:2019xzz}:
    \begin{equation}
        J_{n_1, n_2}^{m_1, m_2} (\tau) = \left\{\begin{aligned}
        \frac{m_1!m_2!}{(m_1+m_2+1)!}\tau^{m_1+m_2+1} + o(\tau^{m_1+m_2+1})\,,\quad m_1+m_2 < n_1+n_2\,, \\
        \frac{n_1!n_2!}{(n_1+n_2+1)!}\tau^{n_1+n_2+1} + o(\tau^{n_1+n_2+1})\,,\quad m_1+m_2 > n_1+n_2\,, \\
        2\frac{m_1!m_2!}{(m_1+m_2+1)!}\tau^{m_1+m_2+1} + o(\tau^{m_1+m_2+1})\,,\quad m_1+m_2 = n_1+n_2\,.
        \end{aligned} \right.
        \label{J_circ}
    \end{equation}
    The coefficients on the {\it r.h.s.} of \eqref{J_circ} can be expressed via beta-function,
    \begin{equation}
        B(k+1, l+1) = \frac{k!l!}{(k+l+1)!} = \int_0^1 \lambda^k (1-\lambda)^l \dd\lambda.
        \label{Beta}
    \end{equation}

    Thus we reduce the integration over $\tau_1$ and $\tau_2$ to that over $\tau$ and $\lambda$. Then we introduce  new variables $\tau_1 = \frac{-\b \tau }{1-\b \tau}\lambda, \tau_2 = \frac{-\b \tau }{1-\b \tau}(1-\lambda)$ (note that these variables differ from $\tau_1$ and $\tau_2$ in \eqref{J(n,m)}).  \par
    As a result, we obtain in the limit,  (with $C(r)$ and $\omega(r)$ implicit):
    \begin{multline}
        \Upsilon_1 = -\dfrac{\eta^2}{4}(p_{\omega_1 \alpha} p_{\omega_2}^{\alpha})\int\limits
        _\mathclap{{[ 0, 1 ]^2}}\dd \sigma_2 \dd \sigma_1 \int\limits
        _\mathclap{{[ 0, 1 ]^2}} \dd \tau_1 \dd \tau_2 \Big( \big[ p_{\omega1}\sigma_1 + p_{\omega2}\sigma_2 \big]_{\alpha}\big[(\tau_1+\tau_2) (-y+p_{\omega1}+p_{\omega2}) -\tau_2 p_1 + \tau_1 p_2 \big]^{\alpha} +\\+ (1-\tau_1-\tau_2)(-y + p_2+p_{\omega1}+p_{\omega2})_{\alpha}(p_1+p_2)^{\alpha} - 2i \Big) \exp i \big [ (p_{\omega1}\sigma_1 + p_{\omega2}\sigma_2 )_{\alpha}((\tau_1+\tau_2) (-y+p_{\omega1}+p_{\omega2}) -\\\!\!\!\!-\tau_2 p_1 + \tau_1 p_2 )^{\alpha} + (1-\tau_1-\tau_2)(-y + p_2+p_{\omega1}+p_{\omega2})_{\alpha}(p_1+p_2)^{\alpha} - y_{\alpha}(p_{\omega1}+p_{\omega2})^{\alpha}-p_{\omega1\alpha}p_{\omega2}^{\alpha} \big]
        \label{A0}
    \end{multline}
    ($\sigma_i = \frac{\sigma_{12}^{\omega i}}{1-\beta}+1$).
    \par
    A pre-exponent of \eqref{vertex-2} is  $A_1+A_2$ with
    \begin{equation}
        A_1 = -2 \dd \tau_1 \dd \sigma_1^1  (1-\tau_1)(\dd \Omega_{12})^2 \Bigg [ (\Omega_{12}+p_2+p_{\omega1}+p_{\omega2}+s)^{\alpha}(p_1+p_2)_{\alpha} \Bigg ],
        \label{preexp21}
    \end{equation}
    \begin{equation}
        A_2 = -2 \dd \tau_1 \dd \sigma_1^1  (1-\tau_1)(\dd \Omega_{12})^2 \Bigg [ (\Omega_{12}+p_2+p_{\omega1}+p_{\omega2}+s)^{\alpha}(-y-p_+-s-u_{12})_{\alpha} \Bigg ].
        \label{preexp22}
    \end{equation}
    According to this decomposition, \eqref{vertex-2} is decomposed as $\Upsilon_2 = \Upsilon_2^1 + \Upsilon_2^2$.\par
    Integration of \eqref{preexp21} over $s, t, u_{12}, v_{12}$  yields
    \begin{equation}
        A_1 = -2 \dd \tau_1 \dd \sigma_1^1  (1-\tau_1)(\dd \Omega_{12})^2 \frac{1}{(1-\beta (1-\tau_1))^3} \Bigg [ (-y+p_2+p_{\omega1}+p_{\omega2})^{\alpha}(p_1+p_2)_{\alpha} \Bigg ]\,.
    \end{equation}
    Taking the limit $\beta \rightarrow -\infty$ we find ($C(r)$ and $\omega(r)$ are implicit)
    \begin{multline}
        \Upsilon_2^1 = \dfrac{\eta^2}{4}(p_{\omega_1 \alpha} p_{\omega_2}^{\alpha})\int\limits
        _\mathclap{{[ 0, 1 ]^2}}\dd \sigma_2 \dd \sigma_1 \int\limits
        _\mathclap{{[ 0, 1 ]^2}} \dd \tau_1 \dd \tau_2  (-y + p_2+p_{\omega1}+p_{\omega2})_{\alpha}(p_1+p_2)^{\alpha} \exp i \big [ (p_{\omega1}\sigma_1 + p_{\omega2}\sigma_2 )_{\alpha}((\tau_1+\\+\tau_2) (-y+p_{\omega1}+p_{\omega2})-\tau_2 p_1 + \tau_1 p_2 )^{\alpha} + (1-\tau_1-\tau_2)(-y + p_2+p_{\omega1}+p_{\omega2})_{\alpha}(p_1+p_2)^{\alpha} - 
        \\
        -y_{\alpha}(p_{\omega1}+p_{\omega2})^{\alpha}-p_{\omega1\alpha}p_{\omega2}^{\alpha} \big].
        \label{A1}
    \end{multline}
    \par
    We can see that \eqref{A1} cancels the term with pre-exponential factor $(-y + p_2+p_{\omega1}+p_{\omega2})_{\alpha}(p_1+p_2)^{\alpha}$ in \eqref{A0}. Analogously to \cite{Didenko:2019xzz} we perform the following steps:
    \begin{multline}
        \Upsilon_1+\Upsilon_2^1 =
        \\
        = -\dfrac{\eta^2}{4}(p_{\omega_1 \alpha} p_{\omega_2}^{\alpha})\int\limits
        _\mathclap{{[ 0, 1 ]^2}}\dd \sigma_2 \dd \sigma_1 \int\limits
        _\mathclap{{[ 0, 1 ]^2}} \dd \tau_1 \dd \tau_2 \Big( \big[ p_{\omega1}\sigma_1 + p_{\omega2}\sigma_2 \big]_{\alpha}\big[(\tau_1+\tau_2) (-y+p_{\omega1}+p_{\omega2}) -\tau_2 p_1 + \tau_1 p_2 \big]^{\alpha} -\\- (\tau_1+\tau_2)(-y + p_2+p_{\omega1}+p_{\omega2})_{\alpha}(p_1+p_2)^{\alpha} - 2i \Big) \exp i \big [ (p_{\omega1}\sigma_1 + p_{\omega2}\sigma_2 )_{\alpha}((\tau_1+\tau_2) (-y+p_{\omega1}+p_{\omega2}) -\\-\tau_2 p_1 + \tau_1 p_2 )^{\alpha} - (1-\tau_1-\tau_2)(-y + p_2+p_{\omega1}+p_{\omega2})_{\alpha}(p_1+p_2)^{\alpha} - y_{\alpha}(p_{\omega1}+p_{\omega2})^{\alpha}-p_{\omega1\alpha}p_{\omega2}^{\alpha} \big] =\\ = \dfrac{i\eta^2}{4} (p_{\omega_1 \alpha} p_{\omega_2}^{\alpha})\int\limits
        _\mathclap{{[ 0, 1 ]^2}}\dd \sigma_2 \dd \sigma_1 \int\limits
        _\mathclap{{[ 0, 1 ]^2}} \dd \tau_1 \dd \tau_2 \Big( \tau_1 \frac{\partial}{\partial \tau_1}+\tau_2 \frac{\partial}{\partial \tau_2} + 2 \Big) \times 
        \\
        \times \exp i \big [ (p_{\omega1}\sigma_1 + p_{\omega2}\sigma_2 )_{\alpha}((\tau_1+\tau_2) (-y+p_{\omega1}+p_{\omega2}) -\\-\tau_2 p_1 + \tau_1 p_2 )^{\alpha} - (1-\tau_1-\tau_2)(-y + p_2+p_{\omega1}+p_{\omega2})_{\alpha}(p_1+p_2)^{\alpha} - y_{\alpha}(p_{\omega1}+p_{\omega2})^{\alpha}-p_{\omega1\alpha}p_{\omega2}^{\alpha} \big] = \\ = \dfrac{i\eta^2}{4} (p_{\omega_1 \alpha} p_{\omega_2}^{\alpha})\int\limits
        _\mathclap{{[ 0, 1 ]^2}}\dd \sigma_2 \dd \sigma_1 \fint{\mathbb{R}^3}\dd \tau_1 \dd \tau_2 \dd \tau_3 \delta(1-\tau_1-\tau_2-\tau_3)\theta(\tau_1)\theta(\tau_2)\theta(\tau_3) \left( \tau_1 \frac{\partial}{\partial \tau_1}+\tau_2 \frac{\partial}{\partial \tau_2} + 2 \right) \E.
    \end{multline}

    Integration by parts yields
    \begin{multline}
        \Upsilon_1+\Upsilon_2^1 =
        \\
        = -\dfrac{i\eta^2}{4}(p_{\omega_1 \alpha} p_{\omega_2}^{\alpha}) \int\limits
        _\mathclap{{[ 0, 1 ]^2}}\dd \sigma_2 \dd \sigma_1\fint{\mathbb{R}^3}\dd \tau_1 \dd \tau_2 \dd \tau_3 \E  \left( \tau_1 \frac{\partial}{\partial \tau_1}+ \tau_2\frac{\partial}{\partial \tau_2} \right) \delta(1-\tau_1-\tau_2-\tau_3)\theta(\tau_1)\theta(\tau_2)\theta(\tau_3)  = \\ = -\dfrac{i\eta^2}{4}(p_{\omega_1 \alpha} p_{\omega_2}^{\alpha}) \int\limits
        _\mathclap{{[ 0, 1 ]^2}}\dd \sigma_2 \dd \sigma_1\fint{\mathbb{R}^3}\dd \tau_1 \dd \tau_2 \dd \tau_3 \E  \left(  (\tau_1 + \tau_2)\frac{\partial}{\partial \tau_3}  \delta(1-\tau_1-\tau_2-\tau_3)\theta(\tau_1)\theta(\tau_2)\right)\theta(\tau_3)  = \\ = \dfrac{i\eta^2}{4}(p_{\omega_1 \alpha} p_{\omega_2}^{\alpha}) \int\limits
        _\mathclap{{[ 0, 1 ]^2}}\dd \sigma_2 \dd \sigma_1\fint{[0, 1]}\dd \tau_1 \fint{[0, 1]} \dd \tau_2 \delta(1-\tau_1-\tau_2) \exp i \big [ (p_{\omega1}\sigma_1 + p_{\omega2}\sigma_2 )_{\alpha}((\tau_1+\tau_2) (-y+p_{\omega1}+p_{\omega2}) -
        \\
        -\tau_2 p_1 + \tau_1 p_2 )^{\alpha} - (1-\tau_1-\tau_2)(-y + p_2+p_{\omega1}+p_{\omega2})_{\alpha}(p_1+p_2)^{\alpha} - y_{\alpha}(p_{\omega1}+p_{\omega2})^{\alpha}-p_{\omega1\alpha}p_{\omega2}^{\alpha} \big].
        \label{rest}
    \end{multline}\par
    Now consider $\Upsilon_2^2$ along with $\Upsilon_3$ (in the following two equations, $\sigma_1^2$ in $\E$ is assumed to be equal to $\sigma_1^1+1$.).
    \begin{multline}
        \Upsilon_2^2 = \dfrac{\eta^2}{8} \fint{\sigma_{12}^{\omega1} \sigma_{12}^{\omega2}}P(-\overrightarrow{\sigma_{12}},\sigma_{12}^{\omega2}, \sigma_{12}^{\omega1},\sigma_{12}^2)  \int_0^1 \dd \tau_1 \int_{-1}^0 \dd \sigma_1^1  (1-\tau_1)(\dd \Omega_{12})^2 \E \times \\ \times \Bigg [ (\Omega_{12}+p_2+p_{\omega1}+p_{\omega2}+s)^{\alpha}(-y-p_+-s-u_{12})_{\alpha} \Bigg ] = \dfrac{\eta^2}{8i} \fint{\sigma_{12}^{\omega1} \sigma_{12}^{\omega2}}P(-\overrightarrow{\sigma_{12}},\sigma_{12}^{\omega2}, \sigma_{12}^{\omega1},\sigma_{12}^2)  \int_0^1 \dd \tau_1  (\dd \Omega_{12})^2 \times \\ \times \Bigg [ \frac{(\Omega_{12}+p_2+p_{\omega1}+p_{\omega2}+s)^{\alpha}(-y-p_+-s-u_{12})_{\alpha}}{(p_1+p_2)_{\alpha}(-y-p_+-s-u_{12})^{\alpha}} \Bigg ](\E|_{\sigma_1^1=0}-\E|_{\sigma_1^1=-1}),
    \end{multline}
    \begin{multline}
        \Upsilon_2^2 + \Upsilon_3  = \dfrac{\eta^2}{8i} \fint{\sigma_{12}^{\omega1} \sigma_{12}^{\omega2}}P(-\overrightarrow{\sigma_{12}},\sigma_{12}^{\omega2}, \sigma_{12}^{\omega1},\sigma_{12}^2)  \int_0^1 \dd \tau_1  (\dd \Omega_{12})^2 \times \\ \times \Bigg [ \frac{(\Omega_{12}+p_2+p_{\omega1}+p_{\omega2}+s)_{\alpha}(-y-p_+-s-u_{12})^{\alpha}}{(p_1+p_2)_{\alpha}(-y-p_+-s-u_{12})^{\alpha}} \E|_{\sigma_1^1=0}-\\- \frac{(\Omega_{12}-p_1+p_{\omega1}+p_{\omega2}+s)_{\alpha}(-y-p_+-s-u_{12})^{\alpha}}{(p_1+p_2)_{\alpha}(-y-p_+-s-u_{12})^{\alpha}}\E|_{\sigma_1^1=-1}\Bigg ] = \\ = \dfrac{\eta^2}{8} \fint{\sigma_{12}^{\omega1} \sigma_{12}^{\omega2}}P(-\overrightarrow{\sigma_{12}},\sigma_{12}^{\omega2}, \sigma_{12}^{\omega1},\sigma_{12}^2)   (\dd \Omega_{12})^2  \Bigg [ \frac{\E|_{\sigma_1^1=0,\tau_1=1}-\E|_{\sigma_1^1=0,\tau_1=0}-\E|_{\sigma_1^1=-1,\tau_1=1}+\E|_{\sigma_1^1=-1,\tau_1=0}}{(p_1+p_2)_{\alpha}(-y-p_+-s-u_{12})^{\alpha}}\Bigg ]=\\=-\dfrac{\eta^2}{8} \fint{\sigma_{12}^{\omega1} \sigma_{12}^{\omega2}}P(-\overrightarrow{\sigma_{12}},\sigma_{12}^{\omega2}, \sigma_{12}^{\omega1},\sigma_{12}^2) (\dd \Omega_{12})^2  \Bigg [ \frac{\E|_{\sigma_1^1=0,\tau_1=0}-\E|_{\sigma_1^1=-1,\tau_1=0}}{(p_1+p_2)_{\alpha}(-y-p_+-s-u_{12})^{\alpha}}\Bigg ] =\\ =-\dfrac{i\eta^2}{8} \fint{\sigma_{12}^{\omega1} \sigma_{12}^{\omega2}}P(-\overrightarrow{\sigma_{12}},\sigma_{12}^{\omega2}, \sigma_{12}^{\omega1},\sigma_{12}^2) \int_{-1}^0 \dd \sigma_1^1 (\dd \Omega_{12})^2  \E|_{\tau_1=0}.
        \label{delta}
    \end{multline}
    The limit $\beta \rightarrow -\infty$ in \eqref{delta} can be taken simply at $\tau_1 = 0$.
    It produces \eqref{rest} with the opposite sign.
    \par
    Thus, only one term \eqref{vertex-4} remains in the vertex, where a pre-exponent is
    \begin{multline}
        4\tau_1 (1-\tau_2) \dd \tau_1 \dd \tau_2  \dd \sigma_{12}^{\omega2} \dd \sigma_2^{\omega1} \Bigg [ \frac{1}{(1-\beta (\tau_1(1-\tau_2)+\tau_2(1-\tau_1)))^4}p_{\omega2\alpha}\Bigg((1-\tau_2)(1-\beta)(\sigma_2^{\omega_1} p_{\omega1})-\\- \tau_2 (y - p_2-p_{\omega1}) \Bigg)^{\alpha} p_{\omega1\beta}\Bigg((1-\tau_1)((1-\beta)+\sigma_{12}^{\omega2})(- p_{\omega2})+ \tau_1 (y + p_1+p_{\omega2} )\Bigg)^{\beta} -\\- \frac{2i}{(1-\beta (\tau_1(1-\tau_2)+\tau_2(1-\tau_1)))^3} \Bigg ],
    \end{multline}
    where $-(1-\beta)\leq \sigma_{12}^{\omega2} \leq 0, -1\leq \sigma_{2}^{\omega1} \leq 0$.
    The leading term at large $-\beta$ is
    \begin{equation}
        \dd \tau_1 \dd \tau_2  \dd \sigma_{12}^{\omega2} \dd \sigma_2^{\omega1} \frac{4\tau_1 (1-\tau_1) (1-\tau_2)^2 (1-\beta)((1-\beta)+\sigma_{12}^{\omega2})\sigma_2^{\omega1}}{(1-\beta (\tau_1(1-\tau_2)+\tau_2(1-\tau_1)))^4}(p_{\omega1\alpha}p_{\omega2}^{\alpha})^2.
        \label{predexp_4}
    \end{equation}
    The rest of terms vanish in the limit $\beta \rightarrow -\infty$.\par
    Taking analogously the limit $\beta \rightarrow -\infty$ in \eqref{predexp_4}
     yields \eqref{vertex-lim}.
\end{appendices}
\noindent

\addcontentsline{toc}{section}{References}
\section*{}

\end{document}